\pdfoutput=1
\documentclass[12pt,a4paper]{article}
\usepackage{hyperref}

\usepackage{ifthen} % for conditional statements
\newboolean{pdflatex}
\setboolean{pdflatex}{true} % False for eps figures 

\newboolean{articletitles}
\setboolean{articletitles}{true} % False removes titles in references

\newboolean{uprightparticles}
\setboolean{uprightparticles}{true} %True for upright particle symbols

\def\paperauthors{LHCb collaboration} % Leave as is for PAPER, CONF and FIGURE
\def\papercopyright{\the\year\ CERN for the benefit of the LHCb collaboration} % new since 9/Apr/2018
\def\paperlicence{CC BY 4.0 licence}
\def\paperlicenceurl{https://creativecommons.org/licenses/by/4.0/}

\usepackage{multirow}
\usepackage{lscape}
\usepackage[normalem]{ulem} 
\usepackage{cancel}
\usepackage[title]{appendix}
\usepackage{rotating}
\usepackage{verbatim} 
\usepackage{afterpage} 
\usepackage{nicefrac} 
\usepackage{mathrsfs} 
\usepackage{graphpap}

\makeatletter
\g@addto@macro\bfseries{\boldmath}
\makeatother

\usepackage[top=1in, bottom=1.25in, left=1in, right=1in]{geometry}

\usepackage{microtype}
\usepackage{lineno}  % for line numbering during review
\usepackage{xspace} % To avoid problems with missing or double spaces after
\usepackage{caption} %these three command get the figure and table captions automatically small
\usepackage{multirow}

\usepackage{graphicx}  % to include figures (can also use other packages)
\usepackage{color}
\usepackage{colortbl}
\graphicspath{{./figs/}} % Make Latex search fig subdir for figures
\usepackage{amsmath} % Adds a large collection of math symbols
\usepackage{amssymb}
\usepackage{amsfonts}
\usepackage{upgreek} % Adds in support for greek letters in roman typeset

\newcommand*\patchAmsMathEnvironmentForLineno[1]{%
\expandafter\let\csname old#1\expandafter\endcsname\csname #1\endcsname
\expandafter\let\csname oldend#1\expandafter\endcsname\csname
end#1\endcsname
 \renewenvironment{#1}%
   {\linenomath\csname old#1\endcsname}%
   {\csname oldend#1\endcsname\endlinenomath}%
}
\newcommand*\patchBothAmsMathEnvironmentsForLineno[1]{%
  \patchAmsMathEnvironmentForLineno{#1}%
  \patchAmsMathEnvironmentForLineno{#1*}%
}
\AtBeginDocument{%
\patchBothAmsMathEnvironmentsForLineno{equation}%
\patchBothAmsMathEnvironmentsForLineno{align}%
\patchBothAmsMathEnvironmentsForLineno{flalign}%
\patchBothAmsMathEnvironmentsForLineno{alignat}%
\patchBothAmsMathEnvironmentsForLineno{gather}%
\patchBothAmsMathEnvironmentsForLineno{multline}%
\patchBothAmsMathEnvironmentsForLineno{eqnarray}%
}

\usepackage{hyperxmp}

\usepackage[%% pdftex,
			]{hyperref}
\usepackage[colorinlistoftodos,textsize=scriptsize]{todonotes}

\usepackage[bottom,flushmargin,hang,multiple]{footmisc}

\usepackage[all]{hypcap} % Internal hyperlinks to floats.

\usepackage{xspace} 
\usepackage{upgreek}

\def\lhcb   {\mbox{LHCb}\xspace}
\def\atlas  {\mbox{ATLAS}\xspace}
\def\cms    {\mbox{CMS}\xspace}

\def\babar  {\mbox{BaBar}\xspace}
\def\belle  {\mbox{Belle}\xspace}
\def\belletwo {\mbox{Belle~II}\xspace}
\def\besiii {\mbox{BESIII}\xspace}

\def\cdf    {\mbox{CDF}\xspace}
\def\dzero  {\mbox{D0}\xspace}

\def\MagUp {\mbox{\em Mag\kern -0.05em Up}\xspace}

\ifthenelse{\boolean{uprightparticles}}%
{
 
 \def\Pgamma      {\ensuremath{\upgamma}\xspace}

 \def\Pvarepsilon {\ensuremath{\upvarepsilon}\xspace}                 
                  
 \def\Peta        {\ensuremath{\upeta}\xspace}

 \def\Pmu         {\ensuremath{\upmu}\xspace}

 \def\Ppi         {\ensuremath{\uppi}\xspace}                 
                  
 \def\Prho        {\ensuremath{\uprho}\xspace}

 \def\Pchi        {\ensuremath{\upchi}\xspace}                 
 \def\Ppsi        {\ensuremath{\uppsi}\xspace}                 
 \def\Pomega      {\ensuremath{\upomega}\xspace}                 

 \def\PDelta      {\ensuremath{\Delta}\xspace}                 
 \def\PXi         {\ensuremath{\Xi}\xspace}                 
 \def\PLambda     {\ensuremath{\Lambda}\xspace}                 
 \def\PSigma      {\ensuremath{\Sigma}\xspace}                 
 \def\POmega      {\ensuremath{\Omega}\xspace}                 
 \def\PUpsilon    {\ensuremath{\Upsilon}\xspace}
 \let\oldPi\Pi
 \def\PPi         {\ensuremath{\oldPi}\xspace}

 \def\PB      {\ensuremath{\mathrm{B}}\xspace}                 
 \def\PC      {\ensuremath{\mathrm{C}}\xspace}                 
 \def\PD      {\ensuremath{\mathrm{D}}\xspace}

 \def\PJ      {\ensuremath{\mathrm{J}}\xspace}                 
 \def\PK      {\ensuremath{\mathrm{K}}\xspace}

 \def\PP      {\ensuremath{\mathrm{P}}\xspace}

 \def\PS      {\ensuremath{\mathrm{S}}\xspace}

 \def\PX      {\ensuremath{\mathrm{X}}\xspace}

 \def\Pb      {\ensuremath{\mathrm{b}}\xspace}                 
 \def\Pc      {\ensuremath{\mathrm{c}}\xspace}                 
                  
 \def\Pe      {\ensuremath{\mathrm{e}}\xspace}

 \def\Pi      {\ensuremath{\mathrm{i}}\xspace}

 \def\Pm      {\ensuremath{\mathrm{m}}\xspace}

 \def\Pp      {\ensuremath{\mathrm{p}}\xspace}                 
 \def\Pq      {\ensuremath{\mathrm{q}}\xspace}                 
                  
 \def\Ps      {\ensuremath{\mathrm{s}}\xspace}

 \def\thebaroffset{0.0em}
}
{
 
 \def\Pgamma      {\ensuremath{\gamma}\xspace}

 \def\Pvarepsilon {\ensuremath{\varepsilon}\xspace}                 
                  
 \def\Peta        {\ensuremath{\eta}\xspace}

 \def\Pmu         {\ensuremath{\mu}\xspace}

 \def\Ppi         {\ensuremath{\pi}\xspace}                 
                  
 \def\Prho        {\ensuremath{\rho}\xspace}

 \def\Pchi        {\ensuremath{\chi}\xspace}                 
 \def\Ppsi        {\ensuremath{\psi}\xspace}                 
 \def\Pomega      {\ensuremath{\omega}\xspace}                 
 \mathchardef\PDelta="7101
 \mathchardef\PXi="7104
 \mathchardef\PLambda="7103
 \mathchardef\PSigma="7106
 \mathchardef\POmega="710A
 \mathchardef\PUpsilon="7107
 \mathchardef\PPi="7105
                  
 \def\PB      {\ensuremath{B}\xspace}                 
 \def\PC      {\ensuremath{C}\xspace}                 
 \def\PD      {\ensuremath{D}\xspace}

 \def\PJ      {\ensuremath{J}\xspace}                 
 \def\PK      {\ensuremath{K}\xspace}

 \def\PP      {\ensuremath{P}\xspace}

 \def\PS      {\ensuremath{S}\xspace}

 \def\PX      {\ensuremath{X}\xspace}

 \def\Pb      {\ensuremath{b}\xspace}                 
 \def\Pc      {\ensuremath{c}\xspace}                 
                  
 \def\Pe      {\ensuremath{e}\xspace}

 \def\Pi      {\ensuremath{i}\xspace}

 \def\Pm      {\ensuremath{m}\xspace}

 \def\Pp      {\ensuremath{p}\xspace}                 
 \def\Pq      {\ensuremath{q}\xspace}                 
                  
 \def\Ps      {\ensuremath{s}\xspace}

 \def\thebaroffset{0.18em}
}
\newcommand{\offsetoverline}[2][\thebaroffset]{\kern #1\overline{\kern -#1 #2}}%

\makeatletter
\ifcase \@ptsize \relax% 10pt
  \newcommand{\miniscule}{\@setfontsize\miniscule{4}{5}}% \tiny: 5/6
\or% 11pt
  \newcommand{\miniscule}{\@setfontsize\miniscule{5}{6}}% \tiny: 6/7
\or% 12pt
  \newcommand{\miniscule}{\@setfontsize\miniscule{5}{6}}% \tiny: 6/7
\fi
\makeatother

\DeclareRobustCommand{\optbar}[1]{\shortstack{{\miniscule (\rule[.5ex]{1.25em}{.18mm})}
  \\ [-.7ex] $#1$}}

\def\epem       {{\ensuremath{\Pe^+\Pe^-}}\xspace}
\def\mumu       {{\ensuremath{\Pmu^+\Pmu^-}}\xspace}

\def\g      {{\ensuremath{\Pgamma}}\xspace}

\def\quark     {{\ensuremath{\Pq}}\xspace}
\def\quarkbar  {{\ensuremath{\overline \quark}}\xspace}

\def\squark    {{\ensuremath{\Ps}}\xspace}

\def\cquark    {{\ensuremath{\Pc}}\xspace}
\def\cquarkbar {{\ensuremath{\overline \cquark}}\xspace}
\def\ccbar     {{\ensuremath{\cquark\cquarkbar}}\xspace}
\def\bquark    {{\ensuremath{\Pb}}\xspace}

\def\pion   {{\ensuremath{\Ppi}}\xspace}
\def\piz    {{\ensuremath{\pion^0}}\xspace}
\def\pip    {{\ensuremath{\pion^+}}\xspace}
\def\pim    {{\ensuremath{\pion^-}}\xspace}

\def\kaon    {{\ensuremath{\PK}}\xspace}
\def\KorKbar {\kern \thebaroffset\optbar{\kern -\thebaroffset \PK}{}\xspace}

\def\Kp      {{\ensuremath{\kaon^+}}\xspace}
\def\Km      {{\ensuremath{\kaon^-}}\xspace}

\def\Kstarz  {{\ensuremath{\kaon^{*0}}}\xspace}

\def\Kstarp  {{\ensuremath{\kaon^{*+}}}\xspace}

\def\Dbar    {{\ensuremath{\offsetoverline{\PD}}}\xspace}
\def\D       {{\ensuremath{\PD}}\xspace}
\def\Db      {{\ensuremath{\Dbar}}\xspace}
\def\DorDbar {\kern \thebaroffset\optbar{\kern -\thebaroffset \PD}\xspace}
\def\Dz      {{\ensuremath{\D^0}}\xspace}
\def\Dzb     {{\ensuremath{\Dbar{}^0}}\xspace}
\def\Dp      {{\ensuremath{\D^+}}\xspace}
\def\Dm      {{\ensuremath{\D^-}}\xspace}

\def\DpDm    {\ensuremath{\Dp {\kern -0.16em \Dm}}\xspace}

\def\Dstarb  {{\ensuremath{\Dbar{}^*}}\xspace}
\def\Dstarz  {{\ensuremath{\D^{*0}}}\xspace}
\def\Dstarzb {{\ensuremath{\Dbar{}^{*0}}}\xspace}

\def\Dstarp  {{\ensuremath{\D^{*+}}}\xspace}

\def\B       {{\ensuremath{\PB}}\xspace}

\def\BorBbar {\kern \thebaroffset\optbar{\kern -\thebaroffset \PB}\xspace}

\def\Bd      {{\ensuremath{\B^0}}\xspace}

\def\BdorBdbar {\kern \thebaroffset\optbar{\kern -\thebaroffset \Bd}\xspace}
\def\Bu      {{\ensuremath{\B^+}}\xspace}

\def\Bp      {{\ensuremath{\Bu}}\xspace}

\def\Bpm     {{\ensuremath{\B^\pm}}\xspace}

\def\Bs      {{\ensuremath{\B^0_\squark}}\xspace}

\def\BsorBsbar {\kern \thebaroffset\optbar{\kern -\thebaroffset \Bs}\xspace}

\def\jpsi     {{\ensuremath{{\PJ\mskip -3mu/\mskip -2mu\Ppsi}}}\xspace}
\def\psitwos  {{\ensuremath{\Ppsi{(2\PS)}}}\xspace}

\def\chicone  {{\ensuremath{\Pchi_{\cquark 1}}}\xspace}

\def\Y#1S{\ensuremath{\PUpsilon{(#1S)}}\xspace}

\def\proton      {{\ensuremath{\Pp}}\xspace}

\def\LorLbar     {\kern \thebaroffset\optbar{\kern -\thebaroffset \PLambda}\xspace}

\def\BF         {{\ensuremath{\mathcal{B}}}\xspace}
\def\BR         {\BF}

\newcommand{\decay}[2]{\ensuremath{#1\!\to #2}\xspace} 

\def\to                 {\ensuremath{\rightarrow}\xspace}

\def\AT#1     {\ensuremath{A_{\mathrm{T}}^{#1}}\xspace}           % 2

\def\C#1      {\ensuremath{\mathcal{C}_{#1}}\xspace}                       % 9
\def\Cp#1     {\ensuremath{\mathcal{C}_{#1}^{'}}\xspace}                    % 7
\def\Ceff#1   {\ensuremath{\mathcal{C}_{#1}^{\mathrm{(eff)}}}\xspace}        % 9  
\def\Cpeff#1  {\ensuremath{\mathcal{C}_{#1}^{'\mathrm{(eff)}}}\xspace}       % 7
\def\Ope#1    {\ensuremath{\mathcal{O}_{#1}}\xspace}                       % 2
\def\Opep#1   {\ensuremath{\mathcal{O}_{#1}^{'}}\xspace}                    % 7

\newcommand{\nospaceunit}[1]{\ensuremath{\text{#1}}}       
\newcommand{\aunit}[1]{\ensuremath{\text{\,#1}}}       
\newcommand{\tev}{\aunit{Te\kern -0.1em V}\xspace}
\newcommand{\gev}{\aunit{Ge\kern -0.1em V}\xspace}
\newcommand{\mev}{\aunit{Me\kern -0.1em V}\xspace}
\newcommand{\kev}{\aunit{ke\kern -0.1em V}\xspace}
\newcommand{\ev}{\aunit{e\kern -0.1em V}\xspace}
 
\newcommand{\mevc}{\ensuremath{\aunit{Me\kern -0.1em V\!/}c}\xspace}
\newcommand{\gevc}{\ensuremath{\aunit{Ge\kern -0.1em V\!/}c}\xspace}
\newcommand{\mevcc}{\ensuremath{\aunit{Me\kern -0.1em V\!/}c^2}\xspace}
\newcommand{\gevcc}{\ensuremath{\aunit{Ge\kern -0.1em V\!/}c^2}\xspace}
\def\mum  {\ensuremath{\,\upmu\nospaceunit{m}}\xspace}

\def\fb   {\ensuremath{\aunit{fb}}\xspace}
\def\invfb   {\ensuremath{\fb^{-1}}\xspace}

\def\gsim{{~\raise.15em\hbox{$>$}\kern-.85em
          \lower.35em\hbox{$\sim$}~}\xspace}
\def\lsim{{~\raise.15em\hbox{$<$}\kern-.85em
          \lower.35em\hbox{$\sim$}~}\xspace}

\def\pt         {\ensuremath{p_{\mathrm{T}}}\xspace}

\def\ptot       {\ensuremath{p}\xspace}

\def\evtgen     {\mbox{\textsc{EvtGen}}\xspace}

\def\geant      {\mbox{\textsc{Geant4}}\xspace}

\def\photos     {\mbox{\textsc{Photos}}\xspace}

\def\pythia     {\mbox{\textsc{Pythia}}\xspace}

\def\tell1  {TELL1\xspace}
\def\ukl1   {UKL1\xspace}

\newcommand{\ie}{\mbox{\itshape i.e.}\xspace}

\newcommand{\lhcborcid}[1]{\href{https://orcid.org/#1}{\hspace*{0.1em}\raisebox{-0.45ex}{\includegraphics[width=1em]{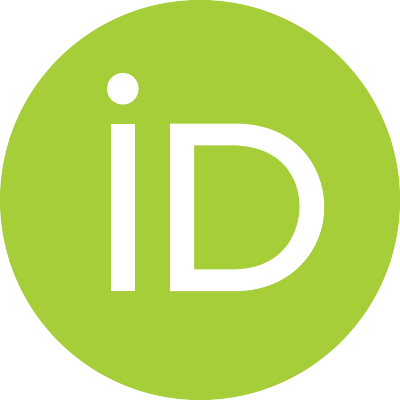}}}}

\usepackage{cite} % Allows for ranges in citations
\usepackage{mciteplus}
\DeclareMathOperator*{\bigplus}{\scalerel*{+}{\sum}}
\definecolor{myviolet}{rgb}{204,0,255}
\usepackage{scalerel}
\usepackage{comment}

\usepackage{rotating}
\usepackage{dashrule}
\usetikzlibrary{patterns}

\def\g{{\ensuremath{\Pgamma}}\xspace}
\def\X{{\ensuremath{\Pchi_{\cquark1}(3872)}}\xspace}
\def\PsiG{{\ensuremath{\psitwos\g}}\xspace}

\def\XPsiG{\decay{\X}{\PsiG}}

\usepackage{longtable} % only for template; not usually to be used in PAPERs
\usetikzlibrary{patterns}

\begin{document}

%%%%%%%%%%%%%%%%%%%%%%%%%
%%%%% Title     %%%%%%%%%
%%%%%%%%%%%%%%%%%%%%%%%%%
\renewcommand{\thefootnote}{\fnsymbol{footnote}}
\setcounter{footnote}{1}

% %%%%%%% CHOOSE TITLE PAGE--------
%\onecolumn
%\input{title-LHCb-INT}
%\input{title-LHCb-ANA}
%\input{title-LHCb-CONF}
%\input{title-LHCb-FIGURE}
% ===============================================================================
% Purpose: LHCb-PAPER journal paper title page template
% Author: 
% Created on: 2010-09-25
% ===============================================================================

%%%%%%%%%%%%%%%%%%%%%%%%%
%%%%%  TITLE PAGE  %%%%%%
%%%%%%%%%%%%%%%%%%%%%%%%%
\begin{titlepage}
\pagenumbering{roman}

% Header ---------------------------------------------------
\vspace*{-1.5cm}
\centerline{\large EUROPEAN ORGANIZATION FOR NUCLEAR RESEARCH (CERN)}
\vspace*{1.5cm}
\noindent
\begin{tabular*}{\linewidth}{lc@{\extracolsep{\fill}}r@{\extracolsep{0pt}}}
\ifthenelse{\boolean{pdflatex}}% Logo format choice
{\vspace*{-1.5cm}\mbox{\!\!\!\includegraphics[width=.14\textwidth]{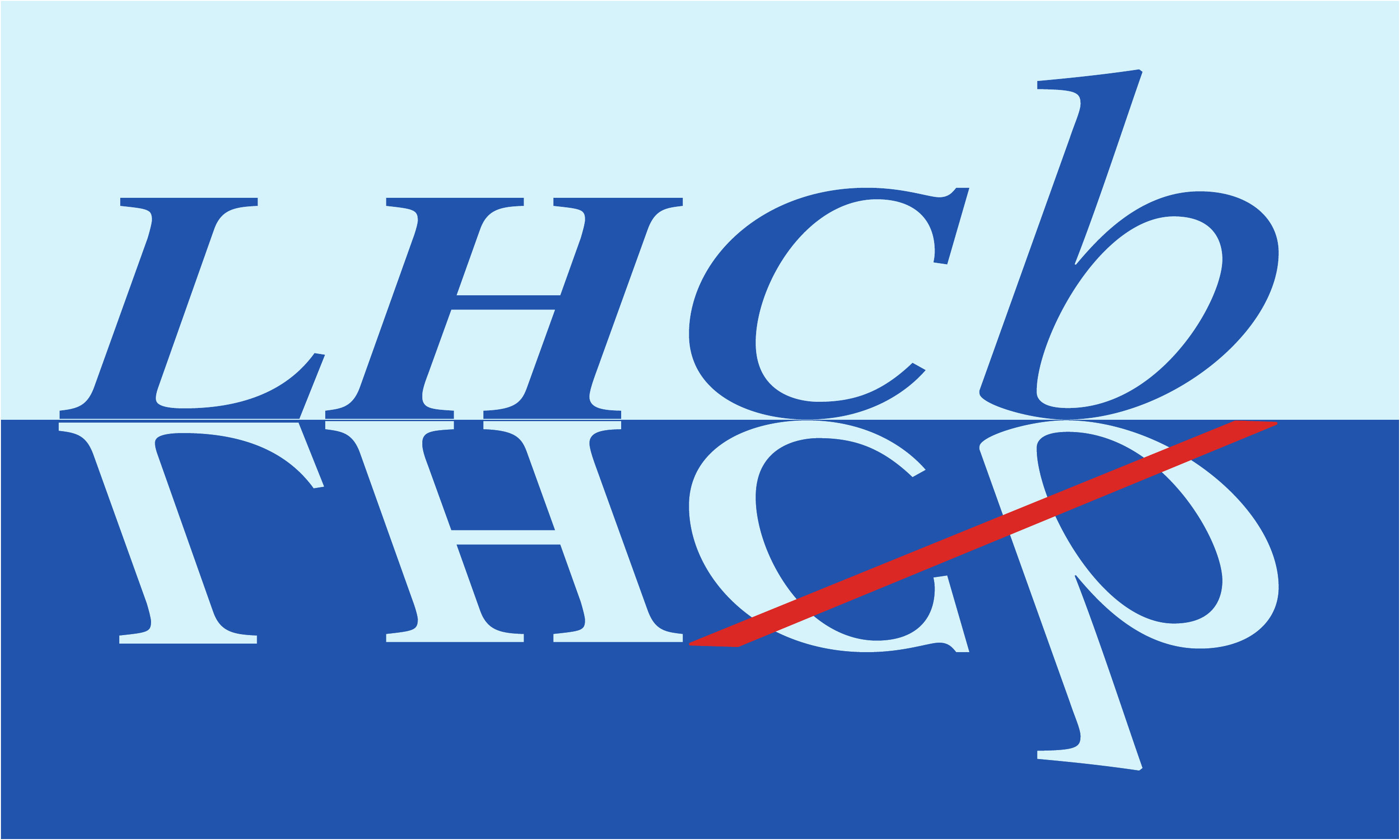}} & &}%
{\vspace*{-1.2cm}\mbox{\!\!\!\includegraphics[width=.12\textwidth]{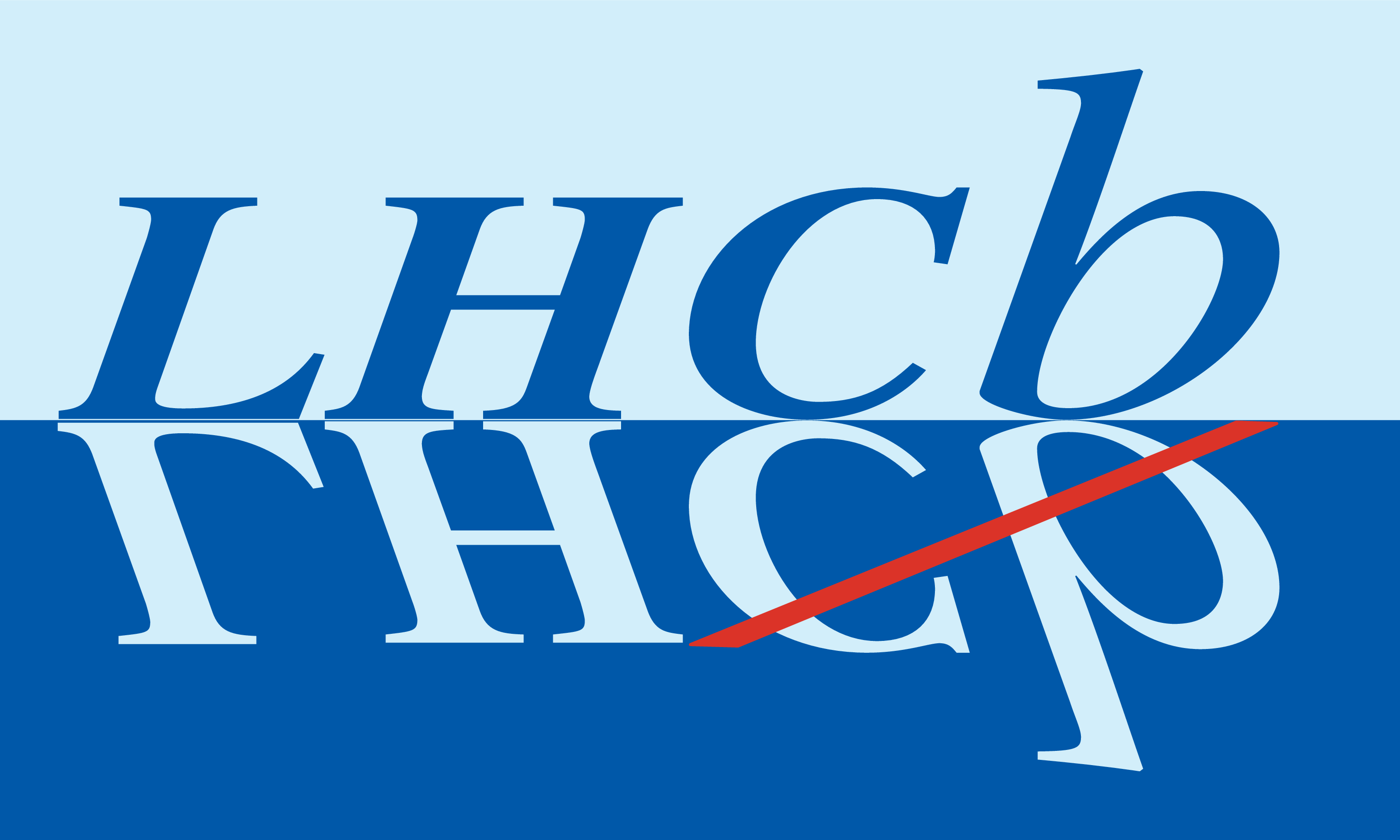}} & &}%
\\
 & & CERN-EP-2024-157 \\  % ID 
 & & LHCb-PAPER-2024-015 \\  % ID 
 %% & & \today \\ % Date - Can also hardwire e.g.: 23 March 2010
 & & June 21, 2024 \\ 
 %& & v1.8 \\ 
 & & \\
% not in paper \hline
\end{tabular*}

\vspace*{1.5cm}

% Title --------------------------------------------------
{\normalfont\bfseries\boldmath\huge
\begin{center}
% DO NOT EDIT HERE. Instead edit macro in main.tex to keep metadata correct
%%  \papertitle
%% tempporarily! 
Probing the~nature of the~$\chicone(3872)$~state
using radiative~decays
\end{center}
}

\vspace*{1.5cm}

% Authors -------------------------------------------------
\begin{center}
%In the footnote, replace 'paper' by 'Letter' in case of submission to PRL or PLB 
% Edit macro in main.tex to keep metadata correct
\paperauthors\footnote{Authors are listed at the end of this paper.}
\end{center}

\vspace{\fill}

% Abstract -----------------------------------------------
\begin{abstract}
\noindent
The radiative decays  ${\chicone(3872)}\to {\Ppsi{(2\PS)}\g}$
and  
${\chicone(3872)} \to {\PJ/\Ppsi\g}$ are used to probe the nature of the $\chicone(3872)$ state
using proton-proton collision data collected with the LHCb detector, corresponding to an integrated luminosity 
of $9\invfb$.
Using the ${\Bu}\to{\chicone(3872)\Kp}$ decay,
the ${\chicone(3872)}\to{\Ppsi{(2\PS)}\g}$ process is observed for the first time 
and the ratio of its partial width to that of the ${\chicone(3872)}\to{\PJ/\Ppsi \g}$ decay 
is measured to be 
$$\dfrac{\Gamma_{{\chicone(3872)}\to{\Ppsi{(2\PS)}\g}}} 
      {\Gamma_{{\chicone(3872)}\to{\PJ/\Ppsi\g}}} = 1.67 \pm 0.21 \pm 0.12 \pm0.04 \,,   
$$
where the first uncertainty is statistical,  the second systematic and the third 
is  due to the uncertainties on  the  branching fractions of the $\Ppsi{(2\PS)}$ and $\PJ/\Ppsi$ mesons. 
The measured ratio makes the interpretation of the $\chicone(3872)$ state as a pure 
$\Dz\overline{D}^{*0}+\overline{\D}^0\Dstarz$ molecule questionable
and strongly indicates   a sizeable compact charmonium or tetraquark 
component within the $\chicone(3872)$ state.
\end{abstract}

\vspace*{1.0cm}

\begin{center}
  %% Submitted to JHEP
  Published in \href{https://doi.org/10.1007/JHEP11(2024)121}{JHEP 11 (2024) 121}
\end{center}

\vspace{\fill}

{\footnotesize 
% Edit macro in main.tex to keep metadata correct
\centerline{\copyright~\papercopyright. \href{\paperlicenceurl}{\paperlicence}.}}
\vspace*{2mm}

\end{titlepage}

%%%%%%%%%%%%%%%%%%%%%%%%%%%%%%%%
%%%%%  EOD OF TITLE PAGE  %%%%%%
%%%%%%%%%%%%%%%%%%%%%%%%%%%%%%%%

%  empty page follows the title page ----
\newpage
\setcounter{page}{2}
\mbox{~}

%\twocolumn
% %%%%%%%%%%%%% ---------

%\listoftodos

\renewcommand{\thefootnote}{\arabic{footnote}}
\setcounter{footnote}{0}

%%%%%%%%%%%%%%%%%%%%%%%%%%%%%%%%
%%%%%  Table of Content   %%%%%%
%%%%%%%%%%%%%%%%%%%%%%%%%%%%%%%%
%%%% Uncomment if desired
%\tableofcontents
\cleardoublepage

%%%%%%%%%%%%%%%%%%%%%%%%%
%%%%% Main text %%%%%%%%%
%%%%%%%%%%%%%%%%%%%%%%%%%

\pagestyle{plain} % restore page numbers for the main text
\setcounter{page}{1}
\pagenumbering{arabic}

%% Uncomment during review phase. 
%% Comment before a final submission.
%% \linenumbers

%% This is the main body
%% It is useful to have a single file so comemnts are not missed in overleaf.
\section{Introduction}
\label{sec:intro}

Decays of beauty hadrons have proven to be a~convenient and 
fruitful tool in the~search and study of 
charmonium\nobreakdash-like  states, whose measured properties point 
to the~presence of a~$\ccbar$~component in  their quark content, 
but
%% do not allow to associate them with any conventional 
%% charmonium resonance
forbid them to be associated with any conventional charmonium
resonance~\cite{
Choi:2007wga,
Mizuk:2009da,
Chilikin:2013tch,
LHCb-PAPER-2014-014,
%% LHCb-PAPER-2015-029, %% penta
LHCb-PAPER-2015-038,
%% LHCb-PAPER-2016-009, %% penta
LHCb-PAPER-2016-015,
LHCb-PAPER-2016-018,
LHCb-PAPER-2016-019,
LHCb-PAPER-2018-034,
LHCb-PAPER-2018-043,
%% LHCb-PAPER-2019-014, %% penta
LHCb-PAPER-2020-035,
LHCb-PAPER-2020-044}.
The~first such state, $\chicone(3872)$, 
was observed 
in 2003 by the~\belle collaboration
in the~$\jpsi\pip\pim$~mass spectrum from 
\mbox{$\decay{\Bu}
{\jpsi\pip\pim\Kp}$}~decays~\cite{Belle:2003nnu}.
For~more than 
two decades
%% twenty years 
%% after this 
since its discovery,  
the~properties of this state have 
been intensively studied in 
$\epem$ collisions by 
the~\babar~\cite{
BaBar:2004iez,
BaBar:2004oro,
BaBar:2006fjg,
BaBar:2007ixp,
BaBar:2007cmo,
BaBar:2008qzi,
BaBar:2008flx,
BaBar:2010wfc},
\belle~\cite{
Belle:2003eeg,
Belle:2006olv,
Belle:2008fma,
Belle:2011vlx,
Belle:2011wdj,
Belle:2015qeg,
Belle:2022puc,
Belle:2023zxm},  
and \besiii~\cite{BESIII:2013fnz,
BESIII:2015klj,
BESIII:2020nbj,
BESIII:2022bse,
BESIII:2023xta,
BESIII:2023hml}
collaborations;
in proton\nobreakdash-antiproton collisions 
by 
the~\cdf~\cite{CDF:2003cab,
CDF:2005cfq,
CDF:2006ocq,
CDF:2009nxk}
and \dzero~\cite{D0:2004zmu} collaborations;
in proton\nobreakdash-proton\,($\proton\proton$)~collisions
by the~\atlas~\cite{ATLAS:2016kwu}, 
CMS~\cite{
CMS:2013fpt,
CMS:2020eiw}
and 
\lhcb~\cite{
LHCb-PAPER-2011-034,
LHCb-PAPER-2013-001,
LHCb-PAPER-2014-008,
LHCb-PAPER-2015-015,
LHCb-PAPER-2016-016,
LHCb-PAPER-2019-023,
LHCb-PAPER-2020-008,
LHCb-PAPER-2020-009,
LHcb-PAPER-2020-023,
LHCb-PAPER-2021-026,
LHCb-PAPER-2021-045,
LHCb-PAPER-2021-047}
collaborations;
in proton\nobreakdash-ion collisions 
by the~\lhcb collaboration~\cite{LHCb-PAPER-2023-026};
and in lead\nobreakdash-lead 
collisions 
by the~\cms~\cite{CMS:2021znk}
collaboration.

The~closeness of the~mass of 
the~$\chicone(3872)$~state~\cite{LHCb-PAPER-2020-008,
LHCb-PAPER-2020-009} 
to the~$\Dz\Dstarzb$~threshold,
%~\cite{PDG2022},
its narrow width~\cite{LHCb-PAPER-2020-008,
LHCb-PAPER-2020-009,BESIII:2023hml}, 
quantum numbers of 
$\PJ^{\PP\PC}=1^{++}$~\cite{LHCb-PAPER-2013-001,
LHCb-PAPER-2015-015}
and a~large coupling to the~$\Dz\Dstarzb$~system~\cite{
Belle:2006olv,
BaBar:2007cmo,
Belle:2008fma,
Belle:2023zxm,
BESIII:2023hml}
provide 
%% strong and 
natural 
arguments to support 
%% supporting 
%% arguments for 
the~interpretation of the~$\chicone(3872)$~state as 
a~loosely\nobreakdash-coupled 
%% \mbox{$\tfrac{1}{\sqrt{2}}\left(\Dz\Dstarzb+\Dzb\Dstarz\right)$}
\mbox{$\Dz\Dstarzb+\Dzb\Dstarz$}
molecular state~\cite{
Braaten:2003he,
Swanson:2003tb,
Wong:2003xk,
Tornqvist:2004qy,
Hanhart:2007yq}, 
%% where 
in which 
the~colourless open\nobreakdash-charm mesons
are 
%% well separated 
spatially well separated.
%% at a~large distance.
%% 
However, the~expected production 
cross\nobreakdash-section
for such a~molecular object 
in high\nobreakdash-energy hadron 
collisions is too small
to explain the~observed 
production of the~$\chicone(3872)$~state~\cite{
Artoisenet:2009wk,
Bignamini:2009sk,
Esposito:2015fsa,
She:2024dqq}.
In~fact, 
the~measured production 
cross\nobreakdash-section,  
transverse momentum and rapidity spectra 
of the~$\chicone(3872)$~state
are close to~those 
observed for 
conventional charmonium states~\mbox{\cite{
CDF:2003cab,
D0:2004zmu,
LHCb-PAPER-2011-034,
ATLAS:2016kwu,
LHCb-PAPER-2021-026}}.
Alternative hypotheses for the~nature of 
the~$\chicone(3872)$~state include
%% in particular 
a~charmonium $\chicone(2\PP)$~state~\cite{
Barnes:2003vb,
Eichten:2004uh,
Barnes:2005pb,
Suzuki:2005ha},
its virtual companion~\cite{Giacosa:2019zxw}
or its mixture with a~hadronic molecule~\cite{
Close:2003sg,
Suzuki:2005ha};
a~hadro\nobreakdash-charmonium state~\cite{Dubynskiy:2008mq};
a~hybrid meson~\cite{
Close:2003mb,
Li:2004sta}
or 
a~tetraquark~\cite{
Maiani:2004vq,
Matheus:2006xi}.
A~large isospin violation
in $\chicone(3872)$~decays~\cite{LHCb-PAPER-2021-045} 
strongly disfavors 
the~interpretation 
of the~$\chicone(3872)$~particle 
as a~pure charmonium~state.
Experimental studies 
of the~$\chicone(3872)$~lineshape 
in the~$\jpsi\pip\pim$~\cite{
LHCb-PAPER-2020-008} and 
$\Dz\Dstarzb$~\cite{Belle:2023zxm}~final 
states, 
and the~combined analysis 
of the~two final states~\cite{BESIII:2023hml} 
provide important 
information about
the~parameters of
the~low\nobreakdash-energy 
$\Dz\Dstarzb$~scattering
amplitude, namely 
the~scattering length
and the~effective range.
These parameters, 
if measured precisely, 
%% could 
provide   %% unambiguous
crucial information
about the~internal structure of the~$\chicone(3872)$~state.
The~sign of the~scattering length, 
the~values of the~effective range, 
and the~compositeness parameter~\cite{Weinberg:1965zz} 
indicate the~presence 
of a~compact component 
in the~wave function of 
the~$\chicone(3872)$~state~\cite{Esposito:2021vhu,Esposito:2023mxw}.
However, the~current experimental
precision of these parameters
is insufficient for drawing 
a~definite conclusion 
about the~nature of 
the~$\chicone(3872)$~state.
In~the~future, 
the~precision could potentially 
be improved  with increased 
data sample size,  %% data size, 
better mass resolution
and, as pointed out in Ref.~\cite{Belle:2023zxm}, 
with a~combined analysis 
of the~datasets accumulated by 
the~\belle, \belletwo, \besiii  and \lhcb experiments.

The~study of the~radiative decays of the~$\chicone(3872)$ state
into the~$\psitwos\g$ and  $\jpsi\g$ final states
provides an alternative way to 
%% identify 
probe 
its nature~\cite{Swanson:2004pp}.
Calculations for the~ratio of the~partial
radiative  decay widths
into $\psitwos\g$ and $\jpsi\g$~final states, 
\begin{equation}
\label{eq:r}   
    %% \mathcal{R}_{\Ppsi\g} 
    \mathscr{R}_{\Ppsi\g} 
    \equiv 
    \dfrac{\Gamma_{\decay{\chicone(3872)}{\psitwos\g}}}
          {\Gamma_{\decay{\chicone(3872)}{\jpsi\g}}}\,,
\end{equation}
vary 
widely 
%% in a~wide range
depending on the~hypothesis considered
for the nature of the~$\chicone(3872)$~state.
Each partial radiative width is subject to a~sizeable 
theoretical uncertainty, 
while a~significant part of this  
uncertainty cancels in the~ratio 
of decay widths.
The~theoretical predictions
for the~ratio  $\mathscr{R}_{\Ppsi\g}$ 
for a~variety of the~considered models
are summarised in Table~\ref{tab:predictions}.
\begin{table}[tb]
\centering
\caption{Compilation of predictions
for the~ratio $\mathscr{R}_{\Ppsi\g}$ of 
radiative partial decay widths  
of the~$\chicone(3872)$~state.
The~last column indicatively marks
the~considered model: 
$\cquark\cquarkbar$ for predominantly 
charmonium $\chicone(2\PP)$~models;
$\cquark\cquarkbar$/vc for 
the~model of a~virtual companion of
the~$\chicone(2\PP)$~state;
$\D\Dstarb$  
for the~predominantly \mbox{$\Dz\Dstarzb+\Dzb\Dstarz$}
molecular model
and 
%% 
%%  $\D\Dstarb+\cquark\cquarkbar$  
%% for models
%% including 
models
%% model
where 
the~molecular
component is mixed 
with 
$\jpsi\Prho$ and/or  
$\jpsi\Pomega$~components 
and  
a~compact component; 
and 
$\cquark\cquarkbar\quark\quarkbar$ 
for the~compact tetraquark model. 
The~value of $\mathscr{R}_{\Ppsi\g}$ 
%% in the~penultimate row
from Ref.~\cite{Guo:2014taa}
is  expressed 
in terms of the~poorly known ratio of 
the~coupling constants, 
$\left(g_2^\prime/g_2\right)$,  for the~\psitwos 
and \jpsi~mesons
with $\Db^{\left(*\right)}\D^{(*)}$~systems, 
which are expected to be 
similar. %% $\sim1$. 
The~same ratio 
of coupling constants 
dictates the~wide 
spread of predictions
%% in the~antepenultimate row. 
from Ref.~\cite{Molnar:2016dbo}. 
The~value in the~last row 
is the~minimal possible 
value of 
the~ratio $\mathscr{R}_{\Ppsi\g}$
in this model; 
the~ratio  %% $\mathscr{R}_{\Ppsi\g}$
can be as big as 12
for large diquark sizes. 
}
\label{tab:predictions}
%% \begin{footnotesize}
\begin{tabular*}{0.99\textwidth}{@{\hspace{1mm}}l@{\extracolsep{\fill}}ccc@{\hspace{1mm}}}
\multicolumn{2}{l}{\ \ \ \ \ Reference}
& $\mathscr{R}_{\Ppsi\g}$ 
& 
\\[1.5mm]
\hline 
\\[-3.5mm]
T.~Barnes and  S.~Godfrey 
& \cite{Barnes:2003vb}   
& 5.8  
& $\cquark\cquarkbar$  
\\
T.~Barnes, S.~Godfrey and S.~Swanson  
& \cite{Barnes:2005pb}
& $2.6$
& $\cquark\cquarkbar$  
\\
F.~De~Fazio
& \cite{DeFazio:2008xq}
& ($1.64\pm0.25$)
& $\cquark\cquarkbar$  
\\
B.-Q.~Li and K.~T.~Chao 
& \cite{Li:2009zu}
& 1.3 
& $\cquark\cquarkbar$  
\\
Y.~Dong {\it{et al.}} 
& \cite{Dong:2009uf}
& $1.3-5.8$
& $\cquark\cquarkbar$  
\\
A.~M.~Badalian {\it{et al.}} 
& \cite{Badalian:2012jz}  
& $(0.8\pm0.2)$                 
& $\cquark\cquarkbar$  
\\
J.~Ferretti, G.~Galata and E.~Santopinto 
& \cite{Ferretti:2014xqa}
& 6.4 
& $\cquark\cquarkbar$  
\\
A.~M.~Badalian, Yu.~A.~Simonov and 
B.~L.~G.~Bakker
%% B.~Bakker
& \cite{Badalian:2015dha}  
%% & $2.38$
& $2.4$
& $\cquark\cquarkbar$  
\\
W.~J.~Deng {\it{et al.}} 
& \cite{Deng:2016stx}
& 1.3
& $\cquark\cquarkbar$  

\\
F.~Giacosa, M.~Piotrowska and S. Goito
& \cite{Giacosa:2019zxw}
& 5.4 
& $\cquark\cquarkbar$/vc
\\
E.~S.~Swanson 
& \cite{Swanson:2004pp}
%% & $0.15\,\%$
%% & $1.5\,\text{\textperthousand}$
& $0.38\,\%$
%% & molecular
& $\D\Dstarb$
\\
Y.~Dong {\it{et al.}} 
& \cite{Dong:2009uf}
%% & $1.5\times10^{-3}$
%% & $1.5\,\text{\textperthousand}$
& $0.33\,\%$
%% & molecular  %% + $\cquark\cquarkbar$  
& $\D\Dstarb$
\\
D.~P.~Rathaud and  A.~K.~Rai 
& \cite{Rathaud:2016tys}
& 0.25
& $\D\Dstarb$
\\
B.~Grinstein, L.~Maiani and A.~D.~Polosa 
& \cite{Grinstein:2024rcu} 
%% & 0.036 
&  $3.6\,\%$
%% & molecular
& $\D\Dstarb$
\\
F.-K.~Guo {\it{et al.}}  
& \cite{Guo:2014taa}
& $0.21(g_2^\prime/g_2)^2$ 
%% & molecular + $\cquark\cquarkbar$ 
%% & $\D\Dstarb+\cquark\cquarkbar$
& $\D\Dstarb$ 
\\
D.~A.-S.~Molnar, R.~F.~Luiz and  R.~Higa  
& \cite{Molnar:2016dbo}
& $2-10$ 
%% & molecular + $\cquark\cquarkbar$ 
%% & $\D\Dstarb+\cquark\cquarkbar$ 
& $\D\Dstarb$ 
\\
P.~G.~Ortega {\it{et al.}}
& \cite{Ortega:2012rs}
& $1.2$
& $\D\Dstarb$ 
\\
E.~Cincioglu {\it{et al.}} 
& \cite{Cincioglu:2016fkm}
%%  & $0-4$
& $<4$
%% & $\D\Dstarb+\cquark\cquarkbar$ 
& $\D\Dstarb$ 
\\
S.~Takeuchi, M.~Takizawa and K.~Shimizu  
& \cite{Takeuchi:2016hat}
& $1.1-3.4$ %% absract
%% & $2.1-6.2$ %% conclusion
%% & molecular + $\cquark\cquarkbar$ 
%% & $\D\Dstarb+\cquark\cquarkbar$ 
& $\D\Dstarb$ 
\\
R.~F.~Lebed and S.~R.~Martinez 
& \cite{Lebed:2022vks}
& $0.33\,\%$
& $\D\Dstarb$
\\
B.~Grinstein, L.~Maiani and A.~D.~Polosa 
& \cite{Grinstein:2024rcu} 
&  $>\left(0.95^{\,+\,0.01}_{\,-\,0.07}\right)$ 
& $\cquark\cquarkbar\quark\quarkbar$ 
\end{tabular*}
%% \end{footnotesize}
\end{table}
As~a~general rule, large values of this ratio, 
$\mathscr{R}_{\Ppsi\g}\gtrsim1$, are expected 
under the~assumption that the~$\chicone(3872)$~state
is a~conventional charmonium 
$\chicone(2\PP)$~state~\cite{
Barnes:2003vb,
Barnes:2005pb,
DeFazio:2008xq,
Li:2009zu,
Dong:2009uf,
Badalian:2012jz,
Ferretti:2014xqa,
Badalian:2015dha,
Deng:2016stx,
Giacosa:2019zxw}, 
while calculations 
based on the~pure $\D\Dstarb$-molecular hypothesis
give much smaller  values,
\mbox{$\mathscr{R}_{\Ppsi\g} \ll 1$}~\cite{
%% Swanson:2003tb,
Swanson:2004pp,
Dong:2009uf,
Rathaud:2016tys,
Grinstein:2024rcu},
unless specific 
assumptions on 
the~poorly known ratio 
of the~coupling constants 
$\left(g_2^\prime/g_2\right)$
for the~$\psitwos$ 
and $\jpsi$~mesons
with $\Db^{\left(*\right)}\D^{(*)}$~systems~\cite{Guo:2014taa,
Molnar:2016dbo} are used. 
Models based on the~mixture of 
a~predominantly 
$\D\Dstarb$~molecular state
(and sometimes also with 
$\jpsi\Prho$ and $\jpsi\Pomega$ 
contributions) and 
a~compact component
%%%
cover a~wide range of  $\mathscr{R}_{\Ppsi\g}$,
but they have reduced predictive power 
due to their large dependency on the~mixing 
parameters~\cite{Ortega:2012rs,
Cincioglu:2016fkm,
Takeuchi:2016hat,
Lebed:2022vks}.
%% and 
%% the~poorly known coupling constants.  
%% 
The~calculations 
based  on the~Born\nobreakdash--Oppenheimer  
approximation for 
the~$\cquark\cquarkbar\quark\quarkbar$~tetraquark~\cite{Grinstein:2024rcu} 
give a~result that is similar to 
expectations from the~charmonium~models.

%% 
%% %% charmonium 
%% 
%% \cite{Li:2009zu,
%% Achasov:2015wea,
%% Achasov:2016vxb,
%% Yu:2023nxk,
%% Deng:2016stx},
%%  

%Experimentally, 
%the~first evidence of the~\XPsiG decay and measurement
%of the~ratio of radiative width $\mathscr{R}_{\Ppsi\g}$ of 
%$3.4\pm1.4$ was reported 
%by the~\babar collaboration~\cite{BaBar:2008flx}. 
Experimentally, the~\babar collaboration reported the~first evidence of the \mbox{$\decay{\chicone(3872)}{\psitwos\g}$}~decay 
and measured 
%% the~ratio of radiative widths, 
\mbox{$\mathscr{R}_{\Ppsi\g}= 
3.4\pm1.4$}~\cite{BaBar:2008flx}.
Subsequently, evidence for the~same decay was found by 
the~\lhcb experiment 
using data 
collected in $\proton\proton$~collisions in 2011-2012 
at center\nobreakdash-of\nobreakdash-mass energies 
of 7 and 8\tev, corresponding to an integrated luminosity of $3\invfb$~\cite{LHCb-PAPER-2014-008}.
%% 
%% using $3\invfb$ of data 
%% collected in $\proton\proton$~collisions 
%% at center\nobreakdash-of\nobreakdash-mass energies 
%% of 7 and 8\tev~\cite{LHCb-PAPER-2014-008}.
%% 
The~measured 
%% ratio of the~radiative 
%% decay widths, 
$\mathscr{R}_{\Ppsi\g}$ value of 
\mbox{$2.46\pm0.64\pm0.29$}
%% ~\cite{
%% LHCb-PAPER-2014-008}
is 
%% found to be 
in good agreement with the~value 
obtained by the~\babar collaboration.
On~the~other hand, the~\belle~\cite{Belle:2011wdj} and 
\besiii~\cite{BESIII:2020nbj} collaborations did  not 
observe significant signals of the~\mbox{\XPsiG}~decay.
The~most stringent upper limit is set by the~\besiii collaboration, 
%%\begin{equation*}
$ \mathscr{R}_{\Ppsi\g} < 0.59$\,(at 90\,\%~CL), 
%% \mathrm{\ at\ }90\,\%\,\mathrm{CL}\,,
%%\end{equation*}
%% which evidently is in a~tension with the~values
and is in 
tension
%% clear contradiction
%% clearly contradicts 
with the~values 
reported by 
the~\babar and \lhcb~collaborations.

This paper reports a~study of  
the~radiative decays 
of the~$\chicone(3872)$~state into 
the~$\psitwos\g$ and $\jpsi\g$ final states using 
the~$\decay{\Bu}{\chicone(3872)\Kp}$~decay.
The~ratio of decay widths
$\mathscr{R}_{\Ppsi\g}$
defined in Eq.~\eqref{eq:r}
is measured as the~ratio of branching fractions
for the \mbox{$\decay{\Bu}{\left(\decay{\chicone(3872)}{\Ppsi\g}\right)\Kp}$}
decays,\footnote{In 
this paper 
inclusion of charge-conjugate
decays is implied throughout, and 
the~symbol $\Ppsi$ 
denotes the~$\jpsi$ and $\psitwos$~states together.}
\begin{equation} \label{eq:rB}
    \mathscr{R}_{\Ppsi\g} =  
    \dfrac{\BR_{\decay{\Bu}{\left(\decay{\chicone(3872)}{\psitwos\g}\right)\Kp}}}
          {\BR_{\decay{\Bu}{\left(\decay{\chicone(3872)}{\jpsi\g}\right)\Kp}}}\,.
\end{equation}
The~analysis is based 
on $\proton\proton$~collision data
collected with the~\lhcb detector 
for two data\nobreakdash-taking periods, 
Run\,1 and Run\,2. 
The~first sample
%collected in 2011 and 2012,
%corresponding to an~integrated 
%luminosity of~$3\invfb$, 
was previously analysed 
in Ref.~\cite{LHCb-PAPER-2014-008}.
The~second sample was collected between 2015 and 2018 
at a~centre\nobreakdash-of\nobreakdash-mass 
energy of 13\tev
and 
corresponds to an~integrated 
luminosity of~$6\invfb$.

\section{Detector and simulation}
\label{sec:detector}

The \lhcb detector~\cite{LHCb-DP-2008-001,LHCb-DP-2014-002} 
is a~single\nobreakdash-arm forward
spectrometer covering 
the~\mbox{pseudorapidity} range 
\mbox{$2<\eta <5$},
designed for the study of particles containing \bquark or \cquark~quarks. 
The~detector includes 
a~high\nobreakdash-precision tracking system
consisting of 
a~silicon\nobreakdash-strip vertex detector 
surrounding 
the~$\proton\proton$
interaction region~\cite{LHCb-DP-2014-001}, 
a~large\nobreakdash-area 
silicon\nobreakdash-strip detector located
upstream of a~dipole magnet with 
a~bending power of about~$4{\mathrm{\,T\,m}}$, 
and three stations of 
silicon\nobreakdash-strip detectors 
and straw drift tubes~\cite{LHCb-DP-2013-003,
LHCb-DP-2017-001}
placed downstream of the~magnet.
The~tracking system provides a~measurement of 
the~momentum, \ptot, of charged particles with
a~relative uncertainty that varies 
from~0.5\,\% at low momentum 
to~1.0\,\% at 200\gevc.
The~minimum distance of a~track 
to a~primary $\proton\proton$~collision 
vertex\,(PV), the~impact parameter,   %% \,(IP), 
is measured with a~resolution of~$(15+29/\pt)\mum$,
where \pt is the~component of 
the~momentum transverse to the~beam, in\,\gevc.
Different types of charged hadrons 
are distinguished using information
from two ring\nobreakdash-imaging 
Cherenkov detectors~\cite{LHCb-DP-2012-003}. 
Photons, electrons and hadrons are 
identified by a~calorimeter system consisting of
scintillating-pad and preshower detectors, 
an electromagnetic
and a~hadronic calorimeter. 
Muons are identified by 
a~system composed of alternating layers 
of iron and multiwire
proportional chambers~\cite{LHCb-DP-2012-002}.

The~online event selection 
is performed by
a~trigger~\cite{LHCb-DP-2012-004,
LHCb-DP-2019-001}, 
which consists of a~hardware stage, based on information 
from the~calorimeter and muon systems, 
followed by a~software stage, 
which applies a~full event reconstruction.
At~the~hardware trigger stage, events are required to have 
a~muon track with high transverse momentum or dimuon candidates 
in which the~product of 
the~\pt of the~muons 
has a~high value. 
In~the~software trigger, two oppositely charged muons are required 
to form a~good\nobreakdash-quality vertex that is significantly 
displaced from any~PV, 
with a~dimuon mass exceeding $2.7\gevcc$.

Simulated events are used to describe signal shapes,  
background from partially reconstructed 
decays of beauty hadrons, 
and to compute the~efficiencies needed to determine 
the~ratio $\mathscr{R}_{\Ppsi\g}$.
In~the~simulation, $\proton\proton$~collisions are 
generated using \pythia~\cite{Sjostrand:2007gs} with 
a~specific \lhcb configuration~\cite{LHCb-PROC-2010-056}.
Decays of unstable particles are 
described by \evtgen~\cite{Lange:2001uf}, 
in which final\nobreakdash-state radiation is generated 
using \photos~\cite{davidson2015photos}.
The~interaction of the~generated particles with the~detector, 
and its response, are implemented using 
the~\geant toolkit~\cite{Agostinelli:2002hh, Allison:2006ve}  
as described in Ref.~\cite{LHCb-PROC-2011-006}.
The~\pt~and rapidity\,($y$)  
spectra of the~$\Bu$~mesons in simulation 
are corrected to match data.
The~correction factors are calculated 
by comparing the~observed
$\pt$ and $y$~spectra
for
a~high\nobreakdash-purity sample of 
reconstructed \decay{\Bu}{\jpsi\Kp}~decays
with the~corresponding simulated samples. 
In simulation, 
the~variables used 
for the~kaon identification
%% the~corresponding  quantities in simulation 
are resampled according to 
%% values obtained  from 
the~calibration data samples 
of \mbox{$\decay{\Dstarp}
{\left( \decay{\Dz}
{\Km\pip}\right)\pip}$}~decays~\cite{LHCb-DP-2018-001}. 
The~procedure accounts for 
%% correlations between 
%% the~variables associated to a~particular track, 
%% as well as 
the~dependence of 
the~kaon identification
on
the~particle $\pt$
and $\eta$
as well as
the~charged particle multiplicity 
in the~event. 
The~track multiplicity for simulated events 
is corrected to match that 
in the~\decay{\Bu}{\jpsi\Kp}~data sample.
The~track reconstruction efficiency
is corrected using 
a~sample of \mbox{$\decay{\jpsi}{\mumu}$}~decays 
in data~\cite{LHCb-DP-2013-002}, 
to~account for imperfections in the~simulation.
%% of 
%% track reconstruction.
%%
Samples of 
%% the~\mbox{$\decay{\Bu}{\jpsi\Kstarp}$}~decays 
%% where 
%% \mbox{$\decay{\Kstarp}{\Kp\left(\decay{\piz}{\g\g}\right)}$}~decays
%% \mbox{$\decay{\Bu}{\jpsi\left(
%% \decay{\Kstarp}{\Kp\left(
%% \decay{\piz}{\g\g}\right) } \right)}$}~decays
%% 
\mbox{$\decay{\Bu}{\jpsi\left(\decay{\Kstarp}{\Kp\piz}\right)}$}~decays with \mbox{$\decay{\piz}{\g\g}$}
are used to correct 
the~photon reconstruction efficiency  
in the~simulation~\cite{LHCb-PAPER-2012-022,
LHCb-PAPER-2012-053,
Govorkova:2015vqa,
Govorkova:2124605,
Belyaev:2015nlt}.
%%

%Large simulated samples of inclusive 
%\mbox{$\decay{\B}{\Ppsi\PX}$} 
%decays have been used to study 
%the~background contributions 
%from partially reconstructed decays. 
Large simulated samples of inclusive
\B~meson decays to final states with 
a~charmonium, \mbox{$\decay{\B}{\Ppsi\PX}$}, 
are used to study 
the~background contributions 
from partially reconstructed decays. 
The~\mbox{$\decay{\Bu}{\jpsi\Kstarp}$}~decay 
%% \mbox{$\decay{\Bu}{\jpsi\left(\decay{\Kstarp}
%%{\Kp\piz}\right)}$}
%% 
with \mbox{$\decay{\Kstarp}{\Kp\piz}$},
and inclusive  
\mbox{$\decay{\B}{\psitwos\Kp\PX}$} decays, 
%%%% where the~\Kp and \psitwos~mesons
%%%% originate in the~decay of the~same \B~meson, 
%% where \PX stands for either a pair 
%% of pions or a light unflavoured meson, 
%% have been identified 
are 
the~most important contributions to the~background.
%% According to this result, dedicated simulation samples
%% are produced.
An~amplitude model determined from \mbox{$\decay{\Bd}{\jpsi\Kstarz}$}~decays~\cite{LHCb-PAPER-2013-023}
is used to produce 
the~simulated sample of 
 \mbox{$\decay{\Bu}{\jpsi\left(\decay{\Kstarp}
{\Kp\piz}\right)}$}~decays.
The~simulated sample of
 \mbox{$\decay{\B}{\psitwos\Kp\PX}$}~decays
 comprises relevant admixture of \Bp and \Bd mesons.
%is composed
%% a~cocktail 
%% a~properly weighted mixture 
%of %the~simulated samples for
%the~corresponding decays of
%% the~\Bu, \Bd and \Bs~mesons.
%beauty mesons. 
%% 
For~each sample the~individual 
decays of beauty mesons 
are simulated 
as an~incoherent mixture of 
the~decays 
via various excited kaons,
namely 
$\kaon^{*}$,
$\kaon^{*}_0(700)$,
$\kaon_1(1270)$,
$\kaon^{*}(1410)$,
$\kaon^{*}_2(1430)$,
$\kaon^{*}(1680)$;
%%  
%% $\psitwos\kaon^{*}$,
%% $\psitwos\kaon^{*}_0(700)$,
%% $\psitwos\kaon_1(1270)$,
%% $\psitwos\kaon^{*} (1410)$,
%% $\psitwos\kaon^{*}_2(1430)$,
%% $\psitwos\kaon^{*} (1680)$,
as well as decays into 
nonresonant 
$\psitwos\Kp\pion$,
$\psitwos\Kp\pion\pion$,
$\psitwos\Kp\Peta$,
$\psitwos\Kp\Pomega$ and 
$\psitwos\kaon^{*}\Peta$~combinations. 
The~relevant branching
fractions %% for the~known decays 
are taken from  
Refs.~\cite{
PDG2022,
Belle:2010wrf,
Belle:2013shl,
LHCb-PAPER-2016-019},
%% For~unknown decays
%% the~branching fractions 
%% are 
inferred 
from 
%% isotopic 
isospin 
symmetry,
%% or 
%% an~educated guess. 
or estimated based on similar decays where no other information exists.

\section{Event selection}
\label{sec:sel}

The~\mbox{$\decay{\Bp}{\left(\decay{\chicone(3872)}{\Ppsi\g}\right)\Kp}$}~decay
candidates
%% , followed by $\chicone(3872)\to\Ppsi\Pgamma$, 
%% where $\Ppsi$ denotes a \jpsi
%% or \psitwos meson, 
are reconstructed using the~\mbox{$\decay{\Ppsi}{\mumu}$}~decay mode.
Signal candidates 
are reconstructed
by first applying 
a~loose initial selection, 
and subsequently using
a~multivariate classifier 
to suppress backgrounds.
%% higher purity sample 
%% of candidates for both signal and normalisation modes.
%% 
To~reduce systematic uncertainties,  
the~initial selection criteria for both decay modes 
%% signal and normalisation channels 
are kept the~same whenever possible
and similar to those used in 
previous \lhcb studies~\cite{LHCb-PAPER-2014-008, LHCb-PAPER-2021-003}. 

The~muon and kaon candidates are identified by combining 
information from the~Cherenkov~detectors,
calorimeters and muon detectors~\cite{LHCb-PROC-2011-008} 
associated to the~reconstructed tracks. 
To~reduce the~combinatorial background, 
only tracks that 
are inconsistent with originating from any 
reconstructed PV in the~event are considered. 
The~transverse momentum of muon candidates is required to
be greater than 550\mevc. Pairs of oppositely charged muons 
consistent with originating from a~common vertex 
are combined to form 
$\Ppsi$~candidates.
The~reconstructed mass 
of the~muon pair is required to be 
\mbox{$3.020<m_{\mumu}<3.125\gevcc$}
and 
\mbox{$3.597<m_{\mumu}<3.730\gevcc$}
for \jpsi~and \psitwos~candidates, respectively,

Charged particles 
%% consistent with 
identified as kaons
%% the~kaon mass hypothesis
are required to 
have 
%% transverse momentum greater than
$\pt>200\mevc$. 
Photons are reconstructed from clusters 
in the~electromagnetic 
calorimeter
%% . These~clusters must not be 
not associated with reconstructed 
tracks~\cite{LHCb-2003-091,
LHCb-2003-092,
LHCb-DP-2020-001}.
%%
%% Photon identification is based on the~combined information 
%% from electromagnetic and hadronic calorimeters, 
%% scintillation pad, 
%% preshower detectors and the~tracking system. 
%5
Each selected \jpsi and \psitwos~candidate is 
combined with a~photon to form a~$\chicone(3872)$ candidate.
The~transverse energy 
of the~photon candidate should exceed 
$1.0\gev$ 
and $0.5\gev$
for the \mbox{$\decay{\chicone(3872)}{\jpsi\g}$}
and \mbox{$\decay{\chicone(3872)}
{\psitwos\g}$}~decays, respectively.
Finally the \Bp~candidate is obtained by combining 
a~kaon and the~$\chicone(3872)$ candidate.

To~suppress 
%% a~cross\nobreakdash-feed 
contributions from 
\mbox{\decay{\Bu}{\jpsi\Kp}} and \mbox{\decay{\Bu}{\psitwos\Kp}}~decays,
$\Ppsi\Kp$~combinations with a~mass within
$\pm40\mevcc$ of the~known \Bu~mass~\cite{PDG2022} are rejected.
%%
%% To~ensure the~reconstruction quality
A~good quality kinematic fit~\cite{Hulsbergen:2005pu} is required, 
which constrains 
the~dimuon mass to the~known \jpsi or \psitwos mass~\cite{PDG2022} 
and requires the~\Bp~candidate to originate from 
its associated~PV.
%% and the~fit is required to be of good quality.
%%
The~\Bp~decay time
is required to exceed 
$100\mum/c$ to suppress 
the~large combinatorial 
background from tracks
originating from~a~PV.

A~multilayer 
perceptron\,({\sc{MLP}}) classifier
is applied to candidate events
to reduce background further.
The~classifier 
is based on an~artificial neural 
network algorithm~\cite{McCulloch,rosenblatt58}, 
configured with a~cross\nobreakdash-entropy cost 
estimator~\cite{Zhong:2011xm}.
%% 
%% It~reduces the~remaining background to a~low level 
%% while retaining a~high signal efficiency. 
%% 
Different classifiers are trained separately for
the~\mbox{$\decay{\Bu}
{\left(\decay{\chicone(3872)}{\psitwos\g}\right)\Kp}$}
and \mbox{$\decay{\Bu}
{\left(\decay{\chicone(3872)}{\jpsi\g}\right)\Kp}$}~decay modes.
%% signal and normalisation modes. 
Both classifiers use the same input variables.
These are 
related to the~reconstruction quality, 
kinematics and decay time of the~\Bp~candidates, 
kinematics of the~final\nobreakdash-state particles 
and variables that characterise the~kaon and photon identification. 
The~classifiers are trained using simulated samples 
of $\Bp\to\chicone(3872)\Kp$~decays as 
a~proxy for the~signal, but different 
types of background samples 
for the~two decay modes.
For~the~\mbox{$\decay{\Bu}
{\left(\decay{\chicone(3872)}{\jpsi\g}\right)\Kp}$}~sample, 
the~background after the~initial selection is large and mainly
%combinatorial, 
consists of random combinations 
of \jpsi and \Kp~mesons with a~photon. %%  \jpsi- and \Kp-mesons,
The~background training sample is taken from 
the~signal 
mass sidebands in data, \ie
the~\mbox{$\Bp\to\chicone(3872)\Kp$}~candidate decays  with 
the~$\jpsi\g\Kp$~mass outside
of the~interval \mbox{$5.25<m_{\jpsi\g\Kp}<5.32\gevcc$}
and the~$\jpsi\g$~mass outside of 
the~interval \mbox{$3.82<m_{\jpsi\g}<3.93\gevcc$}.
%% are used as background proxy for the~training.
%% 
In~the~\mbox{$\decay{\Bu}
{\left(\decay{\chicone(3872)}{\psitwos\g}\right)\Kp}$}~sample,
the~combinatorial background is already largely suppressed 
by the~initial loose selection.
%% A~study with large simulated samples
%% of~\mbox{$\decay{\B}{\psitwos\PX}$}~decays
%% indicates that 
The~remaining background 
is dominated by contributions 
from \mbox{$\decay{\B}{\psitwos\Kp\PX}$}~decays.
%% while the~remaining background mostly consists of various 
%% \decay{\PB}{\psitwos\Kp\PX} decays.
%% According to this observation,  
%% a~dedicated simulated sample is produced
A~simulated sample
of~\mbox{$\decay{\B}{\psitwos\Kp\PX}$}~decays 
%% produced 
%% as described in Sec.~\ref{sec:detector},
is used as 
the~background proxy for training. 
To~avoid introducing a~bias in the~MLP evaluation,
a~$k$-fold cross\nobreakdash-validation
technique~\cite{chopping} with  $k = 7$ 
is used in the~training of both classifiers.

The~requirement on the~response 
of each {\sc{MLP}} classifier 
is chosen to maximise 
the~figure\nobreakdash-of\nobreakdash-merit 
%% defined as 
%% $\tfrac{S}{\sqrt{B+S}}$,
$S/\sqrt{B+S}$,
where $S$~represents 
the~expected signal yield,
and $B$~is the~expected background yield.
%within a~$3\upsigma$ 
%mass window centred
%around the~known mass of 
%the~$\Bu$~meson.
%% under the~signal.
%% 
The~background yield is calculated 
from fits to data, 
while the~expected signal yield is estimated 
as $S=\varepsilon S_0$,
where 
$\PS_0$~is the~signal yield obtained 
from a~fit to 
the~data 
%% when no 
with a~loose 
requirement applied,
and
$\varepsilon$ is the~relative 
efficiency of the~requirement 
on the~response of 
the~{\sc{MLP}}~classifier determined 
from simulation.

\begin{figure}[t]
\setlength{\unitlength}{1mm}
\centering
\begin{picture}(160,122)
    \put( 0, 2){
    \includegraphics*[width=80mm]{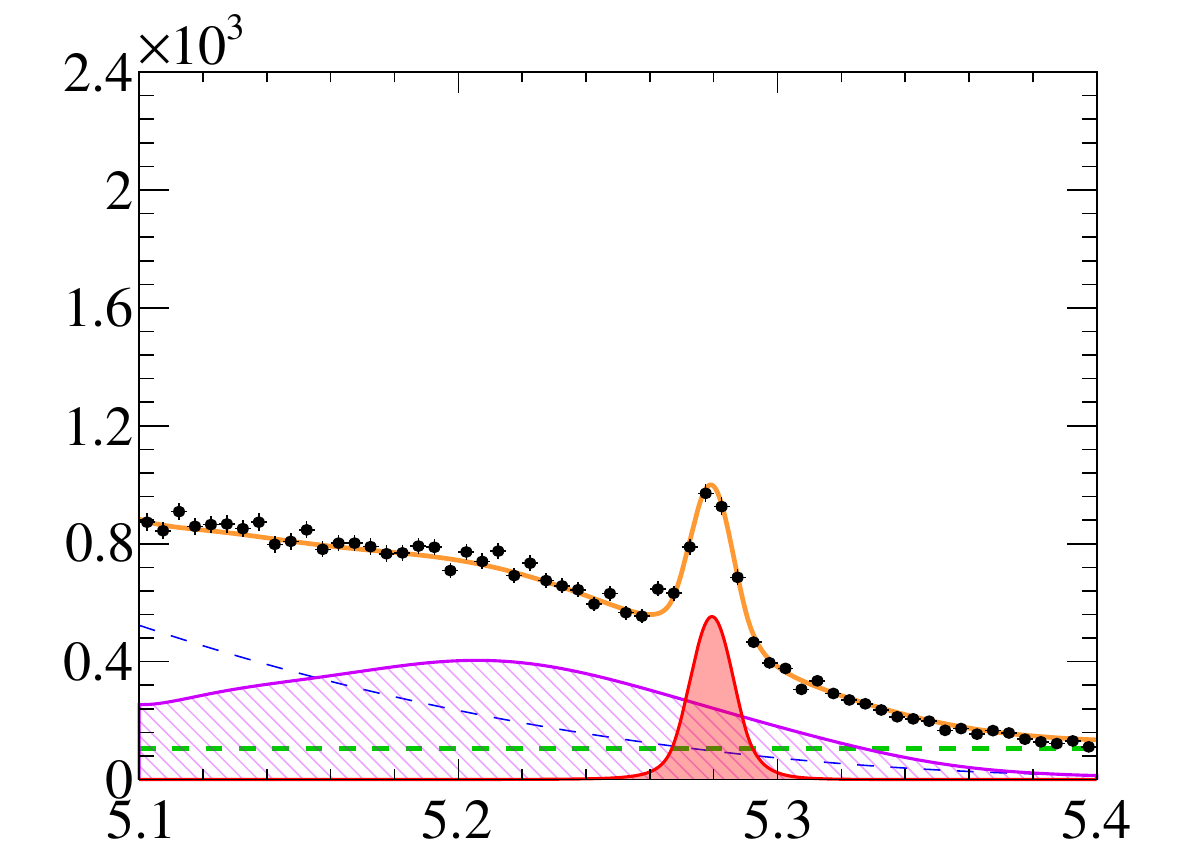}
        }
    \put(80, 2){
	\includegraphics*[width=80mm]{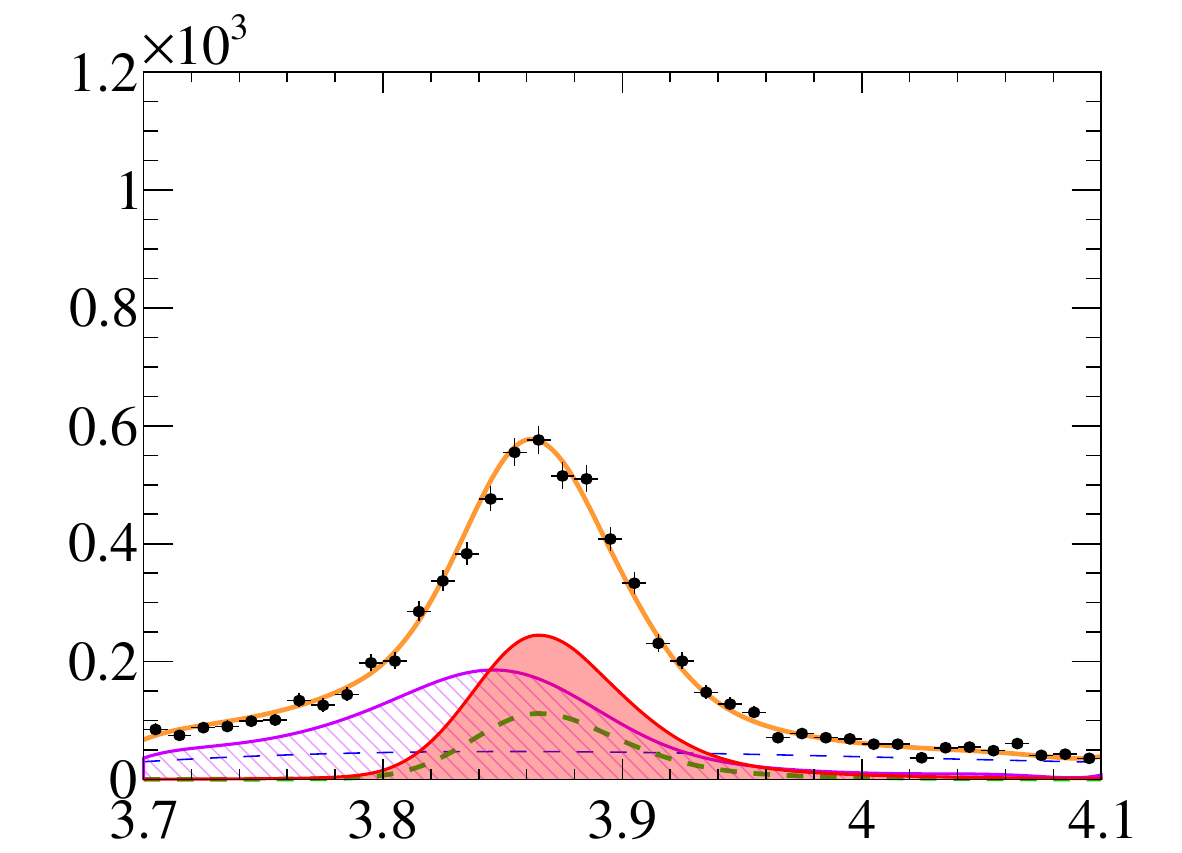}
        }
    \put( 0,62){
	\includegraphics*[width=80mm]{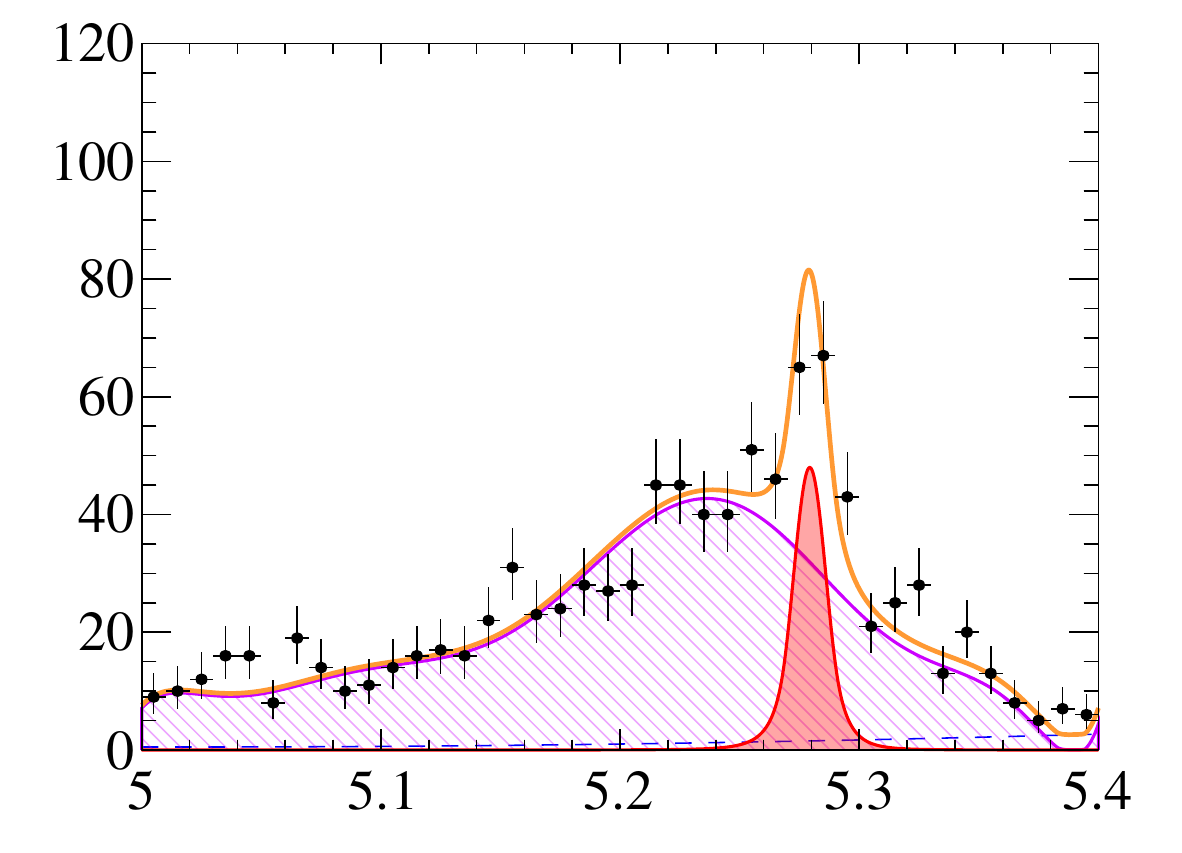}
	}
    \put(80,62){
	\includegraphics*[width=80mm]{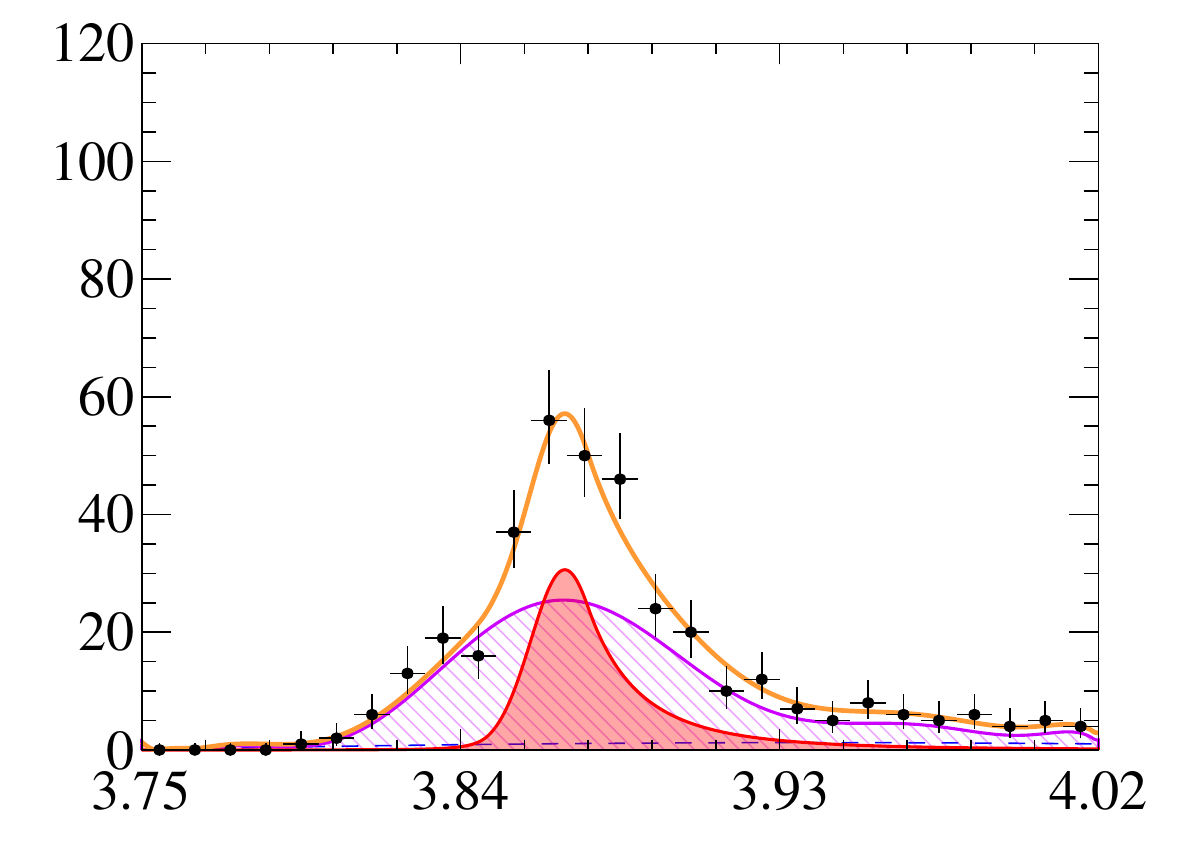}
	}    
    \put( 37,  0){$m_{\jpsi\g\Kp}$}    \put( 60,60){$\left[\!\gevcc\right]$}
    \put(117,  0){$m_{\jpsi\g}$}	   \put(140,60){$\left[\!\gevcc\right]$}
    \put( 37, 60){$m_{\psitwos\g\Kp}$} \put( 60, 0){$\left[\!\gevcc\right]$}
    \put(117, 60){$m_{\psitwos\g}$}	   \put(140, 0){$\left[\!\gevcc\right]$}
    \put(-0,14){\begin{sideways}Candidates/$(5\mevcc)$\end{sideways}}
    \put(80,13){\begin{sideways}Candidates/$(10\mevcc)$\end{sideways}}
    \put(-1,73){\begin{sideways}Candidates/$(10\mevcc)$\end{sideways}}
    \put(79,73){\begin{sideways}Candidates/$(10\mevcc)$\end{sideways}}
    \put( 60, 48){$\begin{array}{l}\lhcb\\9\invfb\end{array}$}
    \put(140, 48){$\begin{array}{l}\lhcb\\9\invfb\end{array}$}
    \put( 60,109){$\begin{array}{l}\lhcb\\9\invfb\end{array}$}
    \put(140,109){$\begin{array}{l}\lhcb\\9\invfb\end{array}$}
    %%
    %% Legend A
    %%
    \put(13, 40){\footnotesize%\small%%\scriptsize
    $\begin{array}{cl}
    %\!\bigplus\mkern-5mu & \text{data} 
    \!\bigplus\mkern-18mu\bullet 
    & \text{Data}
    \\
    \begin{tikzpicture}[x=1mm,y=1mm]\filldraw[fill=red!35!white,draw=red,thick]  (0,0) rectangle (5,3);\end{tikzpicture}  
    & \decay{\Bu}{\X\Kp} 
    \\
    \begin{tikzpicture}[x=1mm,y=1mm]\filldraw[pattern=north west lines, pattern color=violet!35!white,draw=violet,thick]  (0,0) rectangle (5,3);\end{tikzpicture}  
    & \decay{\B}{\jpsi\PX} 
    \\    
    {\color[RGB]{0,204,0}{\hdashrule[0.0ex][x]{5mm}{1.0pt}{2.0mm 0.3mm}}}
    & \text{$\chicone(3872)\Kp$}
    \\
    {\color[RGB]{85,83,246}{\hdashrule[0.0ex][x]{5mm}{1.0pt}{2.0mm 0.3mm}}}
    & \text{Combinatorial}
    \\ 
    {\color[RGB]{255,153,51} {\rule{5mm}{2.0pt}}}
    & \text{Total}
    \end{array}$ }
    %%
    %% Legend B
    %%
    \put(93, 40){\footnotesize%\small%%\scriptsize
    $\begin{array}{cl}
    %\!\bigplus\mkern-5mu & \text{data} 
    \!\bigplus\mkern-18mu\bullet 
    & \text{Data}
    \\
    \begin{tikzpicture}[x=1mm,y=1mm]\filldraw[fill=red!35!white,draw=red,thick]  (0,0) rectangle (5,3);\end{tikzpicture}  
    & \decay{\Bu}{\X\Kp} 
    \\
    \begin{tikzpicture}[x=1mm,y=1mm]\filldraw[pattern=north west lines, pattern color=violet!35!white,draw=violet,thick]  (0,0) rectangle (5,3);\end{tikzpicture}  
    & \decay{\B}{\jpsi\PX} 
    \\    
    {\color[RGB]{0,204,0}{\hdashrule[0.0ex][x]{5mm}{1.0pt}{2.0mm 0.3mm}}}
    & \text{$\chicone(3872)\Kp$}
    \\
    {\color[RGB]{85,83,246}{\hdashrule[0.0ex][x]{5mm}{1.0pt}{2.0mm 0.3mm}}}
    & \text{Combinatorial}
    \\ 
    {\color[RGB]{255,153,51} {\rule{5mm}{2.0pt}}}
    & \text{Total}
    \end{array}$ }
    %%
    %% Legend C 
    %% 
    \put(13,103){\footnotesize%%\scriptsize
    $\begin{array}{cl}
    %\!\bigplus\mkern-5mu & \text{data} 
    \!\bigplus\mkern-18mu\bullet & \text{Data}
    \\
    \begin{tikzpicture}[x=1mm,y=1mm]\filldraw[fill=red!35!white,draw=red,thick]  (0,0) rectangle (5,3);\end{tikzpicture}  
    & \decay{\Bu}{\X\Kp} 
    \\
    \begin{tikzpicture}[x=1mm,y=1mm]\filldraw[pattern=north west lines, pattern color=violet!35!white,draw=violet,thick]  (0,0) rectangle (5,3);\end{tikzpicture}  
    & \decay{\B}{\psitwos\Kp\PX} 
    \\    
    {\color[RGB]{85,83,246}{\hdashrule[0.0ex][x]{5mm}{1.0pt}{2.0mm 0.3mm}}}
    & \text{Combinatorial}
    \\ 
    %{\color[RGB]{0,204,0}{\hdashrule[0.0ex][x]{5mm}{1.0pt}{2.0mm 0.3mm}}}
    %& \text{Peaking}\\
    %{} & \text{background}\\
    {\color[RGB]{255,153,51} {\rule{5mm}{2.0pt}}}
    & \text{Total}
    \end{array}$ }
    %%
    %% Legend D
    %% 
    \put(93,103){\footnotesize%%\scriptsize
    $\begin{array}{cl}
    %\!\bigplus\mkern-5mu & \text{data} 
    \!\bigplus\mkern-18mu\bullet & \text{Data}
    \\
    \begin{tikzpicture}[x=1mm,y=1mm]\filldraw[fill=red!35!white,draw=red,thick]  (0,0) rectangle (5,3);\end{tikzpicture}  
    & \decay{\Bu}{\X\Kp} 
    \\
    \begin{tikzpicture}[x=1mm,y=1mm]\filldraw[pattern=north west lines, pattern color=violet!35!white,draw=violet,thick]  (0,0) rectangle (5,3);\end{tikzpicture}  
    & \decay{\B}{\psitwos\Kp\PX} 
    \\    
    {\color[RGB]{85,83,246}{\hdashrule[0.0ex][x]{5mm}{1.0pt}{2.0mm 0.3mm}}}
    & \text{Combinatorial}
    \\ 
    %{\color[RGB]{0,204,0}{\hdashrule[0.0ex][x]{5mm}{1.0pt}{2.0mm 0.3mm}}}
    %& \text{Peaking}\\
    %{} & \text{background}\\
    {\color[RGB]{255,153,51} {\rule{5mm}{2.0pt}}}
    & \text{Total}
    \end{array}$ }
\end{picture}
\caption{ \small
Distributions 
of the (left) $\Ppsi\g\Kp$ 
and (right) $\Ppsi\g$ mass of selected \Bu candidates
summed over Run\,1 and Run\,2
data-taking periods.
Top and bottom rows correspond to 
the  {${\Bp}\to{\left({\chicone(3872)}\to{\Ppsi{(2\PS)}\g}\right)\Kp}$}
and 
 {${\Bp}\to{\left({\chicone(3872)}\to{\PJ/\Ppsi\g}\right)\Kp}$}
candidates, respectively.
The $\Ppsi\g\Kp$ mass spectra are shown for candidates
within
the narrow $\Ppsi\g$ mass regions around the $\chicone(3872)$ mass,
and vice versa, 
the $\Ppsi\g$ mass spectra are shown for candidates
within
the narrow $\Ppsi\g\Kp$ mass regions around the $\Bu$ mass.
Projections of the fit, described in the text, are overlaid.
}
\label{fig:fits}
\end{figure}

The~distributions 
of the~$\Ppsi\g\Kp$ and $\Ppsi\g$~masses for the~selected \Bu~candidates
corresponding to the full data sample
%% summed over Run\,1 and Run\,2 data\nobreakdash-taking periods 
are shown in Fig.~\ref{fig:fits}.
For better visibility 
the~$\Ppsi\g\Kp$~mass spectra are shown for candidates 
within
the~narrow $\Ppsi\g$~mass regions around the~known 
$\chicone(3872)$~mass,
\mbox{$3.842<m_{\psitwos\g}<3.902\gevcc$} and 
\mbox{$3.782<m_{\jpsi\g}<3.962\gevcc$}. 
Similarly, 
the~$\Ppsi\g$~mass spectra are shown for candidates 
within 
the~narrow $\Ppsi\g\Kp$~mass regions around the~known 
mass of the~$\Bu$~meson,
%% \mbox{$5.258<m_{\psitwos\g\Kp}<5.300\gevcc $} 
%% and~\mbox{$5.258<m_{\jpsi\g\Kp}<5.300\gevcc$}.
\mbox{$5.258<m_{\Ppsi\g\Kp}<5.300\gevcc $}.
%\todo{fill numbers here!}.
%%
To~improve the~mass resolutions 
the~$\Ppsi\g\Kp$ and ~$\Ppsi\g$~masses
are calculated using 
the~kinematic fit
described above~\cite{Hulsbergen:2005pu}.
%% which constrains the~mass of the \jpsi or \psitwos candidate 
%% to its known value~\cite{PDG2022}, 
%% and the~\Bp~candidates to originate from 
%% its associated~PV. 
In~addition, 
%% for calculation of 
to calculate 
the~\mbox{$\Ppsi\g\Kp$}~mass,
the~$\Ppsi\g$~mass is constrained
to the~known value of the~$\chicone(3872)$~state~\cite{LHCb-PAPER-2020-008,
LHCb-PAPER-2020-009,
PDG2022}. 
%% when calculating the~\mbox{$\Ppsi\g\Kp$}~mass.
%% 
Clear signals
corresponding to both \Bp~and~$\chicone(3872)$~states 
are seen in data for both 
%% signal 
%% and normalisation 
channels.

%%%%%% Mass spectra 
%%%%Distributions 
%%%%of the~(left)~$\Ppsi\g\Kp$ 
%%%%and (right)~$\Ppsi\g$~mass of selected \Bu~candidates
%%%%%% \mbox{$\decay{\Bp}{\left(\decay{\chicone(3872)}{\PPsi\g}\right)\Kp}$}~canddates
%%%%%% (top)~\mbox{$\decay{\Bp}{\left(\decay{\chicone(3872)}{\psitwos\g}\right)\Kp}$} 
%%%%%% and  
%%%%%% (bottom)~
%%%%%% \mbox{$\decay{\Bp}{\left(\decay{\chicone(3872)}{\psitwos\g}\right)\Kp}$}
%%%%%% candidates 
%%%%%% \mbox{$\decay{\Bp}{\left(\decay{\chicone(3872)}{\Ppsi\g}\right)\Kp}$}~candidates 
%%%%%%\mbox{$\decay{\Bp}{\chicone(3872)\Kp}$}~candidates 
%%%%summed over Run\,1 and Run\,2
%%%%data\nobreakdash-taking periods.
%%%%%% 
%%%%Top and bottom rows correspond to 
%%%%the~\mbox{$\decay{\Bp}{\left(\decay{\chicone(3872)}{\psitwos\g}\right)\Kp}$}
%%%%and 
%%%%\mbox{$\decay{\Bp}{\left(\decay{\chicone(3872)}{\jpsi\g}\right)\Kp}$}
%%%%candidates, respectively.
%%%%%% 
%%%%The~$\Ppsi\g\Kp$~mass spectra are shown for candidates
%%%%within
%%%%the~narrow $\Ppsi\g$~mass regions around the~$\chicone(3872)$~mass,
%%%%and vice versa, 
%%%%the~$\Ppsi\g$~mass spectra are shown for candidates
%%%%within
%%%%the~narrow $\Ppsi\g\Kp$~mass regions around the~$\Bu$~mass.
%%%%%% 
%%%%Projections of the~fit, described in the~text, are overlaid.
%%%%% 

\section{Signal yield determination}
\label{sec:sig}

Signal yields of both
%% \mbox{$\Bu\to\left(\chicone(3872)\to\psitwos\g\right)\Kp$}
%%  and~\mbox{$\Bu\to\left(\chicone(3872)\to\jpsi\g\right)\Kp$} 
\mbox{$\Bu\to\left(\chicone(3872)\to\Ppsi\g\right)\Kp$}~decay channels
are determined 
using extended unbinned maximum\nobreakdash-likelihood fits
to two\nobreakdash-dimensional distributions 
of $\Pm_{\Ppsi\Pgamma}$ and~$\Pm_{\Ppsi\Pgamma\Kp}$.
%$\left(m_{\psitwos\g},m_{\psitwos\g\Kp}\right)$ and $\left(m_{\jpsi\g},m_{\jpsi\g\Kp}\right)$.
%% 
%%
The~fit model for the~\mbox{$\Bu\to\left(\chicone(3872)\to\psitwos\g\right)\Kp$}~decay channel
consists of three components:
\begin{enumerate}
    \item a~signal component parameterised 
    as a~product of the~\Bu and $\chicone(3872)$~signal shapes, 
    both described by 
    a~modified Gaussian function with power\nobreakdash-law
    tails on both sides of the~Gaussian core~\cite{Skwarnicki:1986xj,LHCb-PAPER-2011-013};
    %where the~\Bu~signal shapes are parameterised by the~modified 
    %Gaussian function with power\nobreakdash-law 
    %tails on the~both sides of the~Gaussian 
    %core~\cite{Skwarnicki:1986xj,LHCb-PAPER-2011-013} 
    %and the~$\chicone(3872)$~signal shape is parameterised 
    %%
    %by sums of the~modified Gaussian function and
    %% a~bifurcated Gaussian function;
    %a~split normal distribution~\cite{fechner1897,Gibbons,JohnS};
    %%
    \item a background 
    component from the~\mbox{$\decay{\B}{\psitwos\Kp\PX}$}~decays, 
    whose shape 
    %% of this component
    is determined from the~simulation %% dedicated simulated sample
    as 
    \begin{subequations}
    \label{eq:legendre}
    \begin{equation}
        %% C\left( m_{\chicone(3872)\Kp} , m_{\psitwos\g}\right) 
        %% \propto 
        \sum\limits_{i=0}^{n}\sum\limits_{j=0}^{n}
        \upbeta_{ij} \mathscr{L}_i(u)\mathscr{L}_j(v) \,, 
     \end{equation}
    where $u$ and $v$ stand for 
    the~$\psitwos\g\Kp$ and  $\psitwos\g$~masses, reduced to 
    the~\mbox{$-1\le u,v \le1$}~interval,
    $n=10$ and 
    $\mathscr{L}_{i}(u)$~and 
    $\mathscr{L}_{j}(v)$~are Legendre polynomials.
    The~$\left(n+1\right)^2$ coefficients $\upbeta_{ij}$ are calculated as 
    \begin{equation}
        \upbeta_{ij} = 
        \sum_{k} \frac{2i+1}{2}\,\frac{2j+1}{2} \mathscr{L}_i(u) \mathscr{L}_j(v) \,, 
    \end{equation}
    \end{subequations} 
    where the~last sum runs over the~events in the~simulated  
    sample that passes all selection criteria;
    \item a~combinatorial background component parameterised with 
    a~two\nobreakdash-dimensional nonfactorisable 
    positive polynomial function
    \begin{equation}
    \label{eq:bernstein}
        %% B \left( m_{\chicone(3872)\Kp} , m_{\psitwos\g}\right) 
        %% \propto 
        \sum\limits_{i=0}^{n}\sum\limits_{j=0}^{n}
        \upalpha^2_{ij}\mathscr{B}^i_n(x)\mathscr{B}^j_n(y) \,, 
    \end{equation}
    where $x$ and $y$ stand for 
    the~$\psitwos\g\Kp$ and  $\psitwos\g$~masses, reduced to 
    the~\mbox{$0\le x,y \le1$}~interval,
    $n=2$,   
    $\mathscr{B}^i_n(x)$ and $\mathscr{B}^j_n(y)$~are the~basic Bernstein~polynomials
    and the~$\left(n+1\right)^2$ fit parameters 
    $\upalpha_{ij}$~are constrained such that
    $\sum_{ij}\upalpha^2_{ij}=1$. 
\end{enumerate} 
The~fit model for the~\mbox{$\Bu\to\left(\chicone(3872)\to\jpsi\g\right)\Kp$}~decays
consists of the~following four components:
\begin{enumerate}
    \item a~signal component parameterised as a~product of the~\Bu 
    and $\chicone(3872)$~signal shapes,
    where the~\Bu~signal shape is parameterised by the~modified 
    Gaussian function 
    %% with power\nobreakdash-law 
    %% tails on the~both sides of the~Gaussian 
    % core~\cite{Skwarnicki:1986xj,LHCb-PAPER-2011-013} 
    and the~$\chicone(3872)$~signal shape is parameterised 
    by a~sum of the~modified Gaussian function and
    %% a~bifurcated Gaussian function;
    a~bifurcated Gaussian distribution~\cite{fechner1897,Gibbons,JohnS};
    %, where
    %the~\Bu and $\chicone(3872)$~signal shapes
    %are parameterised as described above;
    %% 
    %% shapes are parameterised by the~modified Gaussian function  
    %% and the~$\chicone(3872)$~signal shapes are parameterised 
    %% by sums of the~modified Gaussian function and a~bifurcated Gaussian function;
    %%
    \item a~component describing  the background from partially reconstructed
    \mbox{$\decay{\B}{\jpsi\PX}$}~decays, in turn consisting of two 
    components: 
    \begin{itemize}
    \item a~background from the~\mbox{$\decay{\Bp}{\jpsi\left(\decay{\Kstarp}{\Kp\piz}\right)}$}~decay, 
     where the~shape of this component is determined from the~dedicated simulated sample, 
     following the~same approach as used in Eqs.~\eqref{eq:legendre};
   \item a~background from other
     unidentified~\mbox{$\decay{\B}{\jpsi\PX}$}~decays, 
    which is parameterised by 
    a~two\nobreakdash-dimensional Gaussian function;
   \end{itemize} 
    \item a~component describing 
    the~random \mbox{$\chicone(3872)\Kp$}~combinations 
    and parameterised as a~product of the~$\chicone(3872)$~signal shape and 
    a~constant function;
     \item a~combinatorial background described with 
    a~nonfactorisable positive polynomial function 
    similar to Eq.~\eqref{eq:bernstein} with $n=3$.
\end{enumerate}
%% \todo[inline,size=large]{In figure legend there are only 
%% 4 components. For plots (a,b) green dashed line - is (2) ? \\
%% Dasha:\\
%% (1) red shape;\\
%% (2) green dashed line;\\
%% (3) and (4) - merged in violet shaded shape;\\
%% (5) blue dashed line
%% }
The~tail and resolution parameters of the~signal shapes 
are fixed to the~values determined from simulation. 
The~resolution parameters are further corrected 
by scale factors, $s_{\Bu}$ and $s_{\Ppsi\g}$, 
which account for a~small discrepancy between data and 
simulation~\mbox{\cite{LHCb-PAPER-2020-009,
LHCb-PAPER-2020-035,
LHCb-PAPER-2021-034,
LHCb-PAPER-2021-047,
LHCb-PAPER-2022-025,
LHCb-PAPER-2023-039,
LHCb-PAPER-2023-046}}. 
These factors are constrained in the~fit using 
Gaussian constraints
with~values of 
%% \mbox{$s_{\Bu}=1.105\pm0.005$} 
%% and 
%% \mbox{$s_{\Ppsi\g}=1.046\pm0.005$}, 
\mbox{$s_{\Bu}=1.102\pm0.004$} 
and 
\mbox{$s_{\Ppsi\g}=1.027\pm0.004$}, 
obtained from the~analysis of a~large  %% a~high\nobreakdash-statistics 
sample 
of \mbox{$\decay{\Bp}{\left(\decay{\chicone}
{\jpsi\g}\right)\Kp}$}~decays~\cite{LHCb-PAPER-2023-039}.

The~fit is performed simultaneously over the four samples corresponding 
to the~\mbox{$\Bu\to\left(\chicone(3872)\to\psitwos\g\right)\Kp$}
and~\mbox{$\Bu\to\left(\chicone(3872)\to\jpsi\g\right)\Kp$}~decays 
and the~two data\nobreakdash-taking periods.
The~peak position parameters of 
the~signal shapes and the~$s_{\Bu}$ and $s_{\Ppsi\g}$~scale factors
are shared between the~samples. 
%%
%%
%%
%% Distributions 
%%  of the~$\Ppsi\g\Kp$ and ~$\Ppsi\g$~mass of selected \Bu~candidates
%% %% \mbox{$\decay{\Bp}{\left(\decay{\chicone(3872)}{\PPsi\g}\right)\Kp}$}~canddates
%% %% (top)~\mbox{$\decay{\Bp}{\left(\decay{\chicone(3872)}{\psitwos\g}\right)\Kp}$} 
%% %% and  
%% %% (bottom)~
%% %% \mbox{$\decay{\Bp}{\left(\decay{\chicone(3872)}{\psitwos\g}\right)\Kp}$}
%% %% candidates 
%% %% \mbox{$\decay{\Bp}{\left(\decay{\chicone(3872)}{\Ppsi\g}\right)\Kp}$}~candidates 
%% %%\mbox{$\decay{\Bp}{\chicone(3872)\Kp}$}~candidates 
%% summed over Run\,1 and Run\,2 data\nobreakdash-taking periods 
%% together with projections of the results of the~fit 
%% are shown in Fig.~\ref{fig:fits},
%%  
%% 
The~results of the~fit 
are overlaid in Fig.~\ref{fig:fits}
and the~yields of each~fit component are 
summarised in Table~\ref{tab:fits_short}.
The~statistical significance
for the~\mbox{$\decay{\Bu}{\left(\decay{\chicone(3872)}{\psitwos\g}\right)\Kp}$}~signal,
$\mathscr{S}_{\decay{\chicone(3872)}{\psitwos\g}}$, 
is calculated  using Wilks' theorem~\cite{Wilks:1938dza} 
separately for the~Run\,1 and Run\,2 data\nobreakdash-taking periods
and is also listed in~Table~\ref{tab:fits_short}.

\begin{table}[tb!]
\centering
\caption{Yields for the~fit components  
        determined from
        the~simultaneous 
        extended unbinned maximum\nobreakdash-likelihood fit.
    %% to selected 
    %%  \mbox{$\decay{\Bu}{\left(\decay{\chicone(3872)}{\Ppsi\g}\right)\Kp}$}~candidates.
    %%  of the~\mbox{$\decay{\Bu}{\left(\decay{\chicone(3872)}{\psitwos\g}\right)\Kp}$}
    %% and \mbox{$\decay{\Bu}{\left(\decay{\chicone(3872)}{\psitwos\g}\right)\Kp}$}~decays.
        Uncertainties are statistical only.
The~last row shows 
the~statistical significance
%, $\mathscr{S}_{\decay{\chicone(3872)}{\psitwos\g}}$, 
of the~\mbox{$\decay{\Bu}{\left(\decay{\chicone(3872)}{\psitwos\g}\right)\Kp}$}~signal.
	}
\begin{tabular*}{0.70\textwidth}{@{\hspace{3mm}}l@{\extracolsep{\fill}}ccc@{\hspace{3mm}}}
\multirow{2}{*}{Parameter} 
&
& 
\multicolumn{2}{c}{Data\nobreakdash-taking period}
\\
 & 
 &  Run\,1 
 &  Run\,2
\\[1.5mm]
\hline 
\\[-3.5mm]
\multicolumn{3}{c}{$\psitwos\g\Kp$}
\\[1.5mm]
\hline 
\\[-3.5mm]
$N_{\decay{\Bu}{\left(\decay{\chicone(3872)}{\psitwos\g}\right)\Kp}}$ 
&
& $\phantom{00}40  \pm 8\phantom{00}$    
& $\phantom{000}63 \pm 10\phantom{0} $ 
\\
$N_{\decay{\B}{\psitwos\Kp\PX}}$ 
&  
&  $\phantom{0}567  \pm 24\phantom{0}$   
&  $\phantom{00}885 \pm 29\phantom{0}$  
%% &  $\left[10^3\right]$
%% &  $ 0.567  \pm 0.024$   
%% &  $ 0.885  \pm 0.029$  
\\
$N_{\mathrm{comb}}$
&
&  $\phantom{00}55  \pm 17\phantom{0}$   
&  $\phantom{00}132 \pm 19\phantom{0}$  
\\[1.5mm]
\hline 
\\[-3.5mm]
\multicolumn{3}{c}{$\jpsi\g\Kp$}
\\[1.5mm]
\hline 
\\[-3.5mm]
$N_{\decay{\Bu}{\left(\decay{\chicone(3872)}{\jpsi\g}\right)\Kp}}$ 
%% & 
%% & $\phantom{0}426 \pm 28\phantom{0} $
%% & $\phantom{0}1686 \pm 53\phantom{0}$  
& $\left[10^3\right]$
& $ 0.43 \pm 0.03 $
& $ \phantom{0}1.69 \pm 0.05$  

\\
$N_{\decay{\B}{\jpsi\PX}}$ 
%% & 
%% &  $ 3608 \pm 105 $   
%% &  $18719 \pm 263 $  
&  $\left[10^3\right]$
&  $ 3.61 \pm 0.11 $   
&  $18.72 \pm 0.26 $  
\\
$N_{\chicone(3872)\Kp}$ 
%% &
%% &  $ 1178           \pm  59\phantom{0} $   
%% &  $\phantom{0}5527 \pm 229 $  
& $\left[10^3\right]$
&  $ 1.18 \pm  0.06 $   
&  $\phantom{0}5.53 \pm  0.23 $  
\\
$N_{\mathrm{comb}}$ 
%% &
%% &  $ 4051  \pm 105 $   
%% &  $17464  \pm 205 $  
& $\left[10^3\right]$
&  $ 4.05  \pm 0.11 $   
&  $17.46  \pm 0.21 $  
%% 
%%
%%  
%% 
%%        &  $N_{\decay{\Bu}{\left(\decay{\chicone(3872)}{\psitwos\g}\right)\Kp}}$ 
%%        &  $N_{\decay{\Bu}{\left(\decay{\chicone(3872)}{\jpsi\g}\right)\Kp}}$ 
%%        %%
%%    \\[1.5mm]
%%   \hline 
%%   \\[-1.5mm]
%%     %% 
%%     Run\,1  &  $40 \pm 8\phantom{0} $     &  $\phantom{0}426 \pm 28 $ 
%%     \\
%%     Run\,2  &  $65 \pm 10$                &  $1686 \pm 53 $ 
%% 
%% 
%% 
\\[1.5mm]
\hline 
\\[-3.5mm]
$\mathscr{S}_{\decay{\chicone(3872)}{\psitwos\g}}$ 
& 
&  \ \ $5.3\upsigma$ 
&  \ \ $6.7\upsigma$
\end{tabular*} 
	\label{tab:fits_short}
\end{table}

%% {\color{red}{
The~study of the fit projections 
in the~narrow $\Ppsi\Kp\g$ and 
$\Ppsi\g$~mass intervals shows  
a~good description of data, 
supporting the~chosen fit model, 
in particular, the absence of 
the~component describing 
the~random $\chicone(3872)\Kp$~combinations
for the~\psitwos~case.
The~inclusion of this component
is studied and the~effect is accounted for 
by systematic uncertainty 
in Sec.~\ref{sec:systematic}.
%% }}

To validate 
the~observation of 
the~\mbox{$\decay{\Bu}
{\left(\decay{\chicone(3872)}{\psitwos\g}\right)\Kp}$}~decay, 
several cross\nobreakdash-checks are performed. 
The~data are categorised into 
data\nobreakdash-taking periods with 
different polarity of the~\lhcb 
dipole magnet~\cite{LHCb-TDR-001}
and charge of the~\Bpm~candidate.
The~results are found to be consistent 
among all samples. %%  and analysis techniques.
Further, alternative 
techniques are tried 
for the~signal determination.
Namely,  
instead of using fits to~the~two\nobreakdash-dimensional mass 
distributions, fits to the~one\nobreakdash-dimensional 
$\Ppsi\g\Kp$~mass 
distributions are performed for events within
the~narrow region around the~known $\chicone(3872)$~mass.
%% 
%% when the~$\Ppsi\g\Kp$~mass is calculated 
%% with $\Ppsi\g$~mass constrained to the~known mass of 
%% the~$\chicone(3872)$~state. 
%%
The~results are found 
to be in agreement with the~baseline results.
Similarly, consistent results 
have been obtained also  
from the~fits to 
the~one\nobreakdash-dimensional $\Ppsi\g$~mass 
distributions for events 
within
the~narrow $\Ppsi\g\Kp$~mass region around the~known \Bu~mass,
when the~$\Ppsi\g$~mass is calculated 
with $\Ppsi\g\Kp$~mass constrained to the~known 
mass of  the~\Bu~meson.  
%%
%% The~results are found to be consistent 
%% among all samples and analysis techniques. 

\section{Branching 
fraction ratio computation}
\label{sec:bf}

The ratio of the~partial decay widths,
%% branching fractions, 
$\mathscr{R_{\uppsi\g}}$, 
defined in Eq.~\eqref{eq:r} %and~\eqref{eq:rB}, 
is calculated as %% the following
\begin{equation}
\label{eq:ratio}
\mathscr{R}_{\uppsi\g} 
%% =  \dfrac{\BR(\XPsiG)}{\BR(\XJpsiG)} 
=  \dfrac
{N_{\decay{\Bu}{ \left( \decay{\chicone(3872)}{\psitwos\g} \right)\Kp}} }
{N_{\decay{\Bu}{ \left( \decay{\chicone(3872)}{\jpsi\g} \right)\Kp}} }
  \times
   \dfrac
  {\Pvarepsilon_{\decay{\Bu}{ \left( \decay{\chicone(3872)}{\jpsi\g} \right)\Kp}} }
  {\Pvarepsilon_{\decay{\Bu}{ \left( \decay{\chicone(3872)}{\psitwos\g} \right)\Kp}} }
    \times
   \dfrac{\BF_{\decay{\jpsi}{\mumu}}}
         {\BF_{\decay{\psitwos}{\mumu}}}
   \,,    
\end{equation}
where $N$ denotes the signal yield 
from Table~\ref{tab:fits_short}, 
$\Pvarepsilon$ stands for  
the~total efficiency,  
and $\BF$~is %_{\decay{\Ppsi}{\mumu}}$
the~branching fraction of a~dimuon decay of
the~\Ppsi~mesons.
%% The signal yields are given in Table~\ref{tab:fits_short}). 
The~total efficiencies are the~products of detector acceptance, 
reconstruction, selection  and trigger efficiencies, and 
 are calculated using simulated samples, 
 calibrated to match the data as described in Sec.~\ref{sec:detector}. 
 The resulting ratios of efficiencies 
 are found to be 
 $3.51\pm0.08$ and 
 $5.15\pm0.07$ 
 for Run\,1 and   Run\,2 data\nobreakdash-taking periods,  respectively,
%% $5.15\pm0.07$ for Run\,2 data taking periods, 
%% $$\frac{\efftot_{\jpsi\g}}
%% {\efftot_{\psitwos\g}}\biggr|_{\mathrm{Run\,1}} = 3.51\pm0.08,$$
%% $$\frac{\efftot_{\jpsi\g}}
%% {\efftot_{\psitwos\g}}\biggr|_{\mathrm{Run\,2}} = 5.15\pm0.07,$$
where the~uncertainties are due to the~limited size of
the simulated samples. The~difference between the~ratios of efficiencies 
is mainly due to the~tighter requirement on the~response of the~{\sc{MLP}}~classifier 
applied to suppress larger background for Run\,2 data\nobreakdash-taking period.
Under the~assumption of lepton universality,
instead of the~ratio of 
branching fractions for 
the~dimuon decays of the~\jpsi~and 
\psitwos~mesons,
the~corresponding ratio
for their dielectron decays 
of $7.53\pm0.17$ is used~\cite{PDG2022}, 
which is known with a~smaller uncertainty.

%% 
%% The~ratio of branching fractions of charmonium decays 
%% into a~pair of muons is 
%% taken from Ref.~\cite{PDG2022} under assumption of lepton universality
%% \begin{equation}
%%    \dfrac{\BF_{\decay{\jpsi}{\mumu}}}
%%          {\BF_{\decay{\psitwos}{\mumu}}}
%% = 
%%   \dfrac{\BF_{\decay{\jpsi}{\epem}}}
%%          {\BF_{\decay{\psitwos}{\epem}}}
%% = 7.53\pm0.16 \,.
%% \end{equation}
%% 
The~value for the~ratio $\mathscr{R}_{\Ppsi\g}$ is calculated separately for 
the~two data\nobreakdash-taking periods and found to be  
\begin{eqnarray*}  
\mathscr{R}_{\Ppsi\g}^{\mathrm{Run\,1}} & = & 2.50\pm0.52 \,, %% \pm0.06, 
\\
\mathscr{R}_{\Ppsi\g}^{\mathrm{Run\,2}} & = & 1.49\pm0.23 \,, %% \pm0.03,
\end{eqnarray*}
%% where first uncertainty is statistical 
%% and the second one is due to ratio of charmonium decays branching fractions.
where the~uncertainty is statistical only.
Systematic uncertainties are discussed in the~next section.
%% In Sec.~\ref{sec:sum} these values are averaged after 
%% taking into account the systematic uncertainties 
%% to get the result for the whole datasample.

\section{Systematic uncertainties}
\label{sec:systematic}

The~studied channels share 
the~same set of final\nobreakdash-state particles,  
%% and the~same 
trigger and preselection requirements.
%% are applied to both.
%%
This~leads to 
a~significant cancellation 
of many systematic uncertainties 
in the~ratio determination.
The~remaining 
contributions are discussed below and summarised in Table~\ref{tab:syst}.

\begin{table}[hbt]
\centering
\caption{\small 
Relative systematic uncertainties (in~\,\%) in the~ratio of branching fractions.
The~total uncertainty is obtained as the~sum of individual components in quadrature.
}
\begin{tabular*}{0.85\textwidth}{@{\hspace{2mm}}l@{\extracolsep{\fill}}cc@{\hspace{1mm}}}
\multirow{2}{*}{Source}
& \multicolumn{2}{l}{Data\nobreakdash-taking period}
\\
& Run\,1~[\%]
& Run\,2~[\%]
\\[1.5mm]
\hline 
\\[-3.5mm]
Fit model                     &                   &                \\
\ \ \ Signal and combinatorial background 
& $\phantom{<}{^{+5.7}_{-0.1}}$ 
& $\phantom{<}{^{+4.4}_{-2.0}}$\\
\ \ \ \mbox{$\decay{\B}{\psitwos\Kp\PX}$} background
& \\ 
    %% ~~Parameterisation of the~\mbox{$\decay{\B}{\psitwos\Kp\PX}$} background
\ \ \ \ \ \ \ Parameterisation
& $\phantom{<}{^{\,+\,1.6}_{\,-\,4.9}}$ 
& $\phantom{<}{^{\,+\,5.0}_{\,-\,2.9}}$ \\
    %% ~~Content of the~\mbox{$\decay{\B}{\psitwos\Kp\PX}$} background 
\ \ \ \ \ \ \ Composition 
& $\phantom{<}0.9$ 
& $\phantom{<}1.9$ \\     
    %% ~~Simulation sample size of the~\mbox{$\decay{\B}{\psitwos\Kp\PX}$} background
\ \ \ \ \ \ \ Simulation sample size 
& $\phantom{<}4.2$ 
& $\phantom{<}4.3$ \\
    %\qquad uncertainty in $f_{corr}$ values & $\phantom{<}<0.1$ & $\phantom{<}<0.1$ \\
\ \ \ Additional components 
%% & $\phantom{<}2.1$ 
%% & $\phantom{<}2.6$ 
& $\phantom{<}^{\,+\,0.6}_{\,-\,4.4}$ 
& $\phantom{<}^{\,+\,1.2}_{\,-\,2.6}$ 
\\
%% $\Bu$~meson kinematics & $\phantom{<}0.06$  & $\phantom{<}0.08$ 
$\Bu$~meson kinematics 
& $<0.1$  
& $<0.1$ 
\\
%% Track reconstruction           & $\phantom{<}0.02$ & $\phantom{<}0.03$ 
Track reconstruction           
& $<0.1$ 
& $<0.1$ 
\\
%% Photon reconstruction          & $\phantom{<}1.11$  & $\phantom{<}1.14$ 
Photon reconstruction          
& $\phantom{<}1.1$  
& $\phantom{<}1.1$  
\\
Kaon   identification          
& $\phantom{<}1.0$  
& $\phantom{<}1.3$  
\\
Trigger                        
& $\phantom{<}1.1$  
& $\phantom{<}1.1$ 
\\
%% MVA selection                            & $\phantom{<}1.0$  &
Data-simulation (dis)agreement   
& $\phantom{<}1.0$  
& $\phantom{<}^{\,+\,1.0}_{\,-\,1.5}$ 
\\
Simulation sample size for efficiency    
& $\phantom{<}2.3$  
& $\phantom{<}1.4$ 
\\[1.5mm]
\hline 
\\[-3.5mm]
Total 
& $\phantom{<}^{\,+\,8.0}_{\,-\,9.2}$ 
& $\phantom{<}^{\,+\,8.7}_{\,-\,7.9}$\\
\end{tabular*}
\label{tab:syst}
\end{table}

An~important source of systematic uncertainty
is associated with the~fit model.
The~systematic uncertainty related to 
the~description of the~signal fit components 
is estimated  using alternative models. 
A~set of pseudoexperiments based on 
the~baseline fit model is produced and
each pseudoexperiment is fit 
to the~alternative model
and the ratio of 
yields for 
the~\mbox{$\decay{\Bu}{\left(\decay{\chicone(3872)}{\psitwos\g}\right)\Kp}$}
and~\mbox{$\decay{\Bu}{\left(\decay{\chicone(3872)}{\jpsi\g}\right)\Kp}$}
signal components is calculated. 
The~mean value of this ratio 
from the~pseudoexperiments 
%% over the~generated set of pseudoexperiments  
is compared with the~value from the~baseline fit 
and the~largest deviation over the~considered 
list of alternative models
is taken as the corresponding systematic 
uncertainty. 
%The list of alternative models for the~\Bu~signal shape 
%contains
The~alternative models used for the \Bp signal shape are:
\begin{itemize}
\item 
a~modified asymmetric 
Apollonios function~\cite{MartinezSantos:2013ltf};
\item 
a~generalised asymmetric 
Student's $t$-distribution~\cite{Student,Jackman,ZHU2010297,LiNadarjah};
\item
a~modified Novosibirsk function~\cite{BaBar:2011qjw}.
\end{itemize}
The~list of alternative models for 
the~$\chicone(3872)$~signal
consists of the sum of a~model 
from the~list above 
with a~bifurcated Gaussian distribution or 
the~modified Gaussian function, 
and the~sum of 
the~modified Gaussian function
with a~Gaussian function.
The~same technique is applied to 
estimate the~systematic uncertainty
associated with parameterisation of
the~combinatorial background 
components. The~two-dimensional 
nonfactorisable 
positive polynomials 
from Eq.~\eqref{eq:bernstein} 
with $n=1$ or $n=3$ are tested 
as alternative 
models 
for 
%%  the~\mbox{$\decay{\Bu}{\left(\decay{\chicone(3872)}{\psitwos\g}\right)\Kp}$}~case 
the~\mbox{$\psitwos\g\Kp$}~case 
and  with $n=2$
for 
%% the~\mbox{$\decay{\Bu}{\left(\decay{\chicone(3872)}{\jpsi\g}\right)\Kp}$}~case.
the~\mbox{$\jpsi\g\Kp$}~case. 
The~largest positive  and negative 
relative deviations 
over all alternative signal and combinatorial 
background models are found to be 
$^{\,+\,5.7}_{\,-\,0.1}\,\%$  for the~Run\,1 and 
$^{\,+\,4.4}_{\,-\,2.0}\,\%$  for the~Run\,2 data\nobreakdash-taking 
periods, respectively. 
These values are
taken as the~corresponding systematic uncertainty. 

An~additional systematic uncertainty related to the~description of
the~background from the~\mbox{$\decay{\B}{\psitwos\Kp\PX}$}~decays,
is computed 
%% %%
%% %A~systematic uncertainty due to parameterisation %of this fit component
%% It is estimated 
with~alternative models
using  the~technique described above. 
The~list of alternative models consists of 
the~two dimensional function from Eq.~\eqref{eq:legendre}
with $n=12$ and $15$; 
two-dimensional histograms with 
different binning schemes 
with no interpolation, 
bilinear, biquadratic 
and bicubic interpolations. 
The~largest positive and negative 
relative deviations over all alternative
models are found  to be 
$^{\,+\,1.6}_{\,-\,4.9}\,\%$  for the~Run\,1 and 
$^{\,+\,5.0}_{\,-\,2.9}\,\%$  for the~Run\,2 data\nobreakdash-taking 
periods, respectively. 
These values are
taken as the~corresponding systematic uncertainties
due to the~parameterisation of 
the~background from the~\mbox{$\decay{\B}{\psitwos\Kp\PX}$}~decays.

Imprecise knowledge of the~branching fractions for 
the~individual \mbox{$\decay{\B}{\psitwos\Kp\PX}$}~decays
affects the~composition of the~simulated sample, 
and therefore the~shape of the~corresponding 
fit component.
To~estimate the~associated systematic uncertainty, 
a~set of dedicated pseudoexperiments is performed.
%% For~each pseudoexperiment,   
%% a~new simulation sample has been prepared
%% where the~branching fractions
%% for the~\mbox{$\decay{\B}{\psitwos\Kp\PX}$}~decays
%% are sampled from the~known values, 
%% and 
%% %% the~obtained sample 
%% it
%% is
%% used for the parameterisation of 
%% the~corresponding fit component.
%% 
For~each pseudoexperiment, a set of per-event weights 
are derived by sampling random values from a~Gaussian distribution 
of the known \mbox{$\decay{\B}{\psitwos\Kp\PX}$}~branching fraction 
and its uncertainty~\cite{PDG2022,
Belle:2010wrf,
Belle:2013shl,
LHCb-PAPER-2016-019}.
The weights are applied to the~simulation sample and a~new
parameterisation of the corresponding fit component 
is determined for each pseudoexperiment. 
Subsequently, the~fit is performed
and the~ratio of 
the~yields for
the~\mbox{$\decay{\Bu}{\left(\decay{\chicone(3872)}{\psitwos\g}\right)\Kp}$}
and~\mbox{$\decay{\Bu}{\left(\decay{\chicone(3872)}{\jpsi\g}\right)\Kp}$}~signals 
is calculated.
The~root mean square of the~relative deviation 
of the~ratio from the~baseline fit result 
over the~pseudoexperiments is found to 
be $0.9\,\%$ and $1.9\,\%$ for 
the~Run\,1 and Run\,2 data\nobreakdash-taking periods,
which is taken as the~size of 
the~associated systematic uncertainty.
%These values are
%taken as the~corresponding systematic uncertainty
%due to the~composition
%of the~background 
%from the~\mbox{$\decay{\B}{\psitwos%\Kp\PX}$}~decays.

The~uncertainty associated with the~finite size of 
the~simulated sample is estimated using 
similar types of pseudoexperiments.
In~each pseudoexperiment,  
a~new dataset is drawn 
from the~two-dimensional 
%% distribution 
histogram 
based on 
the~statistical uncertainty of the~simulated 
background. 
The~corresponding systematic uncertainty 
is found to be 
$4.2\,\%$  for the~Run\,1 and 
$4.3\,\%$  for Run\,2 data\nobreakdash-taking periods.

The addition of other components in the~fit, 
namely the~components describing 
possible contributions from 
 random \mbox{$\left(\decay{\chicone(3872)}{\psitwos\g}\right)\Kp$}~combinations,
\mbox{$\decay{\Bu}{\psitwos\g\Kp}$} and 
\mbox{$\decay{\Bu}{\jpsi\g\Kp}$}~decays,
as well as alternative parameterisation 
of the~\mbox{$\left(\decay{\chicone(3872)}{\jpsi\g}\right)\Kp$}~fit 
component, causes only a~small change with respect to the~baseline result.
All these components are parameterised 
as a~product of the~corresponding signal function 
%and a~constant function or 
%the~first or second order positive polynomial function.
%% positive polynomial 
%% of the~first or second order.
%%
with plynomial function up to order two.
The~obtained yields for all additional components 
are found to be small and consistent with zero, 
and the~maximal positive and negative  
relative deviations with respect to 
the~baseline fit are found to be 
$^{\,+\,0.6}_{\,-\,4.4}\,\%$  for Run\,1 and 
$^{\,+\,1.2}_{\,-\,2.6}\,\%$  for Run\,2 data\nobreakdash-taking periods.
%%
%These values are taken as systematic uncertainty %related to 
%additional fit component and parameterisation
%of the~\mbox{$\left(\decay{\chicone(3872)}{\jpsi%\g}\right)\Kp$}~fit 
%component.

The~transverse momentum and rapidity spectra for the~\Bu~mesons 
are corrected via an iterative procedure using the~\mbox{$\decay{\Bu}{\jpsi\Kp}$}~control channel.
The~finite size of
the~$\decay{\Bp}{\jpsi\Kp}$ signal sample 
%% used for correction 
%%  of the~simulated
%% transverse momentum and rapidity spectra 
%% of \Bp~mesons 
induces 
an~uncertainty on 
the~\Bu~meson \pt and $y$~spectra.
In~turn, this uncertainty 
induces a corresponding uncertainty
in the~ratio of efficiencies.
It is estimated by using corrections from
a~prior iteration
%% an iteration prior 
%% to the~last one 
and checking the deviation of the~ratio from 
the~baseline value.
This uncertainty is found to be 
%% small, $<0.1\,\%$ 
smaller than~$0.1\,\%$ for both 
Run\,1 and Run\,2 data\nobreakdash-taking periods. %\todo{How it is done?}
%% 
%% The~corresponding spread of these changes amounts to $0.06\,\%$
%% in Run\,1 and $0.08\,\%$ in Run\,2 
%% and is taken as the~systematic uncertainty
%% related to the~\Bu~meson kinematic.

There are residual differences in 
the~tracking reconstruction efficiency 
%% of charged\nobreakdash-particle tracks
that 
do not cancel out in the~efficiency ratio,
given 
the~slightly~different kinematic distributions 
of the~final\nobreakdash-state particles.
The~track\nobreakdash-finding efficiencies 
obtained from simulated samples are corrected using
\mbox{$\decay{\jpsi}{\mumu}$}
calibration channels~\cite{LHCb-DP-2013-002}.
The~uncertainties related to~the~efficiency 
correction factors are propagated to the~ratios of 
the~total efficiencies using pseudoexperiments, 
and are found to be
less than $0.1\,\%$ 
for both
Run\,1 and Run\,2. %%  data\nobreakdash-taking periods.
%% 0.02\,\% for Run\,1 and 0.03\,\% for Run\,2. 
%This value is taken as the~systematic uncertainty 
%due to the~tracking efficiency calibration.

Differences in the~photon reconstruction efficiencies 
between data and simulation 
are studied using a~large sample of 
\mbox{$\decay{\Bu}{\jpsi\Kstarp}$}~decays, 
reconstructed with 
the~\mbox{$\decay{\Kstarp}{\Kp\left(\decay{\piz}{\g\g}\right)}$}
decay mode~\cite{LHCb-PAPER-2012-022,
LHCb-PAPER-2012-053,
Govorkova:2015vqa,
Govorkova:2124605,
Belyaev:2015nlt}.
The~uncertainty 
on the~correction factors 
%%  due to  
%% the~finite size of the~sample
is propagated to the~ratio of 
the~total efficiencies 
using pseudoexperiments
and is found
to be 
$1.1\,\%$
for both
Run\,1 and Run\,2. %%  data\nobreakdash-taking periods.

The~kaon identification variable used for the~{\sc{MLP}}~estimator 
is drawn from 
%% data 
\mbox{$\decay{\Dstarp}{\left(\decay{\Dz}{\Km\pip}\right)\pip}$}
calibration 
samples~\cite{LHCb-DP-2018-001}
and has a~dependence on the~particle
kinematics and track multiplicity in the~event. 
Systematic uncertainties
%% in this procedure
arise from 
the~limited size 
of both the~simulation 
and calibration samples, 
and the~modelling of the~particle identification variable. 
The~limitations due to
the~size of the~simulation 
and calibration samples 
are evaluated by 
using bootstrapping
techniques~\cite{efron:1979,efron:1993}
to create 
%% creating  
multiple samples
and repeating the~procedure 
for each sample. 
The~impact of potential mismodelling of the~kaon identification  
variable is evaluated by describing 
the~corresponding distributions using 
density estimates with different kernel 
widths~\cite{LHCb-DP-2018-001,Poluektov:2014rxa}. 
For~each of these cases, alternative efficiency maps are 
produced to determine the~associated uncertainties. 
Systematic uncertainties of 1.0\,\% for Run\,1 and 1.3\,\% for Run\,2 are assigned from 
the~observed differences obtained with the~alternative~maps. 

A~systematic uncertainty  related 
to the~knowledge of the~trigger efficiencies 
was previously studied using large 
samples of \mbox{$\decay{\Bu}
{\left(\decay{\jpsi}{\mumu}\right)\Kp}$} 
and \mbox{$\decay{\Bu}
{\left(\decay{\psitwos}{\mumu}\right)\Kp}$}~decays
by comparing the 
ratios of the trigger efficiencies in data 
and simulation~\cite{LHCb-PAPER-2012-010}.
Based on this comparison, a~relative uncertainty 
of 1.1\,\% is assigned for 
both data\nobreakdash-taking periods. 

The~remaining discrepancy between 
data and simulation, 
not explicitly covered above,   
% by the~effects discussed above, 
%% in Sec.~\ref{sec:detector}, 
is estimated using 
a~large sample of 
the~\mbox{$\decay{\Bp}{\left( 
\decay{\chicone}{\jpsi\g}\right) \Kp}$}~decays~\cite{LHCb-PAPER-2023-039}.
The~same preselection and {\sc{MLP}} classifier are applied to 
the~control channel as for 
the~\mbox{$\decay{\Bu}{\left(\decay{\chicone(3872)}{\psitwos\g}\right)\Kp}$} channel. 
The~systematic uncertainty is estimated by varying 
the~requirement on the~response
of the~{\sc{MLP}} classifier in the~full range.
The~resulting difference 
in the~data\nobreakdash-simulation efficiency ratio 
is found to be~$1.0\,\%$ in Run\,1 and 
%% from $-1.5\,\%$ to $+1.0\,\%$
$^{\,+\,1.0}_{\,-\,1.5}\,\%$~in~Run\,2. 
%% $+1.0$ to $-1.5\,\%$ 

The~finite size of the~simulation samples 
leads to an~uncertainty on the~ratios 
of total efficiencies, which translates to 
an~uncertainty in the~efficiency ratios 
of $2.3\,\%$ for Run\,1 and $1.4\,\%$ for Run\,2.
%these ones translate to 
%uncertainty in the~efficiency ratios 
%and are found 
%to be $2.3\,\%$ for Run\,1 and $1.4\,\%$ for %Run\,2.
%%
The~total relative systematic uncertainties 
on the~ratio of branching fractions 
$\mathscr{R}_{\Ppsi\Pgamma}$
are calculated as 
the~sum in quadrature of all the~values 
described above and are found 
to be $^{\,+\,8.0}_{\,-\,9.2}\,\%$ 
for Run\,1 and $^{\,+\,8.7}_{\,-\,7.9}\,\%$ for Run\,2.
%$10.8\,\%$ for Run\,1 and $7.9\,\%$ for Run\,2.

The~statistical significance 
of the~\mbox{$\decay{\X}{\psitwos\Pgamma}$}~decay
is recalculated using Wilks' theorem 
for each alternative fit model, 
and the~smallest values of 
$4.8$ and $6.0$~standard
deviations
for the~Run\,1~and
Run\,2 samples, respectively, 
are taken as the~significance 
%% that accounts 
including 
the~systematic uncertainty.

\section{Results and summary}
\label{sec:sum}
\begin{figure}[t]
\setlength{\unitlength}{1mm}
\centering
\begin{picture}(150,135)
%%
%% \graphpaper[5](-10,-10)(170,145)
%% 
\put(0, 5){
\includegraphics*[width=75mm]{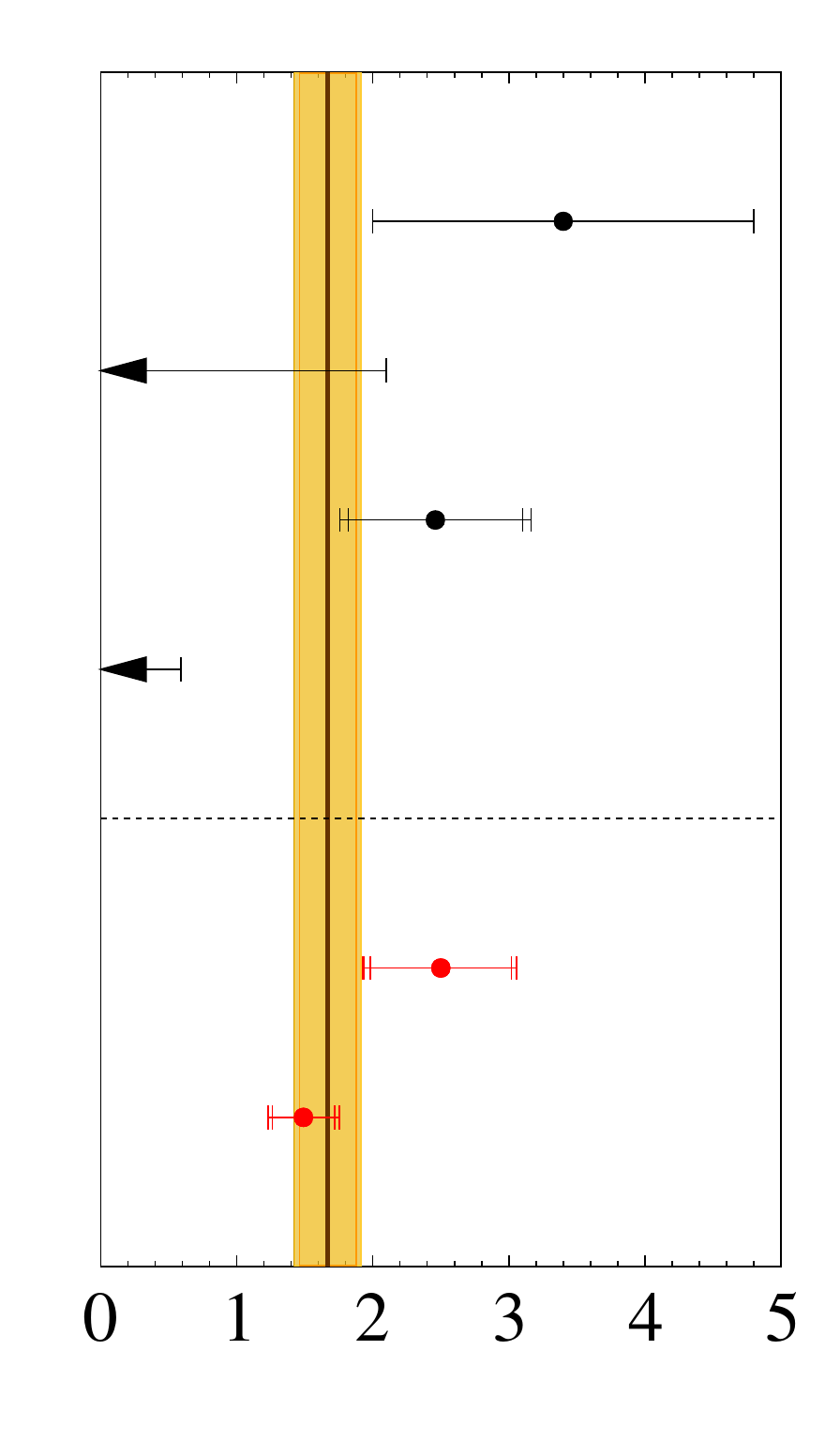} 
}
	%
%% \put(73,72){\Large\begin{Tabular}[1.74]{lll} 
\put(73,72){\Large\begingroup\renewcommand*{\arraystretch}{1.74}\begin{tabular}{lll} 
    \babar        & 2008        &  \cite{BaBar:2008flx}        \\
    \belle        & 2011        &  \cite{Belle:2011wdj}        \\
    \lhcb/Run\,1  & 2014        &  \cite{LHCb-PAPER-2014-008}  \\ 
    \besiii       & 2020        &  \cite{BESIII:2020nbj}       \\
    & & 
    \\
    \lhcb/Run\,1  & 2024        &   \\
    \lhcb/Run\,2  & 2024        & 
\end{tabular}\endgroup}
   \put(15,5){\large$\mathscr{R}_{\Ppsi\g}=\dfrac{\Gamma_{\decay{\chicone(3872)}{\psitwos\g}}}
                           {\Gamma_{\decay{\chicone(3872)}{\jpsi\g}}}$}

\end{picture}
\caption{ \small
 Summary of experimental results for 
 the ratio of the partial decay widths  for
 the radiative {${\chicone(3872)}\to{\Ppsi{(2\PS)}\g}$}
 and {${\chicone(3872)}\to{\PJ/\Ppsi\g}$} decays.
 The results from this analysis
 for the Run\,1 and Run\,2 data sets
 are shown as red points with error bars.
 The coloured band corresponds to the average of 
 the \lhcb results in this paper. Where provided, 
 the outer uncertainties bars correspond to total measurement uncertainties, while inner ones are for statistical uncertainties only.
 }
\label{fig:results}
\end{figure}

The~decay \mbox{$\decay{\Bu}{\chicone(3872)\Kp}$} is  exploited to 
study the~radiative decays  of the~$\chicone(3872)$~state into 
 $\psitwos\g$  and $\jpsi\g$~final states
using data collected by the~\lhcb experiment 
in $\proton\proton$~collisions at 
centre\nobreakdash-of\nobreakdash-mass 
energies of 7, 8 and 13\tev and 
corresponding to an~integrated 
luminosity of~$9\invfb$.
The~significance of the~\mbox{$\decay{\Bu}
{\left(\decay{\chicone(3872)}{\psitwos\g}\right)\Kp}$}~signal
is found to be $4.8$ and $6.0$~standard deviations
for the~Run\,1 and Run\,2~data\nobreakdash-taking periods,
%% corresponding to
which is 
the~first observation 
of the~\mbox{$\decay{\chicone(3872)}{\psitwos\g}$}~decay.
The~ratio of 
branching fractions 
for 
the~\mbox{$\decay{\Bu}{\left(\decay{\chicone(3872)}{\psitwos\g}\right)\Kp}$} 
and~\mbox{$\decay{\Bu}{\left(\decay{\chicone(3872)}{\jpsi\g}\right)\Kp}$}~decays 
is measured separately for the~Run\,1 and  Run\,2~data\nobreakdash-taking periods. 
This ratio is interpreted as the~ratio of the~partial decay widths 
for the~\mbox{$\decay{\chicone(3872)}{\Ppsi\g}$}~decays 
from Eq.~\eqref{eq:r}
%%  and \mbox{$\decay{\chicone(3872)}{\jpsi\g}$}~is 
%%  measured separately for two data taking periods 
\begingroup
\allowdisplaybreaks
\begin{eqnarray*}
\mathscr{R}_{\Ppsi\g}^{\mathrm{Run\,1}} 
& = & 2.50\pm0.52^{\,+\,0.20}_{\,-\,0.23} \pm0.06  \,, 
\\
\mathscr{R}_{\Ppsi\g}^{\mathrm{Run\,2}} 
& = & 1.49\pm0.23^{\,+\,0.13}_{\,-\,0.12} \pm0.03  \,, 
\end{eqnarray*}
\endgroup
where the~first uncertainty is statistical, 
the~second systematic and the~third 
due to the~uncertainties on the~ratio of branching 
fractions of \psitwos and \jpsi~mesons
into the~dilepton final state. 
%% 
%% The~significance of the~\mbox{$\decay{\Bu}
%% {\left(\decay{\chicone(3872)}{\psitwos\g}\right)\Kp}$}~signal
%% is found to be $4.8$ and $6.0$~standard deviations
%% for the~Run\,1 and Run\,2~data\nobreakdash-taking period, respectively,
%% corresponding to the~first observation 
%% of the~\mbox{$\decay{\chicone(3872)}{\psitwos\g}$}~decay.
%% 
The~ratio~$\mathscr{R}_{\Ppsi\g}^{\mathrm{Run\,1}}$
%% of branching fractions obtained for the~Run\,1~dataset 
is in good agreement with
(and supersedes) 
the~value 
of \mbox{$\mathscr{R}_{\Ppsi\Pgamma} = 2.46\pm0.70$}, 
obtained in the~previous 
study~\cite{LHCb-PAPER-2014-008}.
The~results for the~Run\,1 and Run\,2~data\nobreakdash-taking periods are combined 
using the~best linear unbiased estimator~\cite{LYONS1988110}
accounting for the~correlated systematic uncertainties.
The~obtained value for the~ratio $\mathscr{R}_{\Ppsi\g}$ is found to~be 
\begin{equation*}
%% \mathscr{R}_{\Ppsi\g} = 1.6706\pm0.2105\pm0.1189\pm0.0369.
\mathscr{R}_{\Ppsi\g} = 1.67 \pm 0.21 \pm 0.12 \pm0.04 \,.
\end{equation*}
A~summary of experimental results for 
 the~ratio of branching fractions of
 the~\mbox{$\decay{\chicone(3872)}{\psitwos\g}$}
 and \mbox{$\decay{\chicone(3872)}{\jpsi\g}$}~decays
 is presented in Fig.~\ref{fig:results}. 
% 
%% This value is 
The~combined ratio of partial 
radiative widths from this study is 
below the upper limit set by 
the~\belle~\cite{Belle:2011wdj} collaboration
and consistent with the~previous measurements 
by the~\babar~\cite{BaBar:2008flx} 
and \lhcb~\cite{LHCb-PAPER-2014-008}~collaborations.
However, it is notably in tension with
%though it is in a~clear tension with
the~upper 
limit set by the~\besiii~collaboration~\cite{BESIII:2020nbj}.

%%%%\caption{ \small
%%%% Summary of experimental results for 
%%%% the~ratio of the partial decay widths 
%%%% for
%%%% the~radiative  \mbox{$\decay{\chicone(3872)}{\psitwos\g}$}
%%%% and \mbox{$\decay{\chicone(3872)}{\jpsi\g}$}~decays.
%%%% %% 
%%%% The~results from this analysis
%%%% for the Run\,1 and Run\,2~data sets
%%%% are shown as red points with error bars.
%%%% The~coloured band corresponds to the average of 
%%%% the~\lhcb results in this paper. Where provided, 
%%%% the outer uncertainties bars correspond to total measurement uncertainties, while inner ones are for statistical uncertainties only.
%%%% }

%% \todo[inline,size=large]{VB: I need t update Fig.~\ref{fig:results}!}
 
The~large measured value of the~$\mathscr{R}_{\Ppsi\g}$~ratio 
is generally inconsistent with the~calculations 
based on the~pure $\D\Dstarb$~molecular hypothesis 
for the~$\chicone(3872)$~state~\cite{Swanson:2003tb,
Dong:2009uf,
Rathaud:2016tys,
Grinstein:2024rcu}
unless 
some special assumptions are 
made~\cite{Guo:2014taa,Molnar:2016dbo}.
On~the~contrary,
it agrees 
with a~broad range of
predictions based on other hypotheses of the~$\chicone(3872)$~structure,
including  conventional 
$\cquark\cquarkbar$~charmonium~\cite{
Barnes:2003vb,
Barnes:2005pb, 
DeFazio:2008xq,
Li:2009zu, 
Dong:2009uf,
Badalian:2012jz,
Ferretti:2014xqa,
Badalian:2015dha,
Deng:2016stx,
Giacosa:2019zxw},
$\cquark\cquarkbar\quark\quarkbar$~tetraquark~\cite{Grinstein:2024rcu},
and molecules  mixed with a~sizeable compact component~\cite{
Ortega:2012rs,
Takeuchi:2016hat,
Cincioglu:2016fkm,
Molnar:2016dbo,
Guo:2014taa}.
This~measurement 
provides a~strong 
argument 
in favour of 
a~compact component 
in the~$\chicone(3872)$~structure.

% Do not include this in any draft (just for information in the template)
%% \input{acknowledgements_intro}
% Comment this in for paper drafts; do not include this in analysis note, conference and figure reports

\section*{Acknowledgements}
%
% These Acknowledgements valid from 3-May-2019
%
\noindent We express our gratitude to our colleagues in the CERN
accelerator departments for the excellent performance of the LHC. We
thank the technical and administrative staff at the LHCb
institutes.
We acknowledge support from CERN and from the national agencies:
CAPES, CNPq, FAPERJ and FINEP (Brazil); 
MOST and NSFC (China); 
CNRS/IN2P3 (France); 
BMBF, DFG and MPG (Germany); 
INFN (Italy); 
NWO (Netherlands); 
MNiSW and NCN (Poland); 
MCID/IFA (Romania); 
%MSHE (Russia); 
MICINN (Spain); 
SNSF and SER (Switzerland); 
NASU (Ukraine); 
STFC (United Kingdom); 
DOE NP and NSF (USA).
We acknowledge the computing resources that are provided by CERN, IN2P3
(France), KIT and DESY (Germany), INFN (Italy), SURF (Netherlands),
PIC (Spain), GridPP (United Kingdom), 
%RRCKI and Yandex LLC (Russia), 
CSCS (Switzerland), IFIN-HH (Romania), CBPF (Brazil),
and Polish WLCG (Poland).
We are indebted to the communities behind the multiple open-source
software packages on which we depend.
Individual groups or members have received support from
ARC and ARDC (Australia);
Key Research Program of Frontier Sciences of CAS, CAS PIFI, CAS CCEPP, 
Fundamental Research Funds for the Central Universities, 
and Sci. \& Tech. Program of Guangzhou (China);
Minciencias (Colombia);
EPLANET, Marie Sk\l{}odowska-Curie Actions, ERC and NextGenerationEU (European Union);
A*MIDEX, ANR, IPhU and Labex P2IO, and R\'{e}gion Auvergne-Rh\^{o}ne-Alpes (France);
%RFBR, RSF and Yandex LLC (Russia);
AvH Foundation (Germany);
ICSC (Italy); 
GVA, XuntaGal, GENCAT, Inditex, InTalent and Prog.~Atracci\'on Talento, CM (Spain);
SRC (Sweden);
the Leverhulme Trust, the Royal Society
 and UKRI (United Kingdom).

%% \clearpage 
%% \input{supplementary-app}

\clearpage 
\addcontentsline{toc}{section}{References}
%\setboolean{inbibliography}{true}
\bibliographystyle{LHCb}
\bibliography{main,standard,LHCb-PAPER,LHCb-CONF,LHCb-DP,LHCb-TDR}

\ifx\mcitethebibliography\mciteundefinedmacro
\PackageError{LHCb.bst}{mciteplus.sty has not been loaded}
{This bibstyle requires the use of the mciteplus package.}\fi
\providecommand{\href}[2]{#2}
\begin{mcitethebibliography}{100}
\mciteSetBstSublistMode{n}
\mciteSetBstMaxWidthForm{subitem}{\alph{mcitesubitemcount})}
\mciteSetBstSublistLabelBeginEnd{\mcitemaxwidthsubitemform\space}
{\relax}{\relax}

\bibitem{Choi:2007wga}
Belle collaboration, S.~K. Choi {\em et~al.},
  \ifthenelse{\boolean{articletitles}}{\emph{{Observation of a~resonance-like
  structure in the~\mbox{$\pion^\pm \Ppsi^\prime$}~mass distribution in
  exclusive \mbox{$\decay{\B}{\kaon \pion^{\pm}\Ppsi^\prime}$}~decays}},
  }{}\href{https://doi.org/10.1103/PhysRevLett.100.142001}{Phys.\ Rev.\ Lett.\
  \textbf{100} (2008) 142001},
  \href{http://arxiv.org/abs/0708.1790}{{\normalfont\ttfamily
  arXiv:0708.1790}}\relax
\mciteBstWouldAddEndPuncttrue
\mciteSetBstMidEndSepPunct{\mcitedefaultmidpunct}
{\mcitedefaultendpunct}{\mcitedefaultseppunct}\relax
\EndOfBibitem
\bibitem{Mizuk:2009da}
Belle collaboration, R.~Mizuk {\em et~al.},
  \ifthenelse{\boolean{articletitles}}{\emph{{Dalitz analysis of
  \mbox{$\decay{\B}{\kaon \pip \Ppsi^{\prime}}$}~decays and
  the~\mbox{$\PZ(4430)^+$}}},
  }{}\href{https://doi.org/10.1103/PhysRevD.80.031104}{Phys.\ Rev.\
  \textbf{D80} (2009) 031104},
  \href{http://arxiv.org/abs/0905.2869}{{\normalfont\ttfamily
  arXiv:0905.2869}}\relax
\mciteBstWouldAddEndPuncttrue
\mciteSetBstMidEndSepPunct{\mcitedefaultmidpunct}
{\mcitedefaultendpunct}{\mcitedefaultseppunct}\relax
\EndOfBibitem
\bibitem{Chilikin:2013tch}
Belle collaboration, K.~Chilikin {\em et~al.},
  \ifthenelse{\boolean{articletitles}}{\emph{{Experimental constraints on
  the~spin and parity of the~\mbox{$\PZ(4430)^+$}}},
  }{}\href{https://doi.org/10.1103/PhysRevD.88.074026}{Phys.\ Rev.\
  \textbf{D88} (2013) 074026},
  \href{http://arxiv.org/abs/1306.4894}{{\normalfont\ttfamily
  arXiv:1306.4894}}\relax
\mciteBstWouldAddEndPuncttrue
\mciteSetBstMidEndSepPunct{\mcitedefaultmidpunct}
{\mcitedefaultendpunct}{\mcitedefaultseppunct}\relax
\EndOfBibitem
\bibitem{LHCb-PAPER-2014-014}
LHCb collaboration, R.~Aaij {\em et~al.},
  \ifthenelse{\boolean{articletitles}}{\emph{{Observation of the resonant
  character of the~$\PZ(4430)^-$ state}},
  }{}\href{https://doi.org/10.1103/PhysRevLett.112.222002}{Phys.\ Rev.\ Lett.\
  \textbf{112} (2014) 222002},
  \href{http://arxiv.org/abs/1404.1903}{{\normalfont\ttfamily
  arXiv:1404.1903}}\relax
\mciteBstWouldAddEndPuncttrue
\mciteSetBstMidEndSepPunct{\mcitedefaultmidpunct}
{\mcitedefaultendpunct}{\mcitedefaultseppunct}\relax
\EndOfBibitem
\bibitem{LHCb-PAPER-2015-038}
LHCb collaboration, R.~Aaij {\em et~al.},
  \ifthenelse{\boolean{articletitles}}{\emph{{Model-independent confirmation of
  the~$\PZ(4430)^-$ state}},
  }{}\href{https://doi.org/10.1103/PhysRevD.92.112009}{Phys.\ Rev.\
  \textbf{D92} (2015) 112009},
  \href{http://arxiv.org/abs/1510.01951}{{\normalfont\ttfamily
  arXiv:1510.01951}}\relax
\mciteBstWouldAddEndPuncttrue
\mciteSetBstMidEndSepPunct{\mcitedefaultmidpunct}
{\mcitedefaultendpunct}{\mcitedefaultseppunct}\relax
\EndOfBibitem
\bibitem{LHCb-PAPER-2016-015}
LHCb collaboration, R.~Aaij {\em et~al.},
  \ifthenelse{\boolean{articletitles}}{\emph{{Evidence for exotic hadron
  contributions to \mbox{\decay{\Lb}{\jpsi\proton\pim}} decays}},
  }{}\href{https://doi.org/10.1103/PhysRevLett.117.082003}{Phys.\ Rev.\ Lett.\
  \textbf{117} (2016) 082003},
  \href{http://arxiv.org/abs/1606.06999}{{\normalfont\ttfamily
  arXiv:1606.06999}}\relax
\mciteBstWouldAddEndPuncttrue
\mciteSetBstMidEndSepPunct{\mcitedefaultmidpunct}
{\mcitedefaultendpunct}{\mcitedefaultseppunct}\relax
\EndOfBibitem
\bibitem{LHCb-PAPER-2016-018}
LHCb collaboration, R.~Aaij {\em et~al.},
  \ifthenelse{\boolean{articletitles}}{\emph{{Observation of exotic
  $\jpsi\Pphi$ structures from amplitude analysis of
  \mbox{\decay{\Bp}{\jpsi\Pphi\Kp}} decays}},
  }{}\href{https://doi.org/10.1103/PhysRevLett.118.022003}{Phys.\ Rev.\ Lett.\
  \textbf{118} (2017) 022003},
  \href{http://arxiv.org/abs/1606.07895}{{\normalfont\ttfamily
  arXiv:1606.07895}}\relax
\mciteBstWouldAddEndPuncttrue
\mciteSetBstMidEndSepPunct{\mcitedefaultmidpunct}
{\mcitedefaultendpunct}{\mcitedefaultseppunct}\relax
\EndOfBibitem
\bibitem{LHCb-PAPER-2016-019}
LHCb collaboration, R.~Aaij {\em et~al.},
  \ifthenelse{\boolean{articletitles}}{\emph{{Amplitude analysis of
  \mbox{\decay{\Bp}{\jpsi\Pphi\Kp}} decays}},
  }{}\href{https://doi.org/10.1103/PhysRevD.95.012002}{Phys.\ Rev.\
  \textbf{D95} (2017) 012002},
  \href{http://arxiv.org/abs/1606.07898}{{\normalfont\ttfamily
  arXiv:1606.07898}}\relax
\mciteBstWouldAddEndPuncttrue
\mciteSetBstMidEndSepPunct{\mcitedefaultmidpunct}
{\mcitedefaultendpunct}{\mcitedefaultseppunct}\relax
\EndOfBibitem
\bibitem{LHCb-PAPER-2018-034}
LHCb collaboration, R.~Aaij {\em et~al.},
  \ifthenelse{\boolean{articletitles}}{\emph{{Evidence for an
  $\Peta_\cquark(1\PS) \pim$ resonance in \mbox{\decay{\Bz}{\Peta_\cquark(1\PS)
  \Kp\pim}} decays}},
  }{}\href{https://doi.org/10.1140/epjc/s10052-018-6447-z}{Eur.\ Phys.\ J.\
  \textbf{C78} (2018) 1019},
  \href{http://arxiv.org/abs/1809.07416}{{\normalfont\ttfamily
  arXiv:1809.07416}}\relax
\mciteBstWouldAddEndPuncttrue
\mciteSetBstMidEndSepPunct{\mcitedefaultmidpunct}
{\mcitedefaultendpunct}{\mcitedefaultseppunct}\relax
\EndOfBibitem
\bibitem{LHCb-PAPER-2018-043}
LHCb collaboration, R.~Aaij {\em et~al.},
  \ifthenelse{\boolean{articletitles}}{\emph{{Model-independent observation of
  exotic contributions to \mbox{\decay{\Bz}{\jpsi\Kp\pim}} decays}},
  }{}\href{https://doi.org/10.1103/PhysRevLett.122.152002}{Phys.\ Rev.\ Lett.\
  \textbf{122} (2019) 152002},
  \href{http://arxiv.org/abs/1901.05745}{{\normalfont\ttfamily
  arXiv:1901.05745}}\relax
\mciteBstWouldAddEndPuncttrue
\mciteSetBstMidEndSepPunct{\mcitedefaultmidpunct}
{\mcitedefaultendpunct}{\mcitedefaultseppunct}\relax
\EndOfBibitem
\bibitem{LHCb-PAPER-2020-035}
LHCb collaboration, R.~Aaij {\em et~al.},
  \ifthenelse{\boolean{articletitles}}{\emph{{Study of $B_s^0 \to \jpsi \pip
  \pim \Kp \Km$ decays}},
  }{}\href{https://doi.org/10.1007/JHEP02(2021)024}{JHEP \textbf{02} (2021)
  024}, \href{http://arxiv.org/abs/2011.01867}{{\normalfont\ttfamily
  arXiv:2011.01867}}\relax
\mciteBstWouldAddEndPuncttrue
\mciteSetBstMidEndSepPunct{\mcitedefaultmidpunct}
{\mcitedefaultendpunct}{\mcitedefaultseppunct}\relax
\EndOfBibitem
\bibitem{LHCb-PAPER-2020-044}
LHCb collaboration, R.~Aaij {\em et~al.},
  \ifthenelse{\boolean{articletitles}}{\emph{{Observation of new resonances
  decaying to $ \jpsi\Kp$ and $ \jpsi\Pphi$ }},
  }{}\href{https://doi.org/10.1103/PhysRevLett.127.082001}{Phys.\ Rev.\ Lett.\
  \textbf{127} (2021) 082001},
  \href{http://arxiv.org/abs/2103.01803}{{\normalfont\ttfamily
  arXiv:2103.01803}}\relax
\mciteBstWouldAddEndPuncttrue
\mciteSetBstMidEndSepPunct{\mcitedefaultmidpunct}
{\mcitedefaultendpunct}{\mcitedefaultseppunct}\relax
\EndOfBibitem
\bibitem{Belle:2003nnu}
Belle collaboration, S.~K. Choi {\em et~al.},
  \ifthenelse{\boolean{articletitles}}{\emph{{Observation of a~narrow
  charmonium\nobreakdash-like state in exclusive
  \mbox{$\decay{\B^{\pm}}{\kaon^{\pm}\pip\pim\jpsi}$} decays}},
  }{}\href{https://doi.org/10.1103/PhysRevLett.91.262001}{Phys.\ Rev.\ Lett.\
  \textbf{91} (2003) 262001},
  \href{http://arxiv.org/abs/hep-ex/0309032}{{\normalfont\ttfamily
  arXiv:hep-ex/0309032}}\relax
\mciteBstWouldAddEndPuncttrue
\mciteSetBstMidEndSepPunct{\mcitedefaultmidpunct}
{\mcitedefaultendpunct}{\mcitedefaultseppunct}\relax
\EndOfBibitem
\bibitem{BaBar:2004iez}
BaBar collaboration, B.~Aubert {\em et~al.},
  \ifthenelse{\boolean{articletitles}}{\emph{{Observation of the~decay
  \mbox{$\decay{\B}{\jpsi\Peta\kaon}$} and search for
  \mbox{$\decay{\PX(3872)}{\jpsi\Peta}$}}},
  }{}\href{https://doi.org/10.1103/PhysRevLett.93.041801}{Phys.\ Rev.\ Lett.\
  \textbf{93} (2004) 041801},
  \href{http://arxiv.org/abs/hep-ex/0402025}{{\normalfont\ttfamily
  arXiv:hep-ex/0402025}}\relax
\mciteBstWouldAddEndPuncttrue
\mciteSetBstMidEndSepPunct{\mcitedefaultmidpunct}
{\mcitedefaultendpunct}{\mcitedefaultseppunct}\relax
\EndOfBibitem
\bibitem{BaBar:2004oro}
BaBar collaboration, B.~Aubert {\em et~al.},
  \ifthenelse{\boolean{articletitles}}{\emph{{Study of
  the~\mbox{$\decay{\B}{\jpsi\Km\pip\pim}$}~decay and measurement of
  the~\mbox{$\decay{\B}{\PX(3872)\Km}$}~branching fraction}},
  }{}\href{https://doi.org/10.1103/PhysRevD.71.071103}{Phys.\ Rev.\
  \textbf{D71} (2005) 071103},
  \href{http://arxiv.org/abs/hep-ex/0406022}{{\normalfont\ttfamily
  arXiv:hep-ex/0406022}}\relax
\mciteBstWouldAddEndPuncttrue
\mciteSetBstMidEndSepPunct{\mcitedefaultmidpunct}
{\mcitedefaultendpunct}{\mcitedefaultseppunct}\relax
\EndOfBibitem
\bibitem{BaBar:2006fjg}
BaBar collaboration, B.~Aubert {\em et~al.},
  \ifthenelse{\boolean{articletitles}}{\emph{{Search for
  $\decay{\Bu}{\PX(3872)\Kp}$, $\decay{\PX(3872)}{\jpsi\g}$}},
  }{}\href{https://doi.org/10.1103/PhysRevD.74.071101}{Phys.\ Rev.\
  \textbf{D74} (2006) 071101},
  \href{http://arxiv.org/abs/hep-ex/0607050}{{\normalfont\ttfamily
  arXiv:hep-ex/0607050}}\relax
\mciteBstWouldAddEndPuncttrue
\mciteSetBstMidEndSepPunct{\mcitedefaultmidpunct}
{\mcitedefaultendpunct}{\mcitedefaultseppunct}\relax
\EndOfBibitem
\bibitem{BaBar:2007ixp}
BaBar collaboration, B.~Aubert {\em et~al.},
  \ifthenelse{\boolean{articletitles}}{\emph{{Search for prompt production of
  $\chic$ and~$\PX(3872)$ in \epem~annihilations}},
  }{}\href{https://doi.org/10.1103/PhysRevD.76.071102}{Phys.\ Rev.\
  \textbf{D76} (2007) 071102},
  \href{http://arxiv.org/abs/0707.1633}{{\normalfont\ttfamily
  arXiv:0707.1633}}\relax
\mciteBstWouldAddEndPuncttrue
\mciteSetBstMidEndSepPunct{\mcitedefaultmidpunct}
{\mcitedefaultendpunct}{\mcitedefaultseppunct}\relax
\EndOfBibitem
\bibitem{BaBar:2007cmo}
BaBar collaboration, B.~Aubert {\em et~al.},
  \ifthenelse{\boolean{articletitles}}{\emph{{Study of resonances in exclusive
  \B~decays to~\mbox{$\Dbar^{(*)}\D^{(*)}\kaon$} }},
  }{}\href{https://doi.org/10.1103/PhysRevD.77.011102}{Phys.\ Rev.\
  \textbf{D77} (2008) 011102},
  \href{http://arxiv.org/abs/0708.1565}{{\normalfont\ttfamily
  arXiv:0708.1565}}\relax
\mciteBstWouldAddEndPuncttrue
\mciteSetBstMidEndSepPunct{\mcitedefaultmidpunct}
{\mcitedefaultendpunct}{\mcitedefaultseppunct}\relax
\EndOfBibitem
\bibitem{BaBar:2008qzi}
BaBar collaboration, B.~Aubert {\em et~al.},
  \ifthenelse{\boolean{articletitles}}{\emph{{A~Study of
  \mbox{$\decay{\B}{\PX(3872)\kaon}$}, with
  \mbox{$\decay{\PX(3872)}{\jpsi\pip\pim}$} }},
  }{}\href{https://doi.org/10.1103/PhysRevD.77.111101}{Phys.\ Rev.\
  \textbf{D77} (2008) 111101},
  \href{http://arxiv.org/abs/0803.2838}{{\normalfont\ttfamily
  arXiv:0803.2838}}\relax
\mciteBstWouldAddEndPuncttrue
\mciteSetBstMidEndSepPunct{\mcitedefaultmidpunct}
{\mcitedefaultendpunct}{\mcitedefaultseppunct}\relax
\EndOfBibitem
\bibitem{BaBar:2008flx}
BaBar collaboration, B.~Aubert {\em et~al.},
  \ifthenelse{\boolean{articletitles}}{\emph{{Evidence for
  \mbox{$\decay{\PX(3872)}{\psitwos \g}$} in
  \mbox{$\decay{\B^\pm}{\PX(3872)\kaon^\pm}$}~decays, and a~study of
  \mbox{$\decay{\B}{\cquark\cquarkbar \g \kaon}$}}},
  }{}\href{https://doi.org/10.1103/PhysRevLett.102.132001}{Phys.\ Rev.\ Lett.\
  \textbf{102} (2009) 132001},
  \href{http://arxiv.org/abs/0809.0042}{{\normalfont\ttfamily
  arXiv:0809.0042}}\relax
\mciteBstWouldAddEndPuncttrue
\mciteSetBstMidEndSepPunct{\mcitedefaultmidpunct}
{\mcitedefaultendpunct}{\mcitedefaultseppunct}\relax
\EndOfBibitem
\bibitem{BaBar:2010wfc}
BaBar collaboration, P.~del Amo~Sanchez {\em et~al.},
  \ifthenelse{\boolean{articletitles}}{\emph{{Evidence for the~decay
  \mbox{$\decay{\PX(3872)}{\jpsi\Pomega}$} }},
  }{}\href{https://doi.org/10.1103/PhysRevD.82.011101}{Phys.\ Rev.\
  \textbf{D82} (2010) 011101},
  \href{http://arxiv.org/abs/1005.5190}{{\normalfont\ttfamily
  arXiv:1005.5190}}\relax
\mciteBstWouldAddEndPuncttrue
\mciteSetBstMidEndSepPunct{\mcitedefaultmidpunct}
{\mcitedefaultendpunct}{\mcitedefaultseppunct}\relax
\EndOfBibitem
\bibitem{Belle:2003eeg}
Belle collaboration, K.~Abe {\em et~al.},
  \ifthenelse{\boolean{articletitles}}{\emph{{Observation of
  \mbox{$\decay{\Bu}{\Ppsi(3770)\Kp}$} }},
  }{}\href{https://doi.org/10.1103/PhysRevLett.93.051803}{Phys.\ Rev.\ Lett.\
  \textbf{93} (2004) 051803},
  \href{http://arxiv.org/abs/hep-ex/0307061}{{\normalfont\ttfamily
  arXiv:hep-ex/0307061}}\relax
\mciteBstWouldAddEndPuncttrue
\mciteSetBstMidEndSepPunct{\mcitedefaultmidpunct}
{\mcitedefaultendpunct}{\mcitedefaultseppunct}\relax
\EndOfBibitem
\bibitem{Belle:2006olv}
Belle collaboration, G.~Gokhroo {\em et~al.},
  \ifthenelse{\boolean{articletitles}}{\emph{{Observation of
  a~near\nobreakdash-threshold $\Dz\Dzb\piz$~enhancement in
  \mbox{$\decay{\B}{\Dz\Dzb\piz\kaon}$}~decay}},
  }{}\href{https://doi.org/10.1103/PhysRevLett.97.162002}{Phys.\ Rev.\ Lett.\
  \textbf{97} (2006) 162002},
  \href{http://arxiv.org/abs/hep-ex/0606055}{{\normalfont\ttfamily
  arXiv:hep-ex/0606055}}\relax
\mciteBstWouldAddEndPuncttrue
\mciteSetBstMidEndSepPunct{\mcitedefaultmidpunct}
{\mcitedefaultendpunct}{\mcitedefaultseppunct}\relax
\EndOfBibitem
\bibitem{Belle:2008fma}
Belle collaboration, T.~Aushev {\em et~al.},
  \ifthenelse{\boolean{articletitles}}{\emph{{Study of
  the~\mbox{$\decay{\B}{\PX(3872)\left(\decay{}{\Dstarz\Dzb}\right)
  \kaon}$}~decay}}, }{}\href{https://doi.org/10.1103/PhysRevD.81.031103}{Phys.\
  Rev.\  \textbf{D81} (2010) 031103},
  \href{http://arxiv.org/abs/0810.0358}{{\normalfont\ttfamily
  arXiv:0810.0358}}\relax
\mciteBstWouldAddEndPuncttrue
\mciteSetBstMidEndSepPunct{\mcitedefaultmidpunct}
{\mcitedefaultendpunct}{\mcitedefaultseppunct}\relax
\EndOfBibitem
\bibitem{Belle:2011vlx}
Belle collaboration, S.-K. Choi {\em et~al.},
  \ifthenelse{\boolean{articletitles}}{\emph{{Bounds on the~width, mass
  difference and other properties of
  \mbox{$\decay{\PX(3872)}{\pip\pim\jpsi}$}~decays}},
  }{}\href{https://doi.org/10.1103/PhysRevD.84.052004}{Phys.\ Rev.\
  \textbf{D84} (2011) 052004},
  \href{http://arxiv.org/abs/1107.0163}{{\normalfont\ttfamily
  arXiv:1107.0163}}\relax
\mciteBstWouldAddEndPuncttrue
\mciteSetBstMidEndSepPunct{\mcitedefaultmidpunct}
{\mcitedefaultendpunct}{\mcitedefaultseppunct}\relax
\EndOfBibitem
\bibitem{Belle:2011wdj}
Belle collaboration, V.~Bhardwaj {\em et~al.},
  \ifthenelse{\boolean{articletitles}}{\emph{{Observation of $\PX(3872)\to
  \jpsi \g$ and search for $\PX(3872)\to\Ppsi^\prime\g$ in \B~decays}},
  }{}\href{https://doi.org/10.1103/PhysRevLett.107.091803}{Phys.\ Rev.\ Lett.\
  \textbf{107} (2011) 091803},
  \href{http://arxiv.org/abs/1105.0177}{{\normalfont\ttfamily
  arXiv:1105.0177}}\relax
\mciteBstWouldAddEndPuncttrue
\mciteSetBstMidEndSepPunct{\mcitedefaultmidpunct}
{\mcitedefaultendpunct}{\mcitedefaultseppunct}\relax
\EndOfBibitem
\bibitem{Belle:2015qeg}
Belle collaboration, A.~Bala {\em et~al.},
  \ifthenelse{\boolean{articletitles}}{\emph{{Observation of $\PX(3872)$ in
  $\decay{\B}{\PX(3872)\kaon\pion}$~decays}},
  }{}\href{https://doi.org/10.1103/PhysRevD.91.051101}{Phys.\ Rev.\
  \textbf{D91} (2015) 051101},
  \href{http://arxiv.org/abs/1501.06867}{{\normalfont\ttfamily
  arXiv:1501.06867}}\relax
\mciteBstWouldAddEndPuncttrue
\mciteSetBstMidEndSepPunct{\mcitedefaultmidpunct}
{\mcitedefaultendpunct}{\mcitedefaultseppunct}\relax
\EndOfBibitem
\bibitem{Belle:2022puc}
Belle collaboration, J.~H. Yin {\em et~al.},
  \ifthenelse{\boolean{articletitles}}{\emph{{Search for
  $\decay{\PX(3872)}{\pip\pim\piz}$ at Belle}},
  }{}\href{https://doi.org/10.1103/PhysRevD.107.052004}{Phys.\ Rev.\
  \textbf{D107} (2023) 052004},
  \href{http://arxiv.org/abs/2206.08592}{{\normalfont\ttfamily
  arXiv:2206.08592}}\relax
\mciteBstWouldAddEndPuncttrue
\mciteSetBstMidEndSepPunct{\mcitedefaultmidpunct}
{\mcitedefaultendpunct}{\mcitedefaultseppunct}\relax
\EndOfBibitem
\bibitem{Belle:2023zxm}
Belle collaboration, H.~Hirata {\em et~al.},
  \ifthenelse{\boolean{articletitles}}{\emph{{Study of the~lineshape of
  $\PX(3872)$ using \B~decays to~$\Dz\Dstarzb\kaon$}},
  }{}\href{https://doi.org/10.1103/PhysRevD.107.112011}{Phys.\ Rev.\
  \textbf{D107} (2023) 112011},
  \href{http://arxiv.org/abs/2302.02127}{{\normalfont\ttfamily
  arXiv:2302.02127}}\relax
\mciteBstWouldAddEndPuncttrue
\mciteSetBstMidEndSepPunct{\mcitedefaultmidpunct}
{\mcitedefaultendpunct}{\mcitedefaultseppunct}\relax
\EndOfBibitem
\bibitem{BESIII:2013fnz}
BESIII collaboration, M.~Ablikim {\em et~al.},
  \ifthenelse{\boolean{articletitles}}{\emph{{Observation of
  \mbox{$\decay{\epem}{\g\PX(3872)}$} at BESIII}},
  }{}\href{https://doi.org/10.1103/PhysRevLett.112.092001}{Phys.\ Rev.\ Lett.\
  \textbf{112} (2014) 092001},
  \href{http://arxiv.org/abs/1310.4101}{{\normalfont\ttfamily
  arXiv:1310.4101}}\relax
\mciteBstWouldAddEndPuncttrue
\mciteSetBstMidEndSepPunct{\mcitedefaultmidpunct}
{\mcitedefaultendpunct}{\mcitedefaultseppunct}\relax
\EndOfBibitem
\bibitem{BESIII:2015klj}
BESIII collaboration, M.~Ablikim {\em et~al.},
  \ifthenelse{\boolean{articletitles}}{\emph{{An improved limit for~
  $\Gamma_{\Pe\Pe}$ of $\PX(3872)$ and $\Gamma_{\Pe\Pe}$ measurement of
  $\Ppsi(3686)$}},
  }{}\href{https://doi.org/10.1016/j.physletb.2015.08.013}{Phys.\ Lett.\
  \textbf{B749} (2015) 414},
  \href{http://arxiv.org/abs/1505.02559}{{\normalfont\ttfamily
  arXiv:1505.02559}}\relax
\mciteBstWouldAddEndPuncttrue
\mciteSetBstMidEndSepPunct{\mcitedefaultmidpunct}
{\mcitedefaultendpunct}{\mcitedefaultseppunct}\relax
\EndOfBibitem
\bibitem{BESIII:2020nbj}
BESIII collaboration, M.~Ablikim {\em et~al.},
  \ifthenelse{\boolean{articletitles}}{\emph{{Study of open-charm decays and
  radiative transitions of the~$\PX(3872)$}},
  }{}\href{https://doi.org/10.1103/PhysRevLett.124.242001}{Phys.\ Rev.\ Lett.\
  \textbf{124} (2020) 242001},
  \href{http://arxiv.org/abs/2001.01156}{{\normalfont\ttfamily
  arXiv:2001.01156}}\relax
\mciteBstWouldAddEndPuncttrue
\mciteSetBstMidEndSepPunct{\mcitedefaultmidpunct}
{\mcitedefaultendpunct}{\mcitedefaultseppunct}\relax
\EndOfBibitem
\bibitem{BESIII:2022bse}
BESIII collaboration, M.~Ablikim {\em et~al.},
  \ifthenelse{\boolean{articletitles}}{\emph{{Observation of a~new
  $\PX(3872)$~production process \mbox{$\decay{\epem}{\Pomega\PX(3872)}$}}},
  }{}\href{https://doi.org/10.1103/PhysRevLett.130.151904}{Phys.\ Rev.\ Lett.\
  \textbf{130} (2023) 151904},
  \href{http://arxiv.org/abs/2212.07291}{{\normalfont\ttfamily
  arXiv:2212.07291}}\relax
\mciteBstWouldAddEndPuncttrue
\mciteSetBstMidEndSepPunct{\mcitedefaultmidpunct}
{\mcitedefaultendpunct}{\mcitedefaultseppunct}\relax
\EndOfBibitem
\bibitem{BESIII:2023xta}
BESIII collaboration, M.~Ablikim {\em et~al.},
  \ifthenelse{\boolean{articletitles}}{\emph{{Search for the~light hadron decay
  \mbox{$\decay{\chicone(3872)}{\pip\pim\Peta}$} }},
  }{}\href{https://doi.org/10.1103/PhysRevD.109.L011102}{Phys.\ Rev.\
  \textbf{D109} (2024) L011102},
  \href{http://arxiv.org/abs/2308.13980}{{\normalfont\ttfamily
  arXiv:2308.13980}}\relax
\mciteBstWouldAddEndPuncttrue
\mciteSetBstMidEndSepPunct{\mcitedefaultmidpunct}
{\mcitedefaultendpunct}{\mcitedefaultseppunct}\relax
\EndOfBibitem
\bibitem{BESIII:2023hml}
BESIII collaboration, M.~Ablikim {\em et~al.},
  \ifthenelse{\boolean{articletitles}}{\emph{{A coupled-channel analysis of
  the~$\PX(3872)$ lineshape with BESIII data}},
  }{}\href{http://arxiv.org/abs/2309.01502}{{\normalfont\ttfamily
  arXiv:2309.01502}}\relax
\mciteBstWouldAddEndPuncttrue
\mciteSetBstMidEndSepPunct{\mcitedefaultmidpunct}
{\mcitedefaultendpunct}{\mcitedefaultseppunct}\relax
\EndOfBibitem
\bibitem{CDF:2003cab}
CDF collaboration, D.~Acosta {\em et~al.},
  \ifthenelse{\boolean{articletitles}}{\emph{{Observation of the~narrow state
  \mbox{$\decay{\PX(3872)}{\jpsi \pip\pim}$} in $\antiproton\proton$~collisions
  at \mbox{$\sqs=1.96\tev$}}},
  }{}\href{https://doi.org/10.1103/PhysRevLett.93.072001}{Phys.\ Rev.\ Lett.\
  \textbf{93} (2004) 072001},
  \href{http://arxiv.org/abs/hep-ex/0312021}{{\normalfont\ttfamily
  arXiv:hep-ex/0312021}}\relax
\mciteBstWouldAddEndPuncttrue
\mciteSetBstMidEndSepPunct{\mcitedefaultmidpunct}
{\mcitedefaultendpunct}{\mcitedefaultseppunct}\relax
\EndOfBibitem
\bibitem{CDF:2005cfq}
CDF collaboration, A.~Abulencia {\em et~al.},
  \ifthenelse{\boolean{articletitles}}{\emph{{Measurement of the~dipion mass
  spectrum in \mbox{$\decay{\PX(3872)}{\jpsi \pip\pim}$}~decays.}},
  }{}\href{https://doi.org/10.1103/PhysRevLett.96.102002}{Phys.\ Rev.\ Lett.\
  \textbf{96} (2006) 102002},
  \href{http://arxiv.org/abs/hep-ex/0512074}{{\normalfont\ttfamily
  arXiv:hep-ex/0512074}}\relax
\mciteBstWouldAddEndPuncttrue
\mciteSetBstMidEndSepPunct{\mcitedefaultmidpunct}
{\mcitedefaultendpunct}{\mcitedefaultseppunct}\relax
\EndOfBibitem
\bibitem{CDF:2006ocq}
CDF collaboration, A.~Abulencia {\em et~al.},
  \ifthenelse{\boolean{articletitles}}{\emph{{Analysis of the~quantum numbers
  $\PJ^{\PP\PC}$ of the~$\PX(3872)$}},
  }{}\href{https://doi.org/10.1103/PhysRevLett.98.132002}{Phys.\ Rev.\ Lett.\
  \textbf{98} (2007) 132002},
  \href{http://arxiv.org/abs/hep-ex/0612053}{{\normalfont\ttfamily
  arXiv:hep-ex/0612053}}\relax
\mciteBstWouldAddEndPuncttrue
\mciteSetBstMidEndSepPunct{\mcitedefaultmidpunct}
{\mcitedefaultendpunct}{\mcitedefaultseppunct}\relax
\EndOfBibitem
\bibitem{CDF:2009nxk}
CDF collaboration, T.~Aaltonen {\em et~al.},
  \ifthenelse{\boolean{articletitles}}{\emph{{Precision measurement of
  the~$\PX(3872)$~mass in $\jpsi\pip\pim$~decays}},
  }{}\href{https://doi.org/10.1103/PhysRevLett.103.152001}{Phys.\ Rev.\ Lett.\
  \textbf{103} (2009) 152001},
  \href{http://arxiv.org/abs/0906.5218}{{\normalfont\ttfamily
  arXiv:0906.5218}}\relax
\mciteBstWouldAddEndPuncttrue
\mciteSetBstMidEndSepPunct{\mcitedefaultmidpunct}
{\mcitedefaultendpunct}{\mcitedefaultseppunct}\relax
\EndOfBibitem
\bibitem{D0:2004zmu}
D0 collaboration, V.~M. Abazov {\em et~al.},
  \ifthenelse{\boolean{articletitles}}{\emph{{Observation and properties of
  the~$\PX(3872)$ decaying to $\jpsi \pip\pim$ in
  $\proton\antiproton$~collisions at \mbox{$\sqs = 1.96\tev$}}},
  }{}\href{https://doi.org/10.1103/PhysRevLett.93.162002}{Phys.\ Rev.\ Lett.\
  \textbf{93} (2004) 162002},
  \href{http://arxiv.org/abs/hep-ex/0405004}{{\normalfont\ttfamily
  arXiv:hep-ex/0405004}}\relax
\mciteBstWouldAddEndPuncttrue
\mciteSetBstMidEndSepPunct{\mcitedefaultmidpunct}
{\mcitedefaultendpunct}{\mcitedefaultseppunct}\relax
\EndOfBibitem
\bibitem{ATLAS:2016kwu}
ATLAS collaboration, M.~Aaboud {\em et~al.},
  \ifthenelse{\boolean{articletitles}}{\emph{{Measurements of $\psitwos$ and
  \mbox{$\decay{\PX(3872)}{\jpsi\pip\pim}$} production in
  $\proton\proton$~collisions at $\sqs=8\tev$ with the~ATLAS detector}},
  }{}\href{https://doi.org/10.1007/JHEP01(2017)117}{JHEP \textbf{01} (2017)
  117}, \href{http://arxiv.org/abs/1610.09303}{{\normalfont\ttfamily
  arXiv:1610.09303}}\relax
\mciteBstWouldAddEndPuncttrue
\mciteSetBstMidEndSepPunct{\mcitedefaultmidpunct}
{\mcitedefaultendpunct}{\mcitedefaultseppunct}\relax
\EndOfBibitem
\bibitem{CMS:2013fpt}
CMS collaboration, S.~Chatrchyan {\em et~al.},
  \ifthenelse{\boolean{articletitles}}{\emph{{Measurement of
  the~$\PX(3872)$~production cross section via decays to $\jpsi \pip\pim$ in
  $\proton\proton$~collisions at \mbox{$\sqs=7\tev$}}},
  }{}\href{https://doi.org/10.1007/JHEP04(2013)154}{JHEP \textbf{04} (2013)
  154}, \href{http://arxiv.org/abs/1302.3968}{{\normalfont\ttfamily
  arXiv:1302.3968}}\relax
\mciteBstWouldAddEndPuncttrue
\mciteSetBstMidEndSepPunct{\mcitedefaultmidpunct}
{\mcitedefaultendpunct}{\mcitedefaultseppunct}\relax
\EndOfBibitem
\bibitem{CMS:2020eiw}
CMS collaboration, A.~M. Sirunyan {\em et~al.},
  \ifthenelse{\boolean{articletitles}}{\emph{{Observation of
  the~\mbox{$\decay{\Bs}{\PX(3872)\Pphi}$}~decay}},
  }{}\href{https://doi.org/10.1103/PhysRevLett.125.152001}{Phys.\ Rev.\ Lett.\
  \textbf{125} (2020) 152001},
  \href{http://arxiv.org/abs/2005.04764}{{\normalfont\ttfamily
  arXiv:2005.04764}}\relax
\mciteBstWouldAddEndPuncttrue
\mciteSetBstMidEndSepPunct{\mcitedefaultmidpunct}
{\mcitedefaultendpunct}{\mcitedefaultseppunct}\relax
\EndOfBibitem
\bibitem{LHCb-PAPER-2011-034}
LHCb collaboration, R.~Aaij {\em et~al.},
  \ifthenelse{\boolean{articletitles}}{\emph{{Observation of $\PX(3872)$
  production in \proton\proton collisions at \mbox{$\sqs=$7\tev}}},
  }{}\href{https://doi.org/10.1140/epjc/s10052-012-1972-7}{Eur.\ Phys.\ J.\
  \textbf{C72} (2012) 1972},
  \href{http://arxiv.org/abs/1112.5310}{{\normalfont\ttfamily
  arXiv:1112.5310}}\relax
\mciteBstWouldAddEndPuncttrue
\mciteSetBstMidEndSepPunct{\mcitedefaultmidpunct}
{\mcitedefaultendpunct}{\mcitedefaultseppunct}\relax
\EndOfBibitem
\bibitem{LHCb-PAPER-2013-001}
LHCb collaboration, R.~Aaij {\em et~al.},
  \ifthenelse{\boolean{articletitles}}{\emph{{Determination of the $\PX(3872)$
  meson quantum numbers}},
  }{}\href{https://doi.org/10.1103/PhysRevLett.110.222001}{Phys.\ Rev.\ Lett.\
  \textbf{110} (2013) 222001},
  \href{http://arxiv.org/abs/1302.6269}{{\normalfont\ttfamily
  arXiv:1302.6269}}\relax
\mciteBstWouldAddEndPuncttrue
\mciteSetBstMidEndSepPunct{\mcitedefaultmidpunct}
{\mcitedefaultendpunct}{\mcitedefaultseppunct}\relax
\EndOfBibitem
\bibitem{LHCb-PAPER-2014-008}
LHCb collaboration, R.~Aaij {\em et~al.},
  \ifthenelse{\boolean{articletitles}}{\emph{{Evidence for the decay
  \mbox{\decay{\PX(3872)}{\psitwos\gamma}}}},
  }{}\href{https://doi.org/10.1016/j.nuclphysb.2014.06.011}{Nucl.\ Phys.\
  \textbf{B886} (2014) 665},
  \href{http://arxiv.org/abs/1404.0275}{{\normalfont\ttfamily
  arXiv:1404.0275}}\relax
\mciteBstWouldAddEndPuncttrue
\mciteSetBstMidEndSepPunct{\mcitedefaultmidpunct}
{\mcitedefaultendpunct}{\mcitedefaultseppunct}\relax
\EndOfBibitem
\bibitem{LHCb-PAPER-2015-015}
LHCb collaboration, R.~Aaij {\em et~al.},
  \ifthenelse{\boolean{articletitles}}{\emph{{Quantum numbers of
  the~$\PX(3872)$ state and orbital angular momentum in its
  $\rhoz\jpsi$~decays}},
  }{}\href{https://doi.org/10.1103/PhysRevD.92.011102}{Phys.\ Rev.\
  \textbf{D92} (2015) 011102(R)},
  \href{http://arxiv.org/abs/1504.06339}{{\normalfont\ttfamily
  arXiv:1504.06339}}\relax
\mciteBstWouldAddEndPuncttrue
\mciteSetBstMidEndSepPunct{\mcitedefaultmidpunct}
{\mcitedefaultendpunct}{\mcitedefaultseppunct}\relax
\EndOfBibitem
\bibitem{LHCb-PAPER-2016-016}
LHCb collaboration, R.~Aaij {\em et~al.},
  \ifthenelse{\boolean{articletitles}}{\emph{{Observation of
  \mbox{\decay{\etac(2S)}{\proton\antiproton}} and search for
  \mbox{\decay{\PX(3872)}{\proton\antiproton}} decays}},
  }{}\href{https://doi.org/10.1016/j.physletb.2017.03.046}{Phys.\ Lett.\
  \textbf{B769} (2017) 305},
  \href{http://arxiv.org/abs/1607.06446}{{\normalfont\ttfamily
  arXiv:1607.06446}}\relax
\mciteBstWouldAddEndPuncttrue
\mciteSetBstMidEndSepPunct{\mcitedefaultmidpunct}
{\mcitedefaultendpunct}{\mcitedefaultseppunct}\relax
\EndOfBibitem
\bibitem{LHCb-PAPER-2019-023}
LHCb collaboration, R.~Aaij {\em et~al.},
  \ifthenelse{\boolean{articletitles}}{\emph{{Observation of the
  \mbox{\decay{\Lb}{\chicone(3872)\proton\Km}} decay}},
  }{}\href{https://doi.org/10.1007/JHEP09(2019)028}{JHEP \textbf{09} (2019)
  028}, \href{http://arxiv.org/abs/1907.00954}{{\normalfont\ttfamily
  arXiv:1907.00954}}\relax
\mciteBstWouldAddEndPuncttrue
\mciteSetBstMidEndSepPunct{\mcitedefaultmidpunct}
{\mcitedefaultendpunct}{\mcitedefaultseppunct}\relax
\EndOfBibitem
\bibitem{LHCb-PAPER-2020-008}
LHCb collaboration, R.~Aaij {\em et~al.},
  \ifthenelse{\boolean{articletitles}}{\emph{{Study of the line shape of the
  $\chicone(3872)$ state}},
  }{}\href{https://doi.org/10.1103/PhysRevD.102.092005}{Phys.\ Rev.\
  \textbf{D102} (2020) 092005},
  \href{http://arxiv.org/abs/2005.13419}{{\normalfont\ttfamily
  arXiv:2005.13419}}\relax
\mciteBstWouldAddEndPuncttrue
\mciteSetBstMidEndSepPunct{\mcitedefaultmidpunct}
{\mcitedefaultendpunct}{\mcitedefaultseppunct}\relax
\EndOfBibitem
\bibitem{LHCb-PAPER-2020-009}
LHCb collaboration, R.~Aaij {\em et~al.},
  \ifthenelse{\boolean{articletitles}}{\emph{{Study of the $\psires_2(3823)$
  and $\chicone(3872)$ states in $B^+\to (\jpsi \pi^+\pi^-)K^+$ decays}},
  }{}\href{https://doi.org/10.1007/JHEP08(2020)123}{JHEP \textbf{08} (2020)
  123}, \href{http://arxiv.org/abs/2005.13422}{{\normalfont\ttfamily
  arXiv:2005.13422}}\relax
\mciteBstWouldAddEndPuncttrue
\mciteSetBstMidEndSepPunct{\mcitedefaultmidpunct}
{\mcitedefaultendpunct}{\mcitedefaultseppunct}\relax
\EndOfBibitem
\bibitem{LHcb-PAPER-2020-023}
LHCb collaboration, R.~Aaij {\em et~al.},
  \ifthenelse{\boolean{articletitles}}{\emph{{Observation of
  multiplicity-dependent $\chicone(3872)$ and $\psitwos$ production in
  $\proton\proton$~collisions}},
  }{}\href{https://doi.org/10.1103/PhysRevLett.126.092001}{Phys.\ Rev.\ Lett.\
  \textbf{126} (2021) 092001},
  \href{http://arxiv.org/abs/2009.06619}{{\normalfont\ttfamily
  arXiv:2009.06619}}\relax
\mciteBstWouldAddEndPuncttrue
\mciteSetBstMidEndSepPunct{\mcitedefaultmidpunct}
{\mcitedefaultendpunct}{\mcitedefaultseppunct}\relax
\EndOfBibitem
\bibitem{LHCb-PAPER-2021-026}
LHCb collaboration, R.~Aaij {\em et~al.},
  \ifthenelse{\boolean{articletitles}}{\emph{{Measurement of $\chicone(3872)$
  production in proton\nobreakdash-proton collisions at \mbox{$\sqs=8$}
  and~$13\tev$}}, }{}\href{https://doi.org/10.1007/JHEP01(2022)131}{JHEP
  \textbf{01} (2022) 131},
  \href{http://arxiv.org/abs/2109.07360}{{\normalfont\ttfamily
  arXiv:2109.07360}}\relax
\mciteBstWouldAddEndPuncttrue
\mciteSetBstMidEndSepPunct{\mcitedefaultmidpunct}
{\mcitedefaultendpunct}{\mcitedefaultseppunct}\relax
\EndOfBibitem
\bibitem{LHCb-PAPER-2021-045}
LHCb collaboration, R.~Aaij {\em et~al.},
  \ifthenelse{\boolean{articletitles}}{\emph{{Observation of sizeable $\omegaz$
  contribution to \mbox{$\decay{\chicone(3872)}{\pip\pim \jpsi}$}~decays}},
  }{}\href{https://doi.org/10.1103/PhysRevD.108.L011103}{Phys.\ Rev.\
  \textbf{D108} (2023) L011103},
  \href{http://arxiv.org/abs/2204.12597}{{\normalfont\ttfamily
  arXiv:2204.12597}}\relax
\mciteBstWouldAddEndPuncttrue
\mciteSetBstMidEndSepPunct{\mcitedefaultmidpunct}
{\mcitedefaultendpunct}{\mcitedefaultseppunct}\relax
\EndOfBibitem
\bibitem{LHCb-PAPER-2021-047}
LHCb collaboration, R.~Aaij {\em et~al.},
  \ifthenelse{\boolean{articletitles}}{\emph{{Study of charmonium and
  charmonium-like contributions in \mbox{$\decay{\Bp}{\jpsi \Peta
  \Kp}$}~decays}}, }{}\href{https://doi.org/10.1007/JHEP04(2022)046}{JHEP
  \textbf{04} (2021) 46},
  \href{http://arxiv.org/abs/2202.04045}{{\normalfont\ttfamily
  arXiv:2202.04045}}\relax
\mciteBstWouldAddEndPuncttrue
\mciteSetBstMidEndSepPunct{\mcitedefaultmidpunct}
{\mcitedefaultendpunct}{\mcitedefaultseppunct}\relax
\EndOfBibitem
\bibitem{LHCb-PAPER-2023-026}
LHCb collaboration, R.~Aaij {\em et~al.},
  \ifthenelse{\boolean{articletitles}}{\emph{{Modification of $\theX$ and
  $\psitwos$ production in $\proton\mathrm{Pb}$ collisions at $\sqrt{s_{NN}} =
  8.16$ TeV}}, }{}\href{http://arxiv.org/abs/2402.14975}{{\normalfont\ttfamily
  arXiv:2402.14975}}, {submitted to Phys. Rev. Lett.}\relax
\mciteBstWouldAddEndPunctfalse
\mciteSetBstMidEndSepPunct{\mcitedefaultmidpunct}
{}{\mcitedefaultseppunct}\relax
\EndOfBibitem
\bibitem{CMS:2021znk}
CMS collaboration, A.~M. Sirunyan {\em et~al.},
  \ifthenelse{\boolean{articletitles}}{\emph{{Evidence for $\PX(3872)$ in
  $Pb-Pb$~collisions and studies of its prompt production at
  $\sqrt{s_{\PN\PN}}=5.02\tev$}},
  }{}\href{https://doi.org/10.1103/PhysRevLett.128.032001}{Phys.\ Rev.\ Lett.\
  \textbf{128} (2022) 032001},
  \href{http://arxiv.org/abs/2102.13048}{{\normalfont\ttfamily
  arXiv:2102.13048}}\relax
\mciteBstWouldAddEndPuncttrue
\mciteSetBstMidEndSepPunct{\mcitedefaultmidpunct}
{\mcitedefaultendpunct}{\mcitedefaultseppunct}\relax
\EndOfBibitem
\bibitem{Braaten:2003he}
E.~Braaten and M.~Kusunoki,
  \ifthenelse{\boolean{articletitles}}{\emph{{Low-energy universality and the
  new charmonium resonance at 3870\mev}},
  }{}\href{https://doi.org/10.1103/PhysRevD.69.074005}{Phys.\ Rev.\
  \textbf{D69} (2004) 074005},
  \href{http://arxiv.org/abs/hep-ph/0311147}{{\normalfont\ttfamily
  arXiv:hep-ph/0311147}}\relax
\mciteBstWouldAddEndPuncttrue
\mciteSetBstMidEndSepPunct{\mcitedefaultmidpunct}
{\mcitedefaultendpunct}{\mcitedefaultseppunct}\relax
\EndOfBibitem
\bibitem{Swanson:2003tb}
E.~S. Swanson, \ifthenelse{\boolean{articletitles}}{\emph{{Short range
  structure in the~$\PX(3872)$}},
  }{}\href{https://doi.org/10.1016/j.physletb.2004.03.033}{Phys.\ Lett.\
  \textbf{B588} (2004) 189},
  \href{http://arxiv.org/abs/hep-ph/0311229}{{\normalfont\ttfamily
  arXiv:hep-ph/0311229}}\relax
\mciteBstWouldAddEndPuncttrue
\mciteSetBstMidEndSepPunct{\mcitedefaultmidpunct}
{\mcitedefaultendpunct}{\mcitedefaultseppunct}\relax
\EndOfBibitem
\bibitem{Wong:2003xk}
C.-Y. Wong, \ifthenelse{\boolean{articletitles}}{\emph{{Molecular states of
  heavy quark mesons}},
  }{}\href{https://doi.org/10.1103/PhysRevC.69.055202}{Phys.\ Rev.\
  \textbf{C69} (2004) 055202},
  \href{http://arxiv.org/abs/hep-ph/0311088}{{\normalfont\ttfamily
  arXiv:hep-ph/0311088}}\relax
\mciteBstWouldAddEndPuncttrue
\mciteSetBstMidEndSepPunct{\mcitedefaultmidpunct}
{\mcitedefaultendpunct}{\mcitedefaultseppunct}\relax
\EndOfBibitem
\bibitem{Tornqvist:2004qy}
N.~A. Tornqvist, \ifthenelse{\boolean{articletitles}}{\emph{{Isospin breaking
  of the narrow charmonium state of Belle at 3872\mev as a~deuson}},
  }{}\href{https://doi.org/10.1016/j.physletb.2004.03.077}{Phys.\ Lett.\
  \textbf{B590} (2004) 209},
  \href{http://arxiv.org/abs/hep-ph/0402237}{{\normalfont\ttfamily
  arXiv:hep-ph/0402237}}\relax
\mciteBstWouldAddEndPuncttrue
\mciteSetBstMidEndSepPunct{\mcitedefaultmidpunct}
{\mcitedefaultendpunct}{\mcitedefaultseppunct}\relax
\EndOfBibitem
\bibitem{Hanhart:2007yq}
C.~Hanhart, Y.~S. Kalashnikova, A.~E. Kudryavtsev, and A.~V. Nefediev,
  \ifthenelse{\boolean{articletitles}}{\emph{{Reconciling the~$\PX(3872)$ with
  the~near\nobreakdash-threshold enhancement in the~$\Dz\Dstarzb$~final
  state}}, }{}\href{https://doi.org/10.1103/PhysRevD.76.034007}{Phys.\ Rev.\
  \textbf{D76} (2007) 034007},
  \href{http://arxiv.org/abs/0704.0605}{{\normalfont\ttfamily
  arXiv:0704.0605}}\relax
\mciteBstWouldAddEndPuncttrue
\mciteSetBstMidEndSepPunct{\mcitedefaultmidpunct}
{\mcitedefaultendpunct}{\mcitedefaultseppunct}\relax
\EndOfBibitem
\bibitem{Artoisenet:2009wk}
P.~Artoisenet and E.~Braaten,
  \ifthenelse{\boolean{articletitles}}{\emph{{Production of the~$\PX(3872)$ at
  the~Tevatron and the~LHC}},
  }{}\href{https://doi.org/10.1103/PhysRevD.81.114018}{Phys.\ Rev.\
  \textbf{D81} (2010) 114018},
  \href{http://arxiv.org/abs/0911.2016}{{\normalfont\ttfamily
  arXiv:0911.2016}}\relax
\mciteBstWouldAddEndPuncttrue
\mciteSetBstMidEndSepPunct{\mcitedefaultmidpunct}
{\mcitedefaultendpunct}{\mcitedefaultseppunct}\relax
\EndOfBibitem
\bibitem{Bignamini:2009sk}
C.~Bignamini {\em et~al.}, \ifthenelse{\boolean{articletitles}}{\emph{{Is
  the~$\PX(3872)$~production cross section at Tevatron compatible with a~hadron
  molecule interpretation?}},
  }{}\href{https://doi.org/10.1103/PhysRevLett.103.162001}{Phys.\ Rev.\ Lett.\
  \textbf{103} (2009) 162001},
  \href{http://arxiv.org/abs/0906.0882}{{\normalfont\ttfamily
  arXiv:0906.0882}}\relax
\mciteBstWouldAddEndPuncttrue
\mciteSetBstMidEndSepPunct{\mcitedefaultmidpunct}
{\mcitedefaultendpunct}{\mcitedefaultseppunct}\relax
\EndOfBibitem
\bibitem{Esposito:2015fsa}
A.~Esposito {\em et~al.},
  \ifthenelse{\boolean{articletitles}}{\emph{{Observation of light nuclei at
  ALICE and the~$\PX(3872)$~conundrum}},
  }{}\href{https://doi.org/10.1103/PhysRevD.92.034028}{Phys.\ Rev.\
  \textbf{D92} (2015) 034028},
  \href{http://arxiv.org/abs/1508.00295}{{\normalfont\ttfamily
  arXiv:1508.00295}}\relax
\mciteBstWouldAddEndPuncttrue
\mciteSetBstMidEndSepPunct{\mcitedefaultmidpunct}
{\mcitedefaultendpunct}{\mcitedefaultseppunct}\relax
\EndOfBibitem
\bibitem{She:2024dqq}
Z.-L. She {\em et~al.}, \ifthenelse{\boolean{articletitles}}{\emph{{Identifying
  $\PX(3872)$ molecular state or the~tetraquark state in
  $\proton\proton$~collisions with {\sc{PACIAE\,3.0}}}},
  }{}\href{http://arxiv.org/abs/2402.16736}{{\normalfont\ttfamily
  arXiv:2402.16736}}\relax
\mciteBstWouldAddEndPuncttrue
\mciteSetBstMidEndSepPunct{\mcitedefaultmidpunct}
{\mcitedefaultendpunct}{\mcitedefaultseppunct}\relax
\EndOfBibitem
\bibitem{Barnes:2003vb}
T.~Barnes and S.~Godfrey,
  \ifthenelse{\boolean{articletitles}}{\emph{{Charmonium options for
  the~$\PX(3872)$}},
  }{}\href{https://doi.org/10.1103/PhysRevD.69.054008}{Phys.\ Rev.\
  \textbf{D69} (2004) 054008},
  \href{http://arxiv.org/abs/hep-ph/0311162}{{\normalfont\ttfamily
  arXiv:hep-ph/0311162}}\relax
\mciteBstWouldAddEndPuncttrue
\mciteSetBstMidEndSepPunct{\mcitedefaultmidpunct}
{\mcitedefaultendpunct}{\mcitedefaultseppunct}\relax
\EndOfBibitem
\bibitem{Eichten:2004uh}
E.~J. Eichten, K.~Lane, and C.~Quigg,
  \ifthenelse{\boolean{articletitles}}{\emph{{Charmonium levels near threshold
  and the~narrow state $\decay{\PX(3872)}{\pip\pim\jpsi}$}},
  }{}\href{https://doi.org/10.1103/PhysRevD.69.094019}{Phys.\ Rev.\
  \textbf{D69} (2004) 094019},
  \href{http://arxiv.org/abs/hep-ph/0401210}{{\normalfont\ttfamily
  arXiv:hep-ph/0401210}}\relax
\mciteBstWouldAddEndPuncttrue
\mciteSetBstMidEndSepPunct{\mcitedefaultmidpunct}
{\mcitedefaultendpunct}{\mcitedefaultseppunct}\relax
\EndOfBibitem
\bibitem{Barnes:2005pb}
T.~Barnes, S.~Godfrey, and E.~S. Swanson,
  \ifthenelse{\boolean{articletitles}}{\emph{{Higher charmonia}},
  }{}\href{https://doi.org/10.1103/PhysRevD.72.054026}{Phys.\ Rev.\
  \textbf{D72} (2005) 054026},
  \href{http://arxiv.org/abs/hep-ph/0505002}{{\normalfont\ttfamily
  arXiv:hep-ph/0505002}}\relax
\mciteBstWouldAddEndPuncttrue
\mciteSetBstMidEndSepPunct{\mcitedefaultmidpunct}
{\mcitedefaultendpunct}{\mcitedefaultseppunct}\relax
\EndOfBibitem
\bibitem{Suzuki:2005ha}
M.~Suzuki, \ifthenelse{\boolean{articletitles}}{\emph{{The~$\PX(3872)$ boson:
  molecule or charmonium}},
  }{}\href{https://doi.org/10.1103/PhysRevD.72.114013}{Phys.\ Rev.\
  \textbf{D72} (2005) 114013},
  \href{http://arxiv.org/abs/hep-ph/0508258}{{\normalfont\ttfamily
  arXiv:hep-ph/0508258}}\relax
\mciteBstWouldAddEndPuncttrue
\mciteSetBstMidEndSepPunct{\mcitedefaultmidpunct}
{\mcitedefaultendpunct}{\mcitedefaultseppunct}\relax
\EndOfBibitem
\bibitem{Giacosa:2019zxw}
F.~Giacosa, M.~Piotrowska, and S.~Coito,
  \ifthenelse{\boolean{articletitles}}{\emph{{$\PX(3872)$ as virtual companion
  pole of the~charm\textendash{}anticharm state $\chicone(2\PP)$}},
  }{}\href{https://doi.org/10.1142/S0217751X19501732}{Int.\ J.\ Mod.\ Phys.\
  \textbf{A34} (2019) 1950173},
  \href{http://arxiv.org/abs/1903.06926}{{\normalfont\ttfamily
  arXiv:1903.06926}}\relax
\mciteBstWouldAddEndPuncttrue
\mciteSetBstMidEndSepPunct{\mcitedefaultmidpunct}
{\mcitedefaultendpunct}{\mcitedefaultseppunct}\relax
\EndOfBibitem
\bibitem{Close:2003sg}
F.~E. Close and P.~R. Page,
  \ifthenelse{\boolean{articletitles}}{\emph{{The~$\Dz\Dstarzb$~threshold
  resonance}}, }{}\href{https://doi.org/10.1016/j.physletb.2003.10.032}{Phys.\
  Lett.\  \textbf{B578} (2004) 119},
  \href{http://arxiv.org/abs/hep-ph/0309253}{{\normalfont\ttfamily
  arXiv:hep-ph/0309253}}\relax
\mciteBstWouldAddEndPuncttrue
\mciteSetBstMidEndSepPunct{\mcitedefaultmidpunct}
{\mcitedefaultendpunct}{\mcitedefaultseppunct}\relax
\EndOfBibitem
\bibitem{Dubynskiy:2008mq}
S.~Dubynskiy and M.~B. Voloshin,
  \ifthenelse{\boolean{articletitles}}{\emph{{Hadro\nobreakdash-charmonium}},
  }{}\href{https://doi.org/10.1016/j.physletb.2008.07.086}{Phys.\ Lett.\
  \textbf{B666} (2008) 344},
  \href{http://arxiv.org/abs/0803.2224}{{\normalfont\ttfamily
  arXiv:0803.2224}}\relax
\mciteBstWouldAddEndPuncttrue
\mciteSetBstMidEndSepPunct{\mcitedefaultmidpunct}
{\mcitedefaultendpunct}{\mcitedefaultseppunct}\relax
\EndOfBibitem
\bibitem{Close:2003mb}
F.~E. Close and S.~Godfrey,
  \ifthenelse{\boolean{articletitles}}{\emph{{Charmonium hybrid production in
  exclusive \B~meson decays}},
  }{}\href{https://doi.org/10.1016/j.physletb.2003.09.011}{Phys.\ Lett.\
  \textbf{B574} (2003) 210},
  \href{http://arxiv.org/abs/hep-ph/0305285}{{\normalfont\ttfamily
  arXiv:hep-ph/0305285}}\relax
\mciteBstWouldAddEndPuncttrue
\mciteSetBstMidEndSepPunct{\mcitedefaultmidpunct}
{\mcitedefaultendpunct}{\mcitedefaultseppunct}\relax
\EndOfBibitem
\bibitem{Li:2004sta}
B.~A. Li, \ifthenelse{\boolean{articletitles}}{\emph{{Is $\PX(3872)$ a possible
  candidate of hybrid meson}},
  }{}\href{https://doi.org/10.1016/j.physletb.2004.11.062}{Phys.\ Lett.\
  \textbf{B605} (2005) 306},
  \href{http://arxiv.org/abs/hep-ph/0410264}{{\normalfont\ttfamily
  arXiv:hep-ph/0410264}}\relax
\mciteBstWouldAddEndPuncttrue
\mciteSetBstMidEndSepPunct{\mcitedefaultmidpunct}
{\mcitedefaultendpunct}{\mcitedefaultseppunct}\relax
\EndOfBibitem
\bibitem{Maiani:2004vq}
L.~Maiani, F.~Piccinini, A.~D. Polosa, and V.~Riquer,
  \ifthenelse{\boolean{articletitles}}{\emph{{Diquark-antidiquarks with hidden
  or open charm and the nature of~$\PX(3872)$}},
  }{}\href{https://doi.org/10.1103/PhysRevD.71.014028}{Phys.\ Rev.\
  \textbf{D71} (2005) 014028},
  \href{http://arxiv.org/abs/hep-ph/0412098}{{\normalfont\ttfamily
  arXiv:hep-ph/0412098}}\relax
\mciteBstWouldAddEndPuncttrue
\mciteSetBstMidEndSepPunct{\mcitedefaultmidpunct}
{\mcitedefaultendpunct}{\mcitedefaultseppunct}\relax
\EndOfBibitem
\bibitem{Matheus:2006xi}
R.~D. Matheus, S.~Narison, M.~Nielsen, and J.~M. Richard,
  \ifthenelse{\boolean{articletitles}}{\emph{{Can the~$\PX(3872)$ be
  a~$1^{++}$~four\nobreakdash-quark state?}},
  }{}\href{https://doi.org/10.1103/PhysRevD.75.014005}{Phys.\ Rev.\
  \textbf{D75} (2007) 014005},
  \href{http://arxiv.org/abs/hep-ph/0608297}{{\normalfont\ttfamily
  arXiv:hep-ph/0608297}}\relax
\mciteBstWouldAddEndPuncttrue
\mciteSetBstMidEndSepPunct{\mcitedefaultmidpunct}
{\mcitedefaultendpunct}{\mcitedefaultseppunct}\relax
\EndOfBibitem
\bibitem{Weinberg:1965zz}
S.~Weinberg, \ifthenelse{\boolean{articletitles}}{\emph{{Evidence that
  the~deuteron is not an~elementary particle}},
  }{}\href{https://doi.org/10.1103/PhysRev.137.B672}{Phys.\ Rev.\  \textbf{137}
  (1965) B672}\relax
\mciteBstWouldAddEndPuncttrue
\mciteSetBstMidEndSepPunct{\mcitedefaultmidpunct}
{\mcitedefaultendpunct}{\mcitedefaultseppunct}\relax
\EndOfBibitem
\bibitem{Esposito:2021vhu}
A.~Esposito {\em et~al.}, \ifthenelse{\boolean{articletitles}}{\emph{{From the
  line shape of the~$\PX(3872)$ to its structure}},
  }{}\href{https://doi.org/10.1103/PhysRevD.105.L031503}{Phys.\ Rev.\
  \textbf{D105} (2022) L031503},
  \href{http://arxiv.org/abs/2108.11413}{{\normalfont\ttfamily
  arXiv:2108.11413}}\relax
\mciteBstWouldAddEndPuncttrue
\mciteSetBstMidEndSepPunct{\mcitedefaultmidpunct}
{\mcitedefaultendpunct}{\mcitedefaultseppunct}\relax
\EndOfBibitem
\bibitem{Esposito:2023mxw}
A.~Esposito {\em et~al.}, \ifthenelse{\boolean{articletitles}}{\emph{{The~role
  of the~pion in the~lineshape of the~$\PX(3872)$}},
  }{}\href{https://doi.org/10.1016/j.physletb.2023.138285}{Phys.\ Lett.\
  \textbf{847} (2023) 138285},
  \href{http://arxiv.org/abs/2307.11400}{{\normalfont\ttfamily
  arXiv:2307.11400}}\relax
\mciteBstWouldAddEndPuncttrue
\mciteSetBstMidEndSepPunct{\mcitedefaultmidpunct}
{\mcitedefaultendpunct}{\mcitedefaultseppunct}\relax
\EndOfBibitem
\bibitem{Swanson:2004pp}
E.~S. Swanson, \ifthenelse{\boolean{articletitles}}{\emph{{Diagnostic decays of
  the~$\PX(3872)$}},
  }{}\href{https://doi.org/10.1016/j.physletb.2004.07.059}{Phys.\ Lett.\
  \textbf{B598} (2004) 197},
  \href{http://arxiv.org/abs/hep-ph/0406080}{{\normalfont\ttfamily
  arXiv:hep-ph/0406080}}\relax
\mciteBstWouldAddEndPuncttrue
\mciteSetBstMidEndSepPunct{\mcitedefaultmidpunct}
{\mcitedefaultendpunct}{\mcitedefaultseppunct}\relax
\EndOfBibitem
\bibitem{Guo:2014taa}
F.-K. Guo {\em et~al.}, \ifthenelse{\boolean{articletitles}}{\emph{{What can
  radiative decays of the~$\PX(3872)$~teach us about its nature?}},
  }{}\href{https://doi.org/10.1016/j.physletb.2015.02.013}{Phys.\ Lett.\
  \textbf{B742} (2015) 394},
  \href{http://arxiv.org/abs/1410.6712}{{\normalfont\ttfamily
  arXiv:1410.6712}}\relax
\mciteBstWouldAddEndPuncttrue
\mciteSetBstMidEndSepPunct{\mcitedefaultmidpunct}
{\mcitedefaultendpunct}{\mcitedefaultseppunct}\relax
\EndOfBibitem
\bibitem{Molnar:2016dbo}
D.~A.~S. Molnar, R.~F. Luiz, and R.~Higa,
  \ifthenelse{\boolean{articletitles}}{\emph{{Short-distance RG-analysis of
  $\PX(3872)$ radiative decays}},
  }{}\href{http://arxiv.org/abs/1601.03366}{{\normalfont\ttfamily
  arXiv:1601.03366}}\relax
\mciteBstWouldAddEndPuncttrue
\mciteSetBstMidEndSepPunct{\mcitedefaultmidpunct}
{\mcitedefaultendpunct}{\mcitedefaultseppunct}\relax
\EndOfBibitem
\bibitem{DeFazio:2008xq}
F.~De~Fazio, \ifthenelse{\boolean{articletitles}}{\emph{{Radiative transitions
  of heavy quarkonium states}},
  }{}\href{https://doi.org/10.1103/PhysRevD.79.054015}{Phys.\ Rev.\
  \textbf{D79} (2009) 054015}, Erratum
  \href{https://doi.org/10.1103/PhysRevD.83.099901}{ibid.\   \textbf{D83}
  (2011) 099901}, \href{http://arxiv.org/abs/0812.0716}{{\normalfont\ttfamily
  arXiv:0812.0716}}\relax
\mciteBstWouldAddEndPuncttrue
\mciteSetBstMidEndSepPunct{\mcitedefaultmidpunct}
{\mcitedefaultendpunct}{\mcitedefaultseppunct}\relax
\EndOfBibitem
\bibitem{Li:2009zu}
B.-Q. Li and K.-T. Chao, \ifthenelse{\boolean{articletitles}}{\emph{{Higher
  Charmonia and $\PX,\PY,\PZ$ states with screened potential}},
  }{}\href{https://doi.org/10.1103/PhysRevD.79.094004}{Phys.\ Rev.\
  \textbf{D79} (2009) 094004},
  \href{http://arxiv.org/abs/0903.5506}{{\normalfont\ttfamily
  arXiv:0903.5506}}\relax
\mciteBstWouldAddEndPuncttrue
\mciteSetBstMidEndSepPunct{\mcitedefaultmidpunct}
{\mcitedefaultendpunct}{\mcitedefaultseppunct}\relax
\EndOfBibitem
\bibitem{Dong:2009uf}
Y.~Dong, A.~Faessler, T.~Gutsche, and V.~E. Lyubovitskij,
  \ifthenelse{\boolean{articletitles}}{\emph{{$\jpsi\g$ and $\psitwos\g$ decay
  modes of the $\PX(3872)$}},
  }{}\href{https://doi.org/10.1088/0954-3899/38/1/015001}{J.\ Phys.\
  \textbf{G38} (2011) 015001},
  \href{http://arxiv.org/abs/0909.0380}{{\normalfont\ttfamily
  arXiv:0909.0380}}\relax
\mciteBstWouldAddEndPuncttrue
\mciteSetBstMidEndSepPunct{\mcitedefaultmidpunct}
{\mcitedefaultendpunct}{\mcitedefaultseppunct}\relax
\EndOfBibitem
\bibitem{Badalian:2012jz}
A.~M. Badalian, V.~D. Orlovsky, Y.~A. Simonov, and B.~L.~G. Bakker,
  \ifthenelse{\boolean{articletitles}}{\emph{{The~ratio of decay widths
  of~$\PX(3872)$ to $ \Ppsi^{\prime}\g$ and $\jpsi\g$ as a~test of
  the~$\PX(3872)$~dynamical structure}},
  }{}\href{https://doi.org/10.1103/PhysRevD.85.114002}{Phys.\ Rev.\
  \textbf{D85} (2012) 114002},
  \href{http://arxiv.org/abs/1202.4882}{{\normalfont\ttfamily
  arXiv:1202.4882}}\relax
\mciteBstWouldAddEndPuncttrue
\mciteSetBstMidEndSepPunct{\mcitedefaultmidpunct}
{\mcitedefaultendpunct}{\mcitedefaultseppunct}\relax
\EndOfBibitem
\bibitem{Ferretti:2014xqa}
J.~Ferretti, G.~Galat\`a, and E.~Santopinto,
  \ifthenelse{\boolean{articletitles}}{\emph{{Quark structure of the
  $\PX(3872)$ and $\Pchi_\bquark(3\PP)$ resonances}},
  }{}\href{https://doi.org/10.1103/PhysRevD.90.054010}{Phys.\ Rev.\
  \textbf{D90} (2014) 054010},
  \href{http://arxiv.org/abs/1401.4431}{{\normalfont\ttfamily
  arXiv:1401.4431}}\relax
\mciteBstWouldAddEndPuncttrue
\mciteSetBstMidEndSepPunct{\mcitedefaultmidpunct}
{\mcitedefaultendpunct}{\mcitedefaultseppunct}\relax
\EndOfBibitem
\bibitem{Badalian:2015dha}
A.~M. Badalian, Y.~A. Simonov, and B.~L.~G. Bakker,
  \ifthenelse{\boolean{articletitles}}{\emph{{$\cquark\cquarkbar$~interaction
  above threshold and the~radiative decay
  \mbox{$\decay{\PX(3872)}{\jpsi\g}$}}},
  }{}\href{https://doi.org/10.1103/PhysRevD.91.056001}{Phys.\ Rev.\
  \textbf{D91} (2015) 056001},
  \href{http://arxiv.org/abs/1501.01168}{{\normalfont\ttfamily
  arXiv:1501.01168}}\relax
\mciteBstWouldAddEndPuncttrue
\mciteSetBstMidEndSepPunct{\mcitedefaultmidpunct}
{\mcitedefaultendpunct}{\mcitedefaultseppunct}\relax
\EndOfBibitem
\bibitem{Deng:2016stx}
W.-J. Deng, H.~Liu, L.-C. Gui, and X.-H. Zhong,
  \ifthenelse{\boolean{articletitles}}{\emph{{Charmonium spectrum and their
  electromagnetic transitions with higher multipole contributions}},
  }{}\href{https://doi.org/10.1103/PhysRevD.95.034026}{Phys.\ Rev.\
  \textbf{D95} (2017) 034026},
  \href{http://arxiv.org/abs/1608.00287}{{\normalfont\ttfamily
  arXiv:1608.00287}}\relax
\mciteBstWouldAddEndPuncttrue
\mciteSetBstMidEndSepPunct{\mcitedefaultmidpunct}
{\mcitedefaultendpunct}{\mcitedefaultseppunct}\relax
\EndOfBibitem
\bibitem{Rathaud:2016tys}
D.~P. Rathaud and A.~K. Rai,
  \ifthenelse{\boolean{articletitles}}{\emph{{Dimesonic states with the
  heavy-light flavour mesons}},
  }{}\href{https://doi.org/10.1140/epjp/i2017-11641-3}{Eur.\ Phys.\ J.\ Plus
  \textbf{132} (2017) 370},
  \href{http://arxiv.org/abs/1608.03781}{{\normalfont\ttfamily
  arXiv:1608.03781}}\relax
\mciteBstWouldAddEndPuncttrue
\mciteSetBstMidEndSepPunct{\mcitedefaultmidpunct}
{\mcitedefaultendpunct}{\mcitedefaultseppunct}\relax
\EndOfBibitem
\bibitem{Grinstein:2024rcu}
B.~Grinstein, L.~Maiani, and A.~D. Polosa,
  \ifthenelse{\boolean{articletitles}}{\emph{{Radiative decays of
  $\PX(3872)$~discriminate between the~molecular and compact interpretations}},
  }{}\href{https://doi.org/10.1103/PhysRevD.109.074009}{Phys.\ Rev.\
  \textbf{D109} (2024) 074009},
  \href{http://arxiv.org/abs/2401.11623}{{\normalfont\ttfamily
  arXiv:2401.11623}}\relax
\mciteBstWouldAddEndPuncttrue
\mciteSetBstMidEndSepPunct{\mcitedefaultmidpunct}
{\mcitedefaultendpunct}{\mcitedefaultseppunct}\relax
\EndOfBibitem
\bibitem{Ortega:2012rs}
P.~G. Ortega, D.~R. Entem, and F.~Fernandez,
  \ifthenelse{\boolean{articletitles}}{\emph{{Molecular structures in
  charmonium spectrum: The $\PX\PY\PZ$~puzzle}},
  }{}\href{https://doi.org/10.1088/0954-3899/40/6/065107}{J.\ Phys.\
  \textbf{G40} (2013) 065107},
  \href{http://arxiv.org/abs/1205.1699}{{\normalfont\ttfamily
  arXiv:1205.1699}}\relax
\mciteBstWouldAddEndPuncttrue
\mciteSetBstMidEndSepPunct{\mcitedefaultmidpunct}
{\mcitedefaultendpunct}{\mcitedefaultseppunct}\relax
\EndOfBibitem
\bibitem{Cincioglu:2016fkm}
E.~Cincioglu, J.~Nieves, A.~Ozpineci, and A.~U. Yilmazer,
  \ifthenelse{\boolean{articletitles}}{\emph{{Quarkonium contribution to meson
  molecules}}, }{}\href{https://doi.org/10.1140/epjc/s10052-016-4413-1}{Eur.\
  Phys.\ J.\  \textbf{C76} (2016) 576},
  \href{http://arxiv.org/abs/1606.03239}{{\normalfont\ttfamily
  arXiv:1606.03239}}\relax
\mciteBstWouldAddEndPuncttrue
\mciteSetBstMidEndSepPunct{\mcitedefaultmidpunct}
{\mcitedefaultendpunct}{\mcitedefaultseppunct}\relax
\EndOfBibitem
\bibitem{Takeuchi:2016hat}
S.~Takeuchi, M.~Takizawa, and K.~Shimizu,
  \ifthenelse{\boolean{articletitles}}{\emph{{Radiative decays of
  the~$\PX(3872)$ in the~charmonium\nobreakdash-molecule hybrid picture}},
  }{}\href{https://doi.org/10.7566/JPSCP.17.112001}{JPS Conf.\ Proc.\
  \textbf{17} (2017) 112001},
  \href{http://arxiv.org/abs/1602.04297}{{\normalfont\ttfamily
  arXiv:1602.04297}}\relax
\mciteBstWouldAddEndPuncttrue
\mciteSetBstMidEndSepPunct{\mcitedefaultmidpunct}
{\mcitedefaultendpunct}{\mcitedefaultseppunct}\relax
\EndOfBibitem
\bibitem{Lebed:2022vks}
R.~F. Lebed and S.~R. Martinez,
  \ifthenelse{\boolean{articletitles}}{\emph{{Diabatic representation of exotic
  hadrons in the~dynamical diquark model}},
  }{}\href{https://doi.org/10.1103/PhysRevD.106.074007}{Phys.\ Rev.\
  \textbf{D106} (2022) 074007},
  \href{http://arxiv.org/abs/2207.01101}{{\normalfont\ttfamily
  arXiv:2207.01101}}\relax
\mciteBstWouldAddEndPuncttrue
\mciteSetBstMidEndSepPunct{\mcitedefaultmidpunct}
{\mcitedefaultendpunct}{\mcitedefaultseppunct}\relax
\EndOfBibitem
\bibitem{LHCb-DP-2008-001}
LHCb collaboration, A.~A. Alves~Jr.\ {\em et~al.},
  \ifthenelse{\boolean{articletitles}}{\emph{{The \lhcb detector at the LHC}},
  }{}\href{https://doi.org/10.1088/1748-0221/3/08/S08005}{JINST \textbf{3}
  (2008) S08005}\relax
\mciteBstWouldAddEndPuncttrue
\mciteSetBstMidEndSepPunct{\mcitedefaultmidpunct}
{\mcitedefaultendpunct}{\mcitedefaultseppunct}\relax
\EndOfBibitem
\bibitem{LHCb-DP-2014-002}
LHCb collaboration, R.~Aaij {\em et~al.},
  \ifthenelse{\boolean{articletitles}}{\emph{{LHCb detector performance}},
  }{}\href{https://doi.org/10.1142/S0217751X15300227}{Int.\ J.\ Mod.\ Phys.\
  \textbf{A30} (2015) 1530022},
  \href{http://arxiv.org/abs/1412.6352}{{\normalfont\ttfamily
  arXiv:1412.6352}}\relax
\mciteBstWouldAddEndPuncttrue
\mciteSetBstMidEndSepPunct{\mcitedefaultmidpunct}
{\mcitedefaultendpunct}{\mcitedefaultseppunct}\relax
\EndOfBibitem
\bibitem{LHCb-DP-2014-001}
R.~Aaij {\em et~al.}, \ifthenelse{\boolean{articletitles}}{\emph{{Performance
  of the LHCb Vertex Locator}},
  }{}\href{https://doi.org/10.1088/1748-0221/9/09/P09007}{JINST \textbf{9}
  (2014) P09007}, \href{http://arxiv.org/abs/1405.7808}{{\normalfont\ttfamily
  arXiv:1405.7808}}\relax
\mciteBstWouldAddEndPuncttrue
\mciteSetBstMidEndSepPunct{\mcitedefaultmidpunct}
{\mcitedefaultendpunct}{\mcitedefaultseppunct}\relax
\EndOfBibitem
\bibitem{LHCb-DP-2013-003}
R.~Arink {\em et~al.}, \ifthenelse{\boolean{articletitles}}{\emph{{Performance
  of the LHCb Outer Tracker}},
  }{}\href{https://doi.org/10.1088/1748-0221/9/01/P01002}{JINST \textbf{9}
  (2014) P01002}, \href{http://arxiv.org/abs/1311.3893}{{\normalfont\ttfamily
  arXiv:1311.3893}}\relax
\mciteBstWouldAddEndPuncttrue
\mciteSetBstMidEndSepPunct{\mcitedefaultmidpunct}
{\mcitedefaultendpunct}{\mcitedefaultseppunct}\relax
\EndOfBibitem
\bibitem{LHCb-DP-2017-001}
P.~d'Argent {\em et~al.}, \ifthenelse{\boolean{articletitles}}{\emph{{Improved
  performance of the LHCb Outer Tracker in LHC Run\,2}},
  }{}\href{https://doi.org/10.1088/1748-0221/12/11/P11016}{JINST \textbf{12}
  (2017) P11016}, \href{http://arxiv.org/abs/1708.00819}{{\normalfont\ttfamily
  arXiv:1708.00819}}\relax
\mciteBstWouldAddEndPuncttrue
\mciteSetBstMidEndSepPunct{\mcitedefaultmidpunct}
{\mcitedefaultendpunct}{\mcitedefaultseppunct}\relax
\EndOfBibitem
\bibitem{LHCb-DP-2012-003}
M.~Adinolfi {\em et~al.},
  \ifthenelse{\boolean{articletitles}}{\emph{{Performance of the \lhcb RICH
  detector at the LHC}},
  }{}\href{https://doi.org/10.1140/epjc/s10052-013-2431-9}{Eur.\ Phys.\ J.\
  \textbf{C73} (2013) 2431},
  \href{http://arxiv.org/abs/1211.6759}{{\normalfont\ttfamily
  arXiv:1211.6759}}\relax
\mciteBstWouldAddEndPuncttrue
\mciteSetBstMidEndSepPunct{\mcitedefaultmidpunct}
{\mcitedefaultendpunct}{\mcitedefaultseppunct}\relax
\EndOfBibitem
\bibitem{LHCb-DP-2012-002}
A.~A. Alves~Jr.\ {\em et~al.},
  \ifthenelse{\boolean{articletitles}}{\emph{{Performance of the LHCb muon
  system}}, }{}\href{https://doi.org/10.1088/1748-0221/8/02/P02022}{JINST
  \textbf{8} (2013) P02022},
  \href{http://arxiv.org/abs/1211.1346}{{\normalfont\ttfamily
  arXiv:1211.1346}}\relax
\mciteBstWouldAddEndPuncttrue
\mciteSetBstMidEndSepPunct{\mcitedefaultmidpunct}
{\mcitedefaultendpunct}{\mcitedefaultseppunct}\relax
\EndOfBibitem
\bibitem{LHCb-DP-2012-004}
R.~Aaij {\em et~al.}, \ifthenelse{\boolean{articletitles}}{\emph{{The \lhcb
  trigger and its performance in 2011}},
  }{}\href{https://doi.org/10.1088/1748-0221/8/04/P04022}{JINST \textbf{8}
  (2013) P04022}, \href{http://arxiv.org/abs/1211.3055}{{\normalfont\ttfamily
  arXiv:1211.3055}}\relax
\mciteBstWouldAddEndPuncttrue
\mciteSetBstMidEndSepPunct{\mcitedefaultmidpunct}
{\mcitedefaultendpunct}{\mcitedefaultseppunct}\relax
\EndOfBibitem
\bibitem{LHCb-DP-2019-001}
R.~Aaij {\em et~al.}, \ifthenelse{\boolean{articletitles}}{\emph{{Performance
  of the LHCb trigger and full real-time reconstruction in Run\,2 of the~LHC}},
  }{}\href{https://doi.org/10.1088/1748-0221/14/04/P04013}{JINST \textbf{14}
  (2019) P04013}, \href{http://arxiv.org/abs/1812.10790}{{\normalfont\ttfamily
  arXiv:1812.10790}}\relax
\mciteBstWouldAddEndPuncttrue
\mciteSetBstMidEndSepPunct{\mcitedefaultmidpunct}
{\mcitedefaultendpunct}{\mcitedefaultseppunct}\relax
\EndOfBibitem
\bibitem{Sjostrand:2007gs}
T.~Sj\"{o}strand, S.~Mrenna, and P.~Skands,
  \ifthenelse{\boolean{articletitles}}{\emph{{A brief introduction to
  {\sc{Pythia}}\,$8.1$}},
  }{}\href{https://doi.org/10.1016/j.cpc.2008.01.036}{Comput.\ Phys.\ Commun.\
  \textbf{178} (2008) 852},
  \href{http://arxiv.org/abs/0710.3820}{{\normalfont\ttfamily
  arXiv:0710.3820}}\relax
\mciteBstWouldAddEndPuncttrue
\mciteSetBstMidEndSepPunct{\mcitedefaultmidpunct}
{\mcitedefaultendpunct}{\mcitedefaultseppunct}\relax
\EndOfBibitem
\bibitem{LHCb-PROC-2010-056}
I.~Belyaev {\em et~al.}, \ifthenelse{\boolean{articletitles}}{\emph{{Handling
  of the generation of primary events in {\sc{Gauss}}, the~LHCb simulation
  framework}}, }{}\href{https://doi.org/10.1088/1742-6596/331/3/032047}{J.\
  Phys.\ Conf.\ Ser.\  \textbf{331} (2011) 032047}\relax
\mciteBstWouldAddEndPuncttrue
\mciteSetBstMidEndSepPunct{\mcitedefaultmidpunct}
{\mcitedefaultendpunct}{\mcitedefaultseppunct}\relax
\EndOfBibitem
\bibitem{Lange:2001uf}
D.~J. Lange, \ifthenelse{\boolean{articletitles}}{\emph{{The EvtGen particle
  decay simulation package}},
  }{}\href{https://doi.org/10.1016/S0168-9002(01)00089-4}{Nucl.\ Instrum.\
  Meth.\  \textbf{A462} (2001) 152}\relax
\mciteBstWouldAddEndPuncttrue
\mciteSetBstMidEndSepPunct{\mcitedefaultmidpunct}
{\mcitedefaultendpunct}{\mcitedefaultseppunct}\relax
\EndOfBibitem
\bibitem{davidson2015photos}
N.~Davidson, T.~Przedzinski, and Z.~Was,
  \ifthenelse{\boolean{articletitles}}{\emph{{ {\sc{Photos}} interface in C++:
  Technical and physics documentation}},
  }{}\href{https://doi.org/https://doi.org/10.1016/j.cpc.2015.09.013}{Comp.\
  Phys.\ Comm.\  \textbf{199} (2016) 86},
  \href{http://arxiv.org/abs/1011.0937}{{\normalfont\ttfamily
  arXiv:1011.0937}}\relax
\mciteBstWouldAddEndPuncttrue
\mciteSetBstMidEndSepPunct{\mcitedefaultmidpunct}
{\mcitedefaultendpunct}{\mcitedefaultseppunct}\relax
\EndOfBibitem
\bibitem{Agostinelli:2002hh}
Geant4 collaboration, S.~Agostinelli {\em et~al.},
  \ifthenelse{\boolean{articletitles}}{\emph{{ {\sc{Geant4}}: A simulation
  toolkit}}, }{}\href{https://doi.org/10.1016/S0168-9002(03)01368-8}{Nucl.\
  Instrum.\ Meth.\  \textbf{A506} (2003) 250}\relax
\mciteBstWouldAddEndPuncttrue
\mciteSetBstMidEndSepPunct{\mcitedefaultmidpunct}
{\mcitedefaultendpunct}{\mcitedefaultseppunct}\relax
\EndOfBibitem
\bibitem{Allison:2006ve}
Geant4 collaboration, J.~Allison {\em et~al.},
  \ifthenelse{\boolean{articletitles}}{\emph{{ {\sc{Geant4}} developments and
  applications}}, }{}\href{https://doi.org/10.1109/TNS.2006.869826}{IEEE
  Trans.\ Nucl.\ Sci.\  \textbf{53} (2006) 270}\relax
\mciteBstWouldAddEndPuncttrue
\mciteSetBstMidEndSepPunct{\mcitedefaultmidpunct}
{\mcitedefaultendpunct}{\mcitedefaultseppunct}\relax
\EndOfBibitem
\bibitem{LHCb-PROC-2011-006}
M.~Clemencic {\em et~al.}, \ifthenelse{\boolean{articletitles}}{\emph{{The
  \lhcb simulation application, {\sc{Gauss}}: Design, evolution and
  experience}}, }{}\href{https://doi.org/10.1088/1742-6596/331/3/032023}{J.\
  Phys.\ Conf.\ Ser.\  \textbf{331} (2011) 032023}\relax
\mciteBstWouldAddEndPuncttrue
\mciteSetBstMidEndSepPunct{\mcitedefaultmidpunct}
{\mcitedefaultendpunct}{\mcitedefaultseppunct}\relax
\EndOfBibitem
\bibitem{LHCb-DP-2018-001}
R.~Aaij {\em et~al.}, \ifthenelse{\boolean{articletitles}}{\emph{{Selection and
  processing of calibration samples to measure the particle identification
  performance of the LHCb experiment in Run\,2}},
  }{}\href{https://doi.org/10.1140/epjti/s40485-019-0050-z}{Eur.\ Phys.\ J.\
  Tech.\ Instr.\  \textbf{6} (2019) 1},
  \href{http://arxiv.org/abs/1803.00824}{{\normalfont\ttfamily
  arXiv:1803.00824}}\relax
\mciteBstWouldAddEndPuncttrue
\mciteSetBstMidEndSepPunct{\mcitedefaultmidpunct}
{\mcitedefaultendpunct}{\mcitedefaultseppunct}\relax
\EndOfBibitem
\bibitem{LHCb-DP-2013-002}
LHCb collaboration, R.~Aaij {\em et~al.},
  \ifthenelse{\boolean{articletitles}}{\emph{{Measurement of the track
  reconstruction efficiency at LHCb}},
  }{}\href{https://doi.org/10.1088/1748-0221/10/02/P02007}{JINST \textbf{10}
  (2015) P02007}, \href{http://arxiv.org/abs/1408.1251}{{\normalfont\ttfamily
  arXiv:1408.1251}}\relax
\mciteBstWouldAddEndPuncttrue
\mciteSetBstMidEndSepPunct{\mcitedefaultmidpunct}
{\mcitedefaultendpunct}{\mcitedefaultseppunct}\relax
\EndOfBibitem
\bibitem{LHCb-PAPER-2012-022}
LHCb collaboration, R.~Aaij {\em et~al.},
  \ifthenelse{\boolean{articletitles}}{\emph{{Evidence for the decay
  \mbox{\decay{\Bz}{\jpsi\omegaz}} and measurement of the relative branching
  fractions of \Bs meson decays to $\jpsi\etaz$ and $\jpsi\etapr$}},
  }{}\href{https://doi.org/10.1016/j.nuclphysb.2012.10.021}{Nucl.\ Phys.\
  \textbf{B867} (2013) 547},
  \href{http://arxiv.org/abs/1210.2631}{{\normalfont\ttfamily
  arXiv:1210.2631}}\relax
\mciteBstWouldAddEndPuncttrue
\mciteSetBstMidEndSepPunct{\mcitedefaultmidpunct}
{\mcitedefaultendpunct}{\mcitedefaultseppunct}\relax
\EndOfBibitem
\bibitem{LHCb-PAPER-2012-053}
LHCb collaboration, R.~Aaij {\em et~al.},
  \ifthenelse{\boolean{articletitles}}{\emph{{Observations of
  \mbox{\decay{\Bs}{\psitwos\Peta}} and
  \mbox{\decay{\BdorBs}{\psitwos\pip\pim}} decays}},
  }{}\href{https://doi.org/10.1016/j.nuclphysb.2013.03.004}{Nucl.\ Phys.\
  \textbf{B871} (2013) 403},
  \href{http://arxiv.org/abs/1302.6354}{{\normalfont\ttfamily
  arXiv:1302.6354}}\relax
\mciteBstWouldAddEndPuncttrue
\mciteSetBstMidEndSepPunct{\mcitedefaultmidpunct}
{\mcitedefaultendpunct}{\mcitedefaultseppunct}\relax
\EndOfBibitem
\bibitem{Govorkova:2015vqa}
E.~Govorkova, \ifthenelse{\boolean{articletitles}}{\emph{{Study of
  $\piz/\g$~efficiency using $\B$~meson decays in the~LHCb experiment}},
  }{}\href{https://doi.org/10.1134/S1063778816100070}{Phys.\ Atom.\ Nucl.\
  \textbf{79} (2016) 1474},
  \href{http://arxiv.org/abs/1505.02960}{{\normalfont\ttfamily
  arXiv:1505.02960}}\relax
\mciteBstWouldAddEndPuncttrue
\mciteSetBstMidEndSepPunct{\mcitedefaultmidpunct}
{\mcitedefaultendpunct}{\mcitedefaultseppunct}\relax
\EndOfBibitem
\bibitem{Govorkova:2124605}
K.~Govorkova, \ifthenelse{\boolean{articletitles}}{\emph{{Study of photons and
  neutral pions reconstruction efficiency in the~LHCb experiment}}, }{} 2015.
\newblock
  {\href{http://cds.cern.ch/record/2124605}{CERN-THESIS-2015-272}}\relax
\mciteBstWouldAddEndPuncttrue
\mciteSetBstMidEndSepPunct{\mcitedefaultmidpunct}
{\mcitedefaultendpunct}{\mcitedefaultseppunct}\relax
\EndOfBibitem
\bibitem{Belyaev:2015nlt}
I.~M. Belyaev, E.~M. Govorkova, V.~Y. Egorychev, and D.~V. Savrina,
  \ifthenelse{\boolean{articletitles}}{\emph{{Study of $\piz/\g$ reconstruction
  and selection efficiency in the LHCb experiment}},
  }{}\href{https://doi.org/10.3103/S0027134915060077}{Moscow Univ.\ Phys.\
  Bull.\  \textbf{70} (2015) 497}\relax
\mciteBstWouldAddEndPuncttrue
\mciteSetBstMidEndSepPunct{\mcitedefaultmidpunct}
{\mcitedefaultendpunct}{\mcitedefaultseppunct}\relax
\EndOfBibitem
\bibitem{LHCb-PAPER-2013-023}
LHCb collaboration, R.~Aaij {\em et~al.},
  \ifthenelse{\boolean{articletitles}}{\emph{{Measurement of the polarization
  amplitudes in \mbox{\decay{\Bz}{\jpsi\Kstar(892)^0}} decays}},
  }{}\href{https://doi.org/10.1103/PhysRevD.88.052002}{Phys.\ Rev.\
  \textbf{D88} (2013) 052002},
  \href{http://arxiv.org/abs/1307.2782}{{\normalfont\ttfamily
  arXiv:1307.2782}}\relax
\mciteBstWouldAddEndPuncttrue
\mciteSetBstMidEndSepPunct{\mcitedefaultmidpunct}
{\mcitedefaultendpunct}{\mcitedefaultseppunct}\relax
\EndOfBibitem
\bibitem{PDG2022}
Particle Data Group, R.~L. Workman {\em et~al.},
  \ifthenelse{\boolean{articletitles}}{\emph{{\href{http://pdg.lbl.gov/}{Review
  of particle physics}}}, }{}\href{https://doi.org/10.1093/ptep/ptac097}{Prog.\
  Theor.\ Exp.\ Phys.\  \textbf{2022} (2022) 083C01}\relax
\mciteBstWouldAddEndPuncttrue
\mciteSetBstMidEndSepPunct{\mcitedefaultmidpunct}
{\mcitedefaultendpunct}{\mcitedefaultseppunct}\relax
\EndOfBibitem
\bibitem{Belle:2010wrf}
Belle collaboration, H.~Guler {\em et~al.},
  \ifthenelse{\boolean{articletitles}}{\emph{{Study of the~$\Kp\pip\pim$ final
  state in \mbox{$\decay{\Bu}{\jpsi\Kp\pip\pim}$} and
  \mbox{$\decay{\Bu}{\psitwos\Kp\pip\pim}$} }},
  }{}\href{https://doi.org/10.1103/PhysRevD.83.032005}{Phys.\ Rev.\
  \textbf{D83} (2011) 032005},
  \href{http://arxiv.org/abs/1009.5256}{{\normalfont\ttfamily
  arXiv:1009.5256}}\relax
\mciteBstWouldAddEndPuncttrue
\mciteSetBstMidEndSepPunct{\mcitedefaultmidpunct}
{\mcitedefaultendpunct}{\mcitedefaultseppunct}\relax
\EndOfBibitem
\bibitem{Belle:2013shl}
Belle collaboration, K.~Chilikin {\em et~al.},
  \ifthenelse{\boolean{articletitles}}{\emph{{Experimental constraints on the
  spin and parity of the $\PZ(4430)^+$}},
  }{}\href{https://doi.org/10.1103/PhysRevD.88.074026}{Phys.\ Rev.\
  \textbf{D88} (2013) 074026},
  \href{http://arxiv.org/abs/1306.4894}{{\normalfont\ttfamily
  arXiv:1306.4894}}\relax
\mciteBstWouldAddEndPuncttrue
\mciteSetBstMidEndSepPunct{\mcitedefaultmidpunct}
{\mcitedefaultendpunct}{\mcitedefaultseppunct}\relax
\EndOfBibitem
\bibitem{LHCb-PAPER-2021-003}
LHCb collaboration, R.~Aaij {\em et~al.},
  \ifthenelse{\boolean{articletitles}}{\emph{{Observation of the decay
  $\Lb\to\chicone \proton \pim$}},
  }{}\href{https://doi.org/10.1007/JHEP05(2021)095}{JHEP \textbf{05} (2021)
  095}, \href{http://arxiv.org/abs/2103.04949}{{\normalfont\ttfamily
  arXiv:2103.04949}}\relax
\mciteBstWouldAddEndPuncttrue
\mciteSetBstMidEndSepPunct{\mcitedefaultmidpunct}
{\mcitedefaultendpunct}{\mcitedefaultseppunct}\relax
\EndOfBibitem
\bibitem{LHCb-PROC-2011-008}
A.~Powell {\em et~al.}, \ifthenelse{\boolean{articletitles}}{\emph{{Particle
  identification at LHCb}}, }{}PoS \textbf{ICHEP2010} (2010) 020,
  \href{https://cdsweb.cern.ch/record/1322666?ln=en}{LHCb-PROC-2011-008}\relax
\mciteBstWouldAddEndPuncttrue
\mciteSetBstMidEndSepPunct{\mcitedefaultmidpunct}
{\mcitedefaultendpunct}{\mcitedefaultseppunct}\relax
\EndOfBibitem
\bibitem{LHCb-2003-091}
O.~Deschamps {\em et~al.}, \ifthenelse{\boolean{articletitles}}{\emph{{Photon
  and neutral pion reconstruction}}, }{}
  \href{http://cdsweb.cern.ch/search?p=LHCb-2003-091&f=reportnumber&action_search=Search&c=LHCb}
  {LHCb-2003-091}, 2003\relax
\mciteBstWouldAddEndPuncttrue
\mciteSetBstMidEndSepPunct{\mcitedefaultmidpunct}
{\mcitedefaultendpunct}{\mcitedefaultseppunct}\relax
\EndOfBibitem
\bibitem{LHCb-2003-092}
H.~Terrier and I.~Belyaev, \ifthenelse{\boolean{articletitles}}{\emph{{Particle
  identification with LHCb calorimeters}}, }{}
  \href{http://cdsweb.cern.ch/search?p=LHCb-2003-092&f=reportnumber&action_search=Search&c=LHCb}
  {LHCb-2003-092}, 2003\relax
\mciteBstWouldAddEndPuncttrue
\mciteSetBstMidEndSepPunct{\mcitedefaultmidpunct}
{\mcitedefaultendpunct}{\mcitedefaultseppunct}\relax
\EndOfBibitem
\bibitem{LHCb-DP-2020-001}
C.~Abellan~Beteta {\em et~al.},
  \ifthenelse{\boolean{articletitles}}{\emph{{Calibration and performance of
  the LHCb calorimeters in Run\,1 and~2 at the~LHC}},
  }{}\href{http://arxiv.org/abs/2008.11556}{{\normalfont\ttfamily
  arXiv:2008.11556}}, {submitted to JINST}\relax
\mciteBstWouldAddEndPuncttrue
\mciteSetBstMidEndSepPunct{\mcitedefaultmidpunct}
{\mcitedefaultendpunct}{\mcitedefaultseppunct}\relax
\EndOfBibitem
\bibitem{Hulsbergen:2005pu}
W.~D. Hulsbergen, \ifthenelse{\boolean{articletitles}}{\emph{{Decay chain
  fitting with a Kalman filter}},
  }{}\href{https://doi.org/10.1016/j.nima.2005.06.078}{Nucl.\ Instrum.\ Meth.\
  \textbf{A552} (2005) 566},
  \href{http://arxiv.org/abs/physics/0503191}{{\normalfont\ttfamily
  arXiv:physics/0503191}}\relax
\mciteBstWouldAddEndPuncttrue
\mciteSetBstMidEndSepPunct{\mcitedefaultmidpunct}
{\mcitedefaultendpunct}{\mcitedefaultseppunct}\relax
\EndOfBibitem
\bibitem{McCulloch}
W.~S. McCulloch and W.~Pitts, \ifthenelse{\boolean{articletitles}}{\emph{A
  logical calculus of the ideas immanent in nervous activity},
  }{}\href{https://doi.org/10.1007/BF02478259}{The bulletin of mathematical
  biophysics \textbf{5} (1943) 115}\relax
\mciteBstWouldAddEndPuncttrue
\mciteSetBstMidEndSepPunct{\mcitedefaultmidpunct}
{\mcitedefaultendpunct}{\mcitedefaultseppunct}\relax
\EndOfBibitem
\bibitem{rosenblatt58}
F.~Rosenblatt, \ifthenelse{\boolean{articletitles}}{\emph{The perceptron:
  A~probabilistic model for information storage and organization in the brain},
  }{}Psychological Review \textbf{65} (1958) 386\relax
\mciteBstWouldAddEndPuncttrue
\mciteSetBstMidEndSepPunct{\mcitedefaultmidpunct}
{\mcitedefaultendpunct}{\mcitedefaultseppunct}\relax
\EndOfBibitem
\bibitem{Zhong:2011xm}
J.-H. Zhong {\em et~al.}, \ifthenelse{\boolean{articletitles}}{\emph{{A program
  for the~Bayesian neural network in the~{\sc{ROOT}}~framework}},
  }{}\href{https://doi.org/10.1016/j.cpc.2011.07.019}{Comput.\ Phys.\ Commun.\
  \textbf{182} (2011) 2655},
  \href{http://arxiv.org/abs/1103.2854}{{\normalfont\ttfamily
  arXiv:1103.2854}}\relax
\mciteBstWouldAddEndPuncttrue
\mciteSetBstMidEndSepPunct{\mcitedefaultmidpunct}
{\mcitedefaultendpunct}{\mcitedefaultseppunct}\relax
\EndOfBibitem
\bibitem{chopping}
S.~Geisser, {\em Predictive inference: An introduction},
  \href{https://doi.org/10.1201/9780203742310}{ Chapman and Hall/CRC, New York,
  1993}\relax
\mciteBstWouldAddEndPuncttrue
\mciteSetBstMidEndSepPunct{\mcitedefaultmidpunct}
{\mcitedefaultendpunct}{\mcitedefaultseppunct}\relax
\EndOfBibitem
\bibitem{Skwarnicki:1986xj}
T.~Skwarnicki, {\em {A study of the radiative cascade transitions between
  the~$\PUpsilon^\prime$ and~$\PUpsilon$~resonances}}, PhD thesis, Institute of
  Nuclear Physics, Krakow, 1986,
  {\href{http://inspirehep.net/record/230779/}{DESY-F31-86-02}}\relax
\mciteBstWouldAddEndPuncttrue
\mciteSetBstMidEndSepPunct{\mcitedefaultmidpunct}
{\mcitedefaultendpunct}{\mcitedefaultseppunct}\relax
\EndOfBibitem
\bibitem{LHCb-PAPER-2011-013}
LHCb collaboration, R.~Aaij {\em et~al.},
  \ifthenelse{\boolean{articletitles}}{\emph{{Observation of \jpsi-pair
  production in \proton\proton collisions at \mbox{$\sqs=$7\tev}}},
  }{}\href{https://doi.org/10.1016/j.physletb.2011.12.015}{Phys.\ Lett.\
  \textbf{B707} (2012) 52},
  \href{http://arxiv.org/abs/1109.0963}{{\normalfont\ttfamily
  arXiv:1109.0963}}\relax
\mciteBstWouldAddEndPuncttrue
\mciteSetBstMidEndSepPunct{\mcitedefaultmidpunct}
{\mcitedefaultendpunct}{\mcitedefaultseppunct}\relax
\EndOfBibitem
\bibitem{fechner1897}
G.~T. Fechner, {\em {Kollektivmasslehre}}, {Wilhelm Engelmann, Leipzig}, 1897.
\newblock {Published posthumously, completed and edited by G.~F.~Lipps.}\relax
\mciteBstWouldAddEndPunctfalse
\mciteSetBstMidEndSepPunct{\mcitedefaultmidpunct}
{}{\mcitedefaultseppunct}\relax
\EndOfBibitem
\bibitem{Gibbons}
J.~F. Gibbons and S.~Mylroie,
  \ifthenelse{\boolean{articletitles}}{\emph{{Estimation of impurity profiles
  in ion\nobreakdash‐implanted amorphous targets using joined
  half\nobreakdash‐Gaussian distributions}},
  }{}\href{https://doi.org/10.1063/1.1654511}{Applied Phys.\ Lett.\
  \textbf{22} (1973) 568–569}\relax
\mciteBstWouldAddEndPuncttrue
\mciteSetBstMidEndSepPunct{\mcitedefaultmidpunct}
{\mcitedefaultendpunct}{\mcitedefaultseppunct}\relax
\EndOfBibitem
\bibitem{JohnS}
S.~John, \ifthenelse{\boolean{articletitles}}{\emph{{The three-parameter
  two-piece normal family of distributions and its fitting}},
  }{}\href{https://doi.org/10.1080/03610928208828279}{{Communications in
  Statistics} \textbf{11} (1982) 879}\relax
\mciteBstWouldAddEndPuncttrue
\mciteSetBstMidEndSepPunct{\mcitedefaultmidpunct}
{\mcitedefaultendpunct}{\mcitedefaultseppunct}\relax
\EndOfBibitem
\bibitem{LHCb-PAPER-2021-034}
LHCb collaboration, R.~Aaij {\em et~al.},
  \ifthenelse{\boolean{articletitles}}{\emph{{Study of $\Bc$~decays to
  charmonia and three light hadrons}},
  }{}\href{https://doi.org/10.1007/JHEP01(2022)065}{JHEP \textbf{01} (2022)
  065}, \href{http://arxiv.org/abs/2111.03001}{{\normalfont\ttfamily
  arXiv:2111.03001}}\relax
\mciteBstWouldAddEndPuncttrue
\mciteSetBstMidEndSepPunct{\mcitedefaultmidpunct}
{\mcitedefaultendpunct}{\mcitedefaultseppunct}\relax
\EndOfBibitem
\bibitem{LHCb-PAPER-2022-025}
LHCb collaboration, R.~Aaij {\em et~al.},
  \ifthenelse{\boolean{articletitles}}{\emph{{Study of $\Bc$ decays to
  charmonia plus multihadron final states}},
  }{}\href{https://doi.org/10.1007/JHEP07(2023)198}{JHEP \textbf{07} (2023)
  198}, \href{http://arxiv.org/abs/2208.08660}{{\normalfont\ttfamily
  arXiv:2208.08660}}\relax
\mciteBstWouldAddEndPuncttrue
\mciteSetBstMidEndSepPunct{\mcitedefaultmidpunct}
{\mcitedefaultendpunct}{\mcitedefaultseppunct}\relax
\EndOfBibitem
\bibitem{LHCb-PAPER-2023-039}
LHCb collaboration, R.~Aaij {\em et~al.},
  \ifthenelse{\boolean{articletitles}}{\emph{{Study of $\Bcp \to \chic \pip$
  decays}}, }{}\href{https://doi.org/10.1007/JHEP02(2024)173}{JHEP
  \textbf{2024} (2024) 173},
  \href{http://arxiv.org/abs/2312.12987}{{\normalfont\ttfamily
  arXiv:2312.12987}}\relax
\mciteBstWouldAddEndPuncttrue
\mciteSetBstMidEndSepPunct{\mcitedefaultmidpunct}
{\mcitedefaultendpunct}{\mcitedefaultseppunct}\relax
\EndOfBibitem
\bibitem{LHCb-PAPER-2023-046}
LHCb collaboration, R.~Aaij {\em et~al.},
  \ifthenelse{\boolean{articletitles}}{\emph{{Observation of the $\Bcp \to
  \jpsi \pip \piz$ decay}},
  }{}\href{https://doi.org/10.1007/JHEP04(2024)151}{JHEP \textbf{04} (2024)
  151}, \href{http://arxiv.org/abs/2402.05523}{{\normalfont\ttfamily
  arXiv:2402.05523}}\relax
\mciteBstWouldAddEndPuncttrue
\mciteSetBstMidEndSepPunct{\mcitedefaultmidpunct}
{\mcitedefaultendpunct}{\mcitedefaultseppunct}\relax
\EndOfBibitem
\bibitem{Wilks:1938dza}
S.~S. Wilks, \ifthenelse{\boolean{articletitles}}{\emph{{The large-sample
  distribution of the likelihood ratio for testing composite hypotheses}},
  }{}\href{https://doi.org/10.1214/aoms/1177732360}{Ann.\ Math.\ Stat.\
  \textbf{9} (1938) 60}\relax
\mciteBstWouldAddEndPuncttrue
\mciteSetBstMidEndSepPunct{\mcitedefaultmidpunct}
{\mcitedefaultendpunct}{\mcitedefaultseppunct}\relax
\EndOfBibitem
\bibitem{LHCb-TDR-001}
LHCb collaboration, \ifthenelse{\boolean{articletitles}}{\emph{{LHCb magnet:
  Technical Design Report}}, }{}
  \href{http://cdsweb.cern.ch/search?p=CERN-LHCC-2000-007&f=reportnumber&action_search=Search&c=LHCb}
  {CERN-LHCC-2000-007}, 2000\relax
\mciteBstWouldAddEndPuncttrue
\mciteSetBstMidEndSepPunct{\mcitedefaultmidpunct}
{\mcitedefaultendpunct}{\mcitedefaultseppunct}\relax
\EndOfBibitem
\bibitem{MartinezSantos:2013ltf}
D.~Mart\'\i{}nez~Santos and F.~Dupertuis,
  \ifthenelse{\boolean{articletitles}}{\emph{{Mass distributions marginalized
  over per-event errors}},
  }{}\href{https://doi.org/10.1016/j.nima.2014.06.081}{Nucl.\ Instrum.\ Meth.\
  \textbf{A764} (2014) 150},
  \href{http://arxiv.org/abs/1312.5000}{{\normalfont\ttfamily
  arXiv:1312.5000}}\relax
\mciteBstWouldAddEndPuncttrue
\mciteSetBstMidEndSepPunct{\mcitedefaultmidpunct}
{\mcitedefaultendpunct}{\mcitedefaultseppunct}\relax
\EndOfBibitem
\bibitem{Student}
{Student~(W.\ \,S.\ ~Gosset)},
  \ifthenelse{\boolean{articletitles}}{\emph{{The~probable error of a~mean}},
  }{}\href{https://doi.org/10.1093/biomet/6.1.1}{Biometrika \textbf{6} (1908)
  1}\relax
\mciteBstWouldAddEndPuncttrue
\mciteSetBstMidEndSepPunct{\mcitedefaultmidpunct}
{\mcitedefaultendpunct}{\mcitedefaultseppunct}\relax
\EndOfBibitem
\bibitem{Jackman}
S.~Jackman, {\em Bayesian analysis for the social sciences}, John Wiley \&
  Sons, Inc., Hoboken, New Jersey, USA, 2009\relax
\mciteBstWouldAddEndPuncttrue
\mciteSetBstMidEndSepPunct{\mcitedefaultmidpunct}
{\mcitedefaultendpunct}{\mcitedefaultseppunct}\relax
\EndOfBibitem
\bibitem{ZHU2010297}
D.~Zhu and J.~W. Galbraith, \ifthenelse{\boolean{articletitles}}{\emph{{A
  generalized asymmetric Student-t distribution with application to financial
  econometrics}},
  }{}\href{https://doi.org/https://doi.org/10.1016/j.jeconom.2010.01.013}{Journal
  of Econometrics \textbf{157} (2010) 297}\relax
\mciteBstWouldAddEndPuncttrue
\mciteSetBstMidEndSepPunct{\mcitedefaultmidpunct}
{\mcitedefaultendpunct}{\mcitedefaultseppunct}\relax
\EndOfBibitem
\bibitem{LiNadarjah}
R.~Li and S.~Nadarajah, \ifthenelse{\boolean{articletitles}}{\emph{{ A review
  of Student’s $t$\nobreakdash-distribution and its generalizations}},
  }{}\href{https://doi.org/10.1007/s00181-018-1570-0}{Empir.\ Econ.\
  \textbf{48} (2020) 1461}\relax
\mciteBstWouldAddEndPuncttrue
\mciteSetBstMidEndSepPunct{\mcitedefaultmidpunct}
{\mcitedefaultendpunct}{\mcitedefaultseppunct}\relax
\EndOfBibitem
\bibitem{BaBar:2011qjw}
BaBar collaboration, J.~P. Lees {\em et~al.},
  \ifthenelse{\boolean{articletitles}}{\emph{{Branching fraction measurements
  of the color-suppressed decays $\bar{\PB}^0 \to \PD^{(*)0} \Ppi^0$,
  $\PD^{(*)0} \Peta$, $\PD^{(*)0} \Pomega$, and $\PD^{(*)0} \Peta^\prime$ and
  measurement of the polarization in the decay $\bar{\PB}^0 \to \PD^{*0}
  \Pomega$}}, }{}\href{https://doi.org/10.1103/PhysRevD.84.112007}{Phys.\ Rev.\
   \textbf{D84} (2011) 112007}, Erratum
  \href{https://doi.org/10.1103/PhysRevD.87.039901}{ibid.\   \textbf{D87}
  (2013) 039901}, \href{http://arxiv.org/abs/1107.5751}{{\normalfont\ttfamily
  arXiv:1107.5751}}\relax
\mciteBstWouldAddEndPuncttrue
\mciteSetBstMidEndSepPunct{\mcitedefaultmidpunct}
{\mcitedefaultendpunct}{\mcitedefaultseppunct}\relax
\EndOfBibitem
\bibitem{efron:1979}
B.~Efron, \ifthenelse{\boolean{articletitles}}{\emph{Bootstrap methods: Another
  look at the jackknife},
  }{}\href{https://doi.org/10.1214/aos/1176344552}{Ann.\ Statist.\  \textbf{7}
  (1979) 1}\relax
\mciteBstWouldAddEndPuncttrue
\mciteSetBstMidEndSepPunct{\mcitedefaultmidpunct}
{\mcitedefaultendpunct}{\mcitedefaultseppunct}\relax
\EndOfBibitem
\bibitem{efron:1993}
B.~Efron and R.~J. Tibshirani, {\em An~introduction to the~bootstrap},
  \href{https://doi.org/10.1201/9780429246593}{ Chapman and Hall, New York,
  1994}\relax
\mciteBstWouldAddEndPuncttrue
\mciteSetBstMidEndSepPunct{\mcitedefaultmidpunct}
{\mcitedefaultendpunct}{\mcitedefaultseppunct}\relax
\EndOfBibitem
\bibitem{Poluektov:2014rxa}
A.~Poluektov, \ifthenelse{\boolean{articletitles}}{\emph{{Kernel density
  estimation of a~multidimensional efficiency profile}},
  }{}\href{https://doi.org/10.1088/1748-0221/10/02/P02011}{JINST \textbf{10}
  (2015) P02011}, \href{http://arxiv.org/abs/1411.5528}{{\normalfont\ttfamily
  arXiv:1411.5528}}\relax
\mciteBstWouldAddEndPuncttrue
\mciteSetBstMidEndSepPunct{\mcitedefaultmidpunct}
{\mcitedefaultendpunct}{\mcitedefaultseppunct}\relax
\EndOfBibitem
\bibitem{LHCb-PAPER-2012-010}
LHCb collaboration, R.~Aaij {\em et~al.},
  \ifthenelse{\boolean{articletitles}}{\emph{{Measurement of relative branching
  fractions of \B~decays to \psitwos and \jpsi mesons}},
  }{}\href{https://doi.org/10.1140/epjc/s10052-012-2118-7}{Eur.\ Phys.\ J.\
  \textbf{C72} (2012) 2118},
  \href{http://arxiv.org/abs/1205.0918}{{\normalfont\ttfamily
  arXiv:1205.0918}}\relax
\mciteBstWouldAddEndPuncttrue
\mciteSetBstMidEndSepPunct{\mcitedefaultmidpunct}
{\mcitedefaultendpunct}{\mcitedefaultseppunct}\relax
\EndOfBibitem
\bibitem{LYONS1988110}
L.~Lyons, D.~Gibaut, and P.~Clifford,
  \ifthenelse{\boolean{articletitles}}{\emph{{How to combine correlated
  estimates of a~single physical quantity}},
  }{}\href{https://doi.org/10.1016/0168-9002(88)90018-6}{Nucl.\ Instrum.\
  Meth.\  \textbf{A270} (1988) 110}\relax
\mciteBstWouldAddEndPuncttrue
\mciteSetBstMidEndSepPunct{\mcitedefaultmidpunct}
{\mcitedefaultendpunct}{\mcitedefaultseppunct}\relax
\EndOfBibitem
\end{mcitethebibliography}

\newpage
% LHCb collaboration author list
% Data extracted on June 21st, 2024 at 2:00pm for paper reference LHCb-PAPER-2024-015
\centerline
{\large\bf LHCb collaboration}
\begin
{flushleft}
\small
R.~Aaij$^{36}$\lhcborcid{0000-0003-0533-1952},
A.S.W.~Abdelmotteleb$^{55}$\lhcborcid{0000-0001-7905-0542},
C.~Abellan~Beteta$^{49}$,
F.~Abudin{\'e}n$^{55}$\lhcborcid{0000-0002-6737-3528},
T.~Ackernley$^{59}$\lhcborcid{0000-0002-5951-3498},
A. A. ~Adefisoye$^{67}$\lhcborcid{0000-0003-2448-1550},
B.~Adeva$^{45}$\lhcborcid{0000-0001-9756-3712},
M.~Adinolfi$^{53}$\lhcborcid{0000-0002-1326-1264},
P.~Adlarson$^{80}$\lhcborcid{0000-0001-6280-3851},
C.~Agapopoulou$^{13}$\lhcborcid{0000-0002-2368-0147}
% ...

{\footnotesize \it

% ...
$^{13}$Universit{\'e} Paris-Saclay, CNRS/IN2P3, IJCLab, Orsay, France\\
% ...
$^{36}$Nikhef National Institute for Subatomic Physics, Amsterdam, Netherlands\\
% ...
$^{45}$Instituto Galego de F{\'\i}sica de Altas Enerx{\'\i}as (IGFAE), Universidade de Santiago de Compostela, Santiago de Compostela, Spain\\
% ...
$^{49}$Physik-Institut, Universit{\"a}t Z{\"u}rich, Z{\"u}rich, Switzerland\\
% ...
$^{53}$H.H. Wills Physics Laboratory, University of Bristol, Bristol, United Kingdom\\
% ...
$^{55}$Department of Physics, University of Warwick, Coventry, United Kingdom\\
% ...
$^{58}$School of Physics and Astronomy, University of Glasgow, Glasgow, United Kingdom\\
$^{59}$Oliver Lodge Laboratory, University of Liverpool, Liverpool, United Kingdom\\
% ...
$^{67}$Syracuse University, Syracuse, NY, United States\\
% ...
$^{80}$Department of Physics and Astronomy, Uppsala University, Uppsala, Sweden, associated to $^{58}$\\
% ...
}
\end{flushleft}

\end{document}

%%%%%%%%%%%%%%%%%%%%%%%%%
%%%%% Title     %%%%%%%%%
%%%%%%%%%%%%%%%%%%%%%%%%%
\renewcommand{\thefootnote}{\fnsymbol{footnote}}
\setcounter{footnote}{1}

% %%%%%%% CHOOSE TITLE PAGE--------
%\onecolumn
%\input{title-LHCb-INT}
%\input{title-LHCb-ANA}
%\input{title-LHCb-CONF}
%\input{title-LHCb-FIGURE}
%\input{title-LHCb-PAPER}
%\twocolumn
% %%%%%%%%%%%%% ---------

%\listoftodos

\renewcommand{\thefootnote}{\arabic{footnote}}
\setcounter{footnote}{0}

%%%%%%%%%%%%%%%%%%%%%%%%%%%%%%%%
%%%%%  Table of Content   %%%%%%
%%%%%%%%%%%%%%%%%%%%%%%%%%%%%%%%
%%%% Uncomment if desired
%\tableofcontents
\cleardoublepage

%%%%%%%%%%%%%%%%%%%%%%%%%
%%%%% Main text %%%%%%%%%
%%%%%%%%%%%%%%%%%%%%%%%%%

\pagestyle{plain} % restore page numbers for the main text
\setcounter{page}{1}
\pagenumbering{arabic}

%% Uncomment during review phase. 
%% Comment before a final submission.
%% \linenumbers

\section*{Supplementary material for LHCb-PAPER-2024-157}
%% \label{sec:supp}

 \renewcommand{\thetable}{S\arabic{table}}
 \renewcommand{\thefigure}{S\arabic{figure}}
 \renewcommand{\theequation}{S\arabic{equation}}
 \setcounter{figure}{0}
 \setcounter{table}{0}
 \setcounter{equation}{0}

% Yhis appendix contains supplementary material that will be posted
%% on the public CDS record but will not appear in the paper.

\begin{comment}
A summary of experimental results 
 and theoretical predictions for 
 the~ratio of the partial decay widths 
 for
 the~radiative  $\chicone(3872)$~decays 
%%   \mbox{$\decay{\chicone(3872)}{\psitwos\g}$}
%%  and \mbox{$\decay{\chicone(3872)}{\jpsi\g}$}~decays
is shown in Figs.~\ref{fig:compare_lin} 
and~\ref{fig:compare_log}.
\end{comment}

\begin{figure}[h]
\setlength{\unitlength}{1mm}
\centering
\begin{picture}(150,180)
%%
%% \graphpaper[5](-10,-10)(170,200)
\definecolor{gr}{rgb}{0.35, 0.83, 0.33}
\put(-5,0){
\includegraphics*[width=65mm,height=180mm]{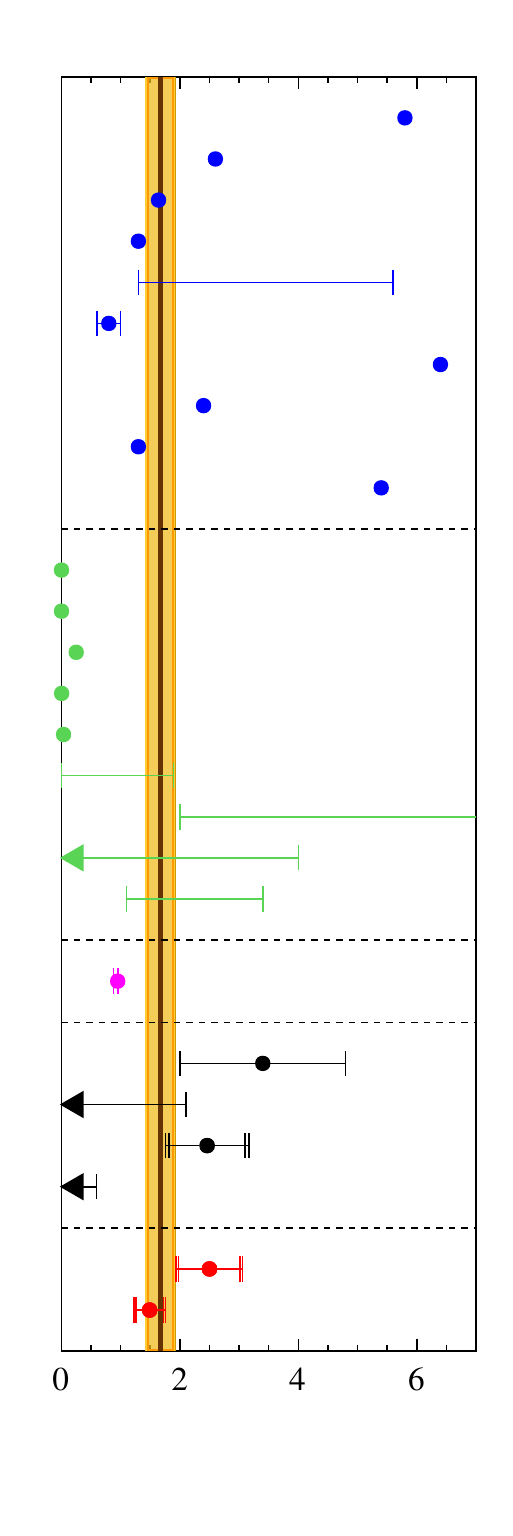} 
}
%
%% \put(58,98){\scriptsize\begin{Tabular}[1.45]{lll} 
\put(58,98){\scriptsize\begingroup\renewcommand*{\arraystretch}{1.45}\begin{tabular}{lll} 
\\
 \multicolumn{2}{l}{
T.~Barnes and  S.~Godfrey %% & 2003  
 } 
& \cite{Barnes:2003vb}   
\\
\multicolumn{2}{l}{
T.~Barnes, S.~Godfrey and S.~Swanson 
}
& \cite{Barnes:2005pb}
\\
\multicolumn{2}{l}{
F.~De Fazio 
}
& \cite{DeFazio:2008xq}
\\
\multicolumn{2}{l}{
B.-Q.~Li and K.~T.~Chao 
}
& \cite{Li:2009zu}
\\
\multicolumn{2}{l}{
Y.~Dong {\it{et al.}} 
}
& \cite{Dong:2009uf}
\\
\multicolumn{2}{l}{
A.~M.~Badalian {\it{et al.}} 
}
& \cite{Badalian:2012jz}  
\\
\multicolumn{2}{l}{
J.~Ferretti, G.~Galata and E.~Santopinto
} 
& \cite{Ferretti:2014xqa}
\\
\multicolumn{2}{l}{
A.~M.~Badalian, Yu.~A.~Simonov and 
B.~L.~G.~Bakker
}
& \cite{Badalian:2015dha}  
\\
\multicolumn{2}{l}{
W.~J.~Deng {\it{et al.}} 
}
& \cite{Deng:2016stx}
\\
\multicolumn{2}{l}{
F.~Giacosa, M.~Piotrowska and S. Goito
}
& \cite{Giacosa:2019zxw}
\\ 
& & 
\\
\multicolumn{2}{l}{
E.~S.~Swanson
} 
& \cite{Swanson:2004pp}
\\
\multicolumn{2}{l}{
Y.~Dong {\it{et al.}}
} 
& \cite{Dong:2009uf}
\\
\multicolumn{2}{l}{
D.~P.~Rathaud and  A.~K.~Rai 
} 
& \cite{Rathaud:2016tys}
\\
\multicolumn{2}{l}{
R.~F.~Lebed and S.~R.~Martinez 
} 
& \cite{Lebed:2022vks}
\\
\multicolumn{2}{l}{
B.~Grinstein, L.~Maiani and A.~D.~Polosa 
}
& \cite{Grinstein:2024rcu}
\\
\multicolumn{2}{l}{
F.-K.~Guo {\it{et al.}}  
}
& \cite{Guo:2014taa}
\\ 
\multicolumn{2}{l}{
D.~A.-S.~Molnar, R.~F.~Luiz and  R.~Higa  
}
& \cite{Molnar:2016dbo}
\\
\multicolumn{2}{l}{
E.~Cincioglu {\it{et al.}} 
}
& \cite{Cincioglu:2016fkm}
\\
\multicolumn{2}{l}{
S.~Takeuchi, M.~Takizawa and K.~Shimizu  
}
& \cite{Takeuchi:2016hat}
\\
& & 
\\ 
\multicolumn{2}{l}{
B.~Grinstein, L.~Maiani and A.~D.~Polosa 
}
& \cite{Grinstein:2024rcu} 
\\
& & 
\\
    \babar        & 2008        &  \cite{BaBar:2008flx}        \\
    \belle        & 2011        &  \cite{Belle:2011wdj}        \\
    \lhcb/Run\,1  & 2014        &  \cite{LHCb-PAPER-2014-008}  \\ 
    \besiii       & 2020        &  \cite{BESIII:2020nbj}       \\
    & & \\
    \lhcb/Run\,1  & 2024        &   \\
    \lhcb/Run\,2  & 2024        & 
\end{tabular}\endgroup}
   \put( 6,8){\large$\mathscr{R}_{\Ppsi\g}=\dfrac{\Gamma_{\decay{\chicone(3872)}{\psitwos\g}}}
                           {\Gamma_{\decay{\chicone(3872)}{\jpsi\g}}}$}

 \put( 6,165){\large{\color{blue}{$\cquark\cquarkbar$}}}
 \put(45,111){\large{\color{gr}$\D\Dstarb$}}
 \put(43, 64){\large{\color{magenta}{$\cquark\cquarkbar\quark\quarkbar$}}}
\end{picture}
\caption{ \small
 Summary of experimental results 
 and theoretical predictions for 
 the~ratio of the partial decay widths 
 for
 the~radiative  \mbox{$\decay{\chicone(3872)}{\psitwos\g}$}
 and \mbox{$\decay{\chicone(3872)}{\jpsi\g}$}~decays.
 The~results from this analysis
 for Run\,1 and Run\,2~data sets
 are shown with red points with error bars,
 and the~coloured band corresponds to average of  
 these LHCb results.
  The~inner error bars (and the~band)
 indicate the~statistical 
 uncertainty whilst 
 the~outer error bars (and the~band)
 show  
 the~sum of 
 the~statistical and systematic uncertainties in quadrature. 
 Previous experimental 
 results are shown in black.
 Predictions from 
 $\cquark\cquarkbar$~charmonium, 
 $\D\Dstarb$~molecular  
 and compact tetraquark models 
 are shown in blue, green  and magenta, 
 respectively. 
 }
\label{fig:compare_lin}
\end{figure}

\begin{figure}[t]
\setlength{\unitlength}{1mm}
\centering
\begin{picture}(150,180)
%%
%% \graphpaper[5](-10,-10)(170,200)
\definecolor{gr}{rgb}{0.35, 0.83, 0.33}
\put(-5,0){
\includegraphics*[width=65mm,height=180mm]{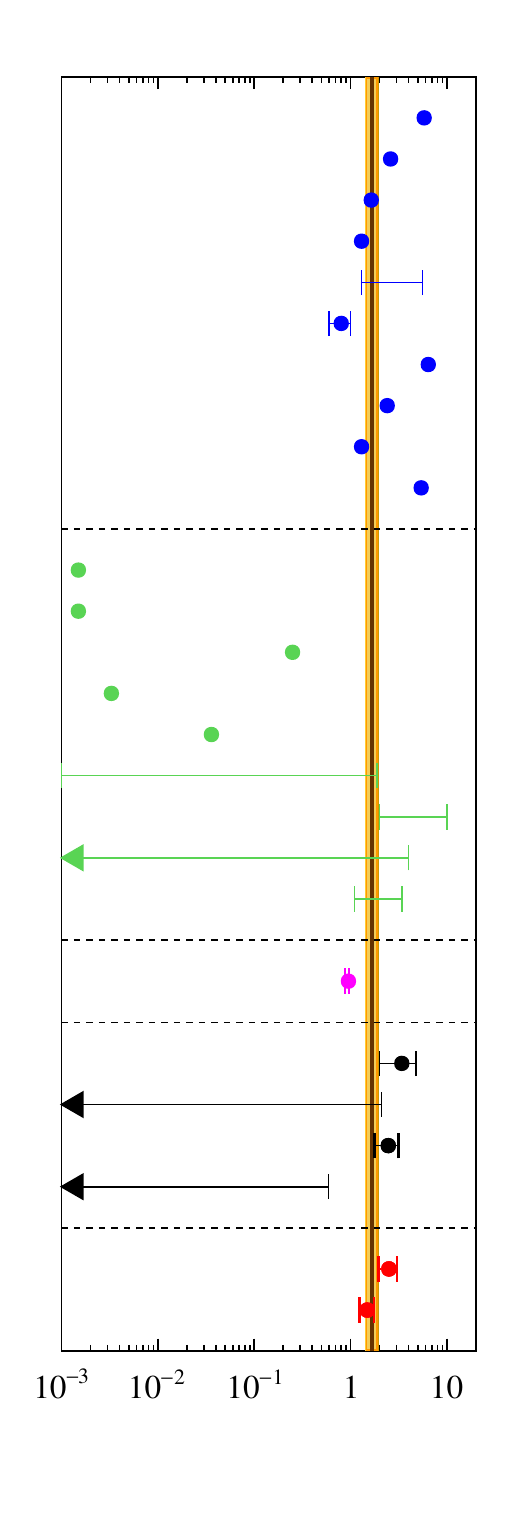} 
}
%
%% \put(58,98){\scriptsize\begin{Tabular}[1.45]{lll} 
\put(58,98){\scriptsize\begingroup\renewcommand*{\arraystretch}{1.45}\begin{tabular}{lll} 
\\
\multicolumn{2}{l}{
T.~Barnes and  S.~Godfrey 
} 
& \cite{Barnes:2003vb}   
\\
\multicolumn{2}{l}{
T.~Barnes, S.~Godfrey and S.~Swanson 
}
& \cite{Barnes:2005pb}
\\
\multicolumn{2}{l}{
F.~De Fazio 
}
& \cite{DeFazio:2008xq}
\\
\multicolumn{2}{l}{
B.-Q.~Li and K.~T.~Chao 
}
& \cite{Li:2009zu}
\\
\multicolumn{2}{l}{
Y.~Dong {\it{et al.}} 
}
& \cite{Dong:2009uf}
\\
\multicolumn{2}{l}{
A.~M.~Badalian {\it{et al.}} 
}
& \cite{Badalian:2012jz}  
\\
\multicolumn{2}{l}{
J.~Ferretti, G.~Galata and E.~Santopinto
} 
& \cite{Ferretti:2014xqa}
\\
\multicolumn{2}{l}{
A.~M.~Badalian, Yu.~A.~Simonov and 
B.~L.~G.~Bakker
}
& \cite{Badalian:2015dha}  
\\
\multicolumn{2}{l}{
W.~J.~Deng {\it{et al.}} 
}
& \cite{Deng:2016stx}
\\
\multicolumn{2}{l}{
F.~Giacosa, M.~Piotrowska and S. Goito
}
& \cite{Giacosa:2019zxw}
\\ 
& & 
\\
\multicolumn{2}{l}{
E.~S.~Swanson
} 
& \cite{Swanson:2004pp}
\\
\multicolumn{2}{l}{
Y.~Dong {\it{et al.}}
} 
& \cite{Dong:2009uf}
\\
\multicolumn{2}{l}{
D.~P.~Rathaud and  A.~K.~Rai 
} 
& \cite{Rathaud:2016tys}
\\
\multicolumn{2}{l}{
R.~F.~Lebed and S.~R.~Martinez 
} 
& \cite{Lebed:2022vks}
\\
\multicolumn{2}{l}{
B.~Grinstein, L.~Maiani and A.~D.~Polosa 
}
& \cite{Grinstein:2024rcu}
\\
\multicolumn{2}{l}{
F.-K.~Guo {\it{et al.}}  
}
& \cite{Guo:2014taa}
\\ 
\multicolumn{2}{l}{
D.~A.-S.~Molnar, R.~F.~Luiz and  R.~Higa  
}
& \cite{Molnar:2016dbo}
\\
\multicolumn{2}{l}{
E.~Cincioglu {\it{et al.}} 
}
& \cite{Cincioglu:2016fkm}
\\
\multicolumn{2}{l}{
S.~Takeuchi, M.~Takizawa and K.~Shimizu  
}
& \cite{Takeuchi:2016hat}
\\
& & 
\\ 
\multicolumn{2}{l}{
B.~Grinstein, L.~Maiani and A.~D.~Polosa 
}
& \cite{Grinstein:2024rcu} 
\\
& & 
\\
    \babar        & 2008        &  \cite{BaBar:2008flx}        \\
    \belle        & 2011        &  \cite{Belle:2011wdj}        \\
    \lhcb/Run\,1  & 2014        &  \cite{LHCb-PAPER-2014-008}  \\ 
    \besiii       & 2020        &  \cite{BESIII:2020nbj}       \\
    & & \\
    \lhcb/Run\,1  & 2024        &   \\
    \lhcb/Run\,2  & 2024        & 
\end{tabular}\endgroup}
   \put( 6,8){\large$\mathscr{R}_{\Ppsi\g}=\dfrac{\Gamma_{\decay{\chicone(3872)}{\psitwos\g}}}
                           {\Gamma_{\decay{\chicone(3872)}{\jpsi\g}}}$}
  
 \put( 6,165){\large{\color{blue}{$\cquark\cquarkbar$}}}
 \put(45,111){\large{\color{gr}$\D\Dstarb$}}
 \put( 6, 64){\large{\color{magenta}{$\cquark\cquarkbar\quark\quarkbar$}}}
%%%% 
\end{picture}
\caption{ \small
 Summary of experimental results 
 and theoretical predictions for 
 the~ratio of the partial decay widths 
 for
 the~radiative  \mbox{$\decay{\chicone(3872)}{\psitwos\g}$}
 and \mbox{$\decay{\chicone(3872)}{\jpsi\g}$}~decays.
 The~results from this analysis
 for Run\,1 and Run\,2~data sets
 are shown with red points with error bars,
 and the~coloured band corresponds to average of  
 these LHCb results.
  The~inner error bars (and the~band)
 indicate the~statistical 
 uncertainty whilst 
 the~outer error bars (and the~band)
 show  
 the~sum of 
 the~statistical and systematic uncertainties in quadrature. 
 Previous experimental 
 results are shown in black.
 Predictions from 
  $\cquark\cquarkbar$~charmonium, 
  $\D\Dstarb$~molecular  
   and compact tetraquark models 
 are shown in blue, green  and magenta, 
 respectively. 
 }
\label{fig:compare_log}
\end{figure}

\clearpage 
\addcontentsline{toc}{section}{References}
%\setboolean{inbibliography}{true}
\bibliographystyle{LHCb}
\bibliography{main,standard,LHCb-PAPER,LHCb-CONF,LHCb-DP,LHCb-TDR}